\documentclass[acmsmall,screen,nonacm]{acmart}
\usepackage{color, colortbl}
\definecolor{Gray}{gray}{0.9}

\definecolor{ui1}{HTML}{c26f6d}
\definecolor{ui2}{HTML}{d9a6a4}
\definecolor{ui3}{HTML}{f2dfdd}
\definecolor{ui4}{HTML}{e4e5eb}
\definecolor{ui5}{HTML}{aab8d0}
\definecolor{ui6}{HTML}{6e90bf}

\usepackage{subcaption}
\usepackage{float}
\usepackage{graphicx}
\usepackage{wrapfig}
\usepackage{enumitem}
\usepackage{xr}
\makeatletter
\newcommand{\setword}[2]{#1\def\@currentlabel{\unexpanded{#1}}\label{#2}}
\makeatother

\usepackage[yyyymmdd]{datetime}

\usepackage{multirow}
\usepackage{tabularx}

\usepackage{dcolumn}
\newcolumntype{d}[1]{D{.}{.}{#1}}
\makeatletter
\newcolumntype{B}[3]{>{\boldmath\DC@{#1}{#2}{#3}}c<{\DC@end}}

\usepackage{todonotes}
\let\xtodo\todo
\renewcommand{\todo}[1]{\xtodo[inline,color=green!50]{TODO: #1}}

\usepackage[yyyymmdd]{datetime}

\setcopyright{acmlicensed}
\copyrightyear{2023}
\acmYear{2023}
\acmDOI{XXXXXXX.XXXXXXX}
\acmJournal{TOCHI}
\acmVolume{1}
\acmNumber{1}
\acmMonth{1}

\begin{document}
\title{Rethinking Opinion Measurement Interfaces for Human-in-the-Loop Optimization}
\titlenote{First author's note (2026): This paper was written in 2022 and is published here as last revised. Its central claim, that measuring human preferences is an interface design problem before it is a modeling problem, now matters far beyond human-in-the-loop optimization, including reinforcement learning from human feedback.}

\author{Changkun Ou}
\orcid{0000-0002-4595-7485}
\affiliation{
  \institution{LMU Munich}
  \country{Germany}
}
\email{research@changkun.de}

\author{Sven Mayer}
\orcid{0000-0001-5462-8782}
\affiliation{
  \institution{LMU Munich}
  \country{Germany}
}
\email{info@sven-mayer.com}

\author{Daniel Buschek}
\orcid{0000-0002-0013-715X}
\affiliation{
  \institution{University of Bayreuth}
  \country{Germany}
}
\email{daniel.buschek@uni-bayreuth.de}

\author{Andreas Butz}
\orcid{0000-0002-9007-9888}
\affiliation{
  \institution{LMU Munich}
  \country{Germany}
}
\email{butz@ifi.lmu.de}

\renewcommand{\shortauthors}{Ou et al.}

\begin{teaserfigure}
\centering
\includegraphics[width=0.97\columnwidth]{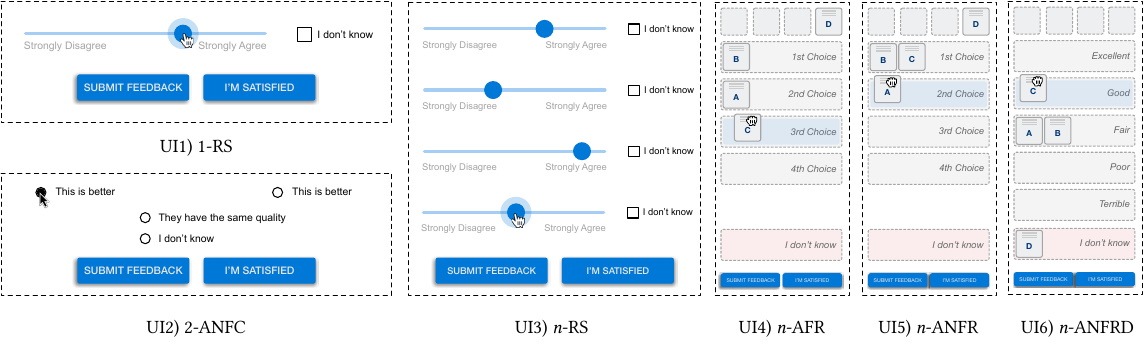}
\caption{
Different opinion measurement interfaces: UI1)~Single item Rating Scale~(1-RS); UI2)~Two Alternative Non-Forced Choice~(2-ANFC); UI3)~$n$-items Rating Scales~($n$-RS); UI4)~$n$-Alternative Forced Ranking~($n$-AFR); UI5)~$n$-Alternative Non-Forced Ranking~($n$-ANFR); UI6)~$n$-Alternative Non-Forced Ranking with Distance~($n$-ANFRD). The 1-RS and $n$-RS interfaces are rating scale-based measurements, and the 2-ANFC, $n$-AFR, and $n$-ANFR are preference-based ranking interfaces. As a combination, the $n$-ANFRD mixes absolute ratings and preferential rankings, which permits expressing not only ranking orders but also local ordinal ranking distance. All interfaces in this gallery also permit users to express their incomplete preference through ``I don't know,'' and to signal when a satisfactory result is achieved through ``I'm satisfied.''
}\label{fig:teaser}
\end{teaserfigure}

\begin{abstract}
Human-in-the-loop optimization refers to systems that present choices to a human decision-maker and query their judgment to optimize system parameters iteratively. The user interface typically provides either an absolute rating scale or asks for a preferred option. However, these approaches can potentially suffer from calibration problems, inconsistencies, and completeness violations of user preferences in such a sequential evaluation. This paper revisits the design space of opinion measurement interfaces for human-in-the-loop optimization. We conducted a study (N=360) using text summarization and image color enhancement tasks to evaluate human sequential feedback in six representative opinion measurement interfaces. Based on our analysis, we recommend using a listwise approach to support sequential opinion measurement, which counterweighs individual rating and preferential choice limitations and conveys more information for subsequent use. We further discuss the trade-offs in different user interface designs and provide guidelines to inform future research.
\end{abstract}


\maketitle

\section{Introduction}
\label{sec:intro}
Human-in-the-loop~\cite{monarch2021hitl} (HITL) optimization has received increasing attention in HCI~\cite{zhou2021interactive, chan2022chi}. Such systems query decision makers for their judgment among machine proposals, e.g., using \emph{Bayesian optimization} (BO)~\cite{kushner1964bo, shahriari2016bo-review} as the background AI and then optimize underlying system parameters toward user preferences according to the feedback. One of the benefits is that this adaptation can support users to explore a design space~\cite{koyama2020sequential, koyama2021ui, chan2022chi} defined by finite criteria or retain more control over pre-configured system decision strategies~\cite{russell2019human}.
However, the \emph{opinion measurement interface}~\cite{nobarany2012opinion} used as input component in HITL optimization has many design parameters itself. Depending on the types of human feedback required, the whole process has different limitations due to the design of the input interface.
For example, a commonly used \emph{pointwise} user interface (UI) asks the user to rate a given option as \emph{utility-based direct feedback}, see \autoref{fig:examples}a. Although, literature has shown that humans are not good at absolute judgments~\cite{tversky1974judgment} and tend to give local judgments~\cite{brochu2007active}. Therefore, a rating interface could suffer from miscalibration and noisy human feedback, thus reducing human-AI cooperation performance.
Another popular \emph{pairwise} design~\cite{brochu2007active, christiano2017deep} is to ask humans which of the two given system outcomes is preferred (\emph{preference-based indirect feedback}). However, indirect preferential feedback is less efficient in measuring human opinions and requires querying more preferential feedback to model a latent preference function~\cite{gonzalez2017pbo, siivola2021preferential, nguyen2021topk}. Additionally, it might suffer from interpretation ambiguity, intransitive and circular preferences (e.g., prefer A over B, and prefer B over C, but prefer C over A, known as \emph{rock-paper-scissors relation}), which also presents a challenge for input precision and overall system performance.

To overcome the limited number of evaluations at a time and collect more feedback, one can extend pointwise ($n$=1) and pairwise ($n$=2) feedback to listwise ranking~\cite{cao2007learning, siivola2021preferential} ($n>2$), where $n$ is the number of alternative options. However, many input interface design variations depend on the queried feedback type. Specifically, for direct rating feedback, the listwise approach needs to use $n$ different rating scales; for indirect preferential feedback, the listwise approach requires a UI to allow users to rank the given outcome alternatives. Despite different queried feedback types, a listwise interface always implicitly facilitates comparisons, as it presents more options simultaneously.
Recently, \citet{ou2023expertise} presented a new approach to overcome current limitations using a $n$-alternative ranking approach which provides a ``I don't know'' option for signal items. However, the effectiveness is not yet understood. Thus, the main objective of this paper is to understand the following two questions:\vspace{4pt}

\noindent\textbf{RQ1}~\emph{What is the impact of using different opinion measurement interfaces for HITL optimization?}

\noindent\textbf{RQ2}~\emph{What is the best-suited interface to use in a HITL optimization to leverage human opinions?}\vspace{4pt}

Due to the many design choices influencing opinion measurement interfaces, we first derived the design space for opinion measurement interfaces on HITL optimization. We identified six design variations, including \textit{pointwise}, \textit{pairwise}, and \textit{listwise} approaches depending on the type of feedback queried. They can also be varied to support expressing indifference and incomplete preferences. Based on our design space, we conducted an online between-groups study (N=360) to understand the impact of six opinion measurement interfaces on HITL optimization. We selected a \emph{text summarization} task and an \emph{image color enhancement} task to evaluate the different opinion measurement interfaces in two domains. In both tasks, humans express their feedback using different user interfaces regarding the quality of text summarization or color enhancement. 

As a key result, we recommend using a listwise approach and allowing users to express ranking distance to support sequential opinion measurement. This design counterweighs individual ratings and preferential limitations and conveys more direct and effective information for subsequent usage in HITL optimization.
The main contribution of this paper is an empirical investigation of how the design of opinion measurement interfaces can influence HITL optimization. We will also discuss design implications and trade-offs for future HITL optimization research and applications.

\section{Related Work}
In this section, we will discuss the assumptions made when measuring human opinions and how they relate to actual human decisions and preferences. Then, we will discuss how these assumptions could impact modeling human feedback with BO. Lastly, we will revisit the opinion measurement interfaces commonly used today in non-sequential measuring contexts and how existing HITL optimization systems design their opinion measurement interfaces.

\subsection{Assumptions in Human Preferences and Decisions}
Behaviorism argues that objective decisions and judgments of a human can be observed through their behavior. However, this observation may be limited by the ambiguity of interpretation~\cite{skinner2011about} due to a lack of observation and impermeable states of mind~\cite{overgaard2006the}. Opinion measurement is an important complement to behaviorism~\cite{nobarany2012opinion}. An opinion measurement UI measures ``tendencies to evaluate an entity with some degree of favor or disfavor, ordinarily expressed in cognitive, affective, and behavioral responses''~\cite{eagly1993the}, which reflect human preferences or decisions under uncertainty.

Traditional behavioral economics has widely studied rational choice~\cite{dreier1996rational}, and the corresponding algorithmic modeling to infer preferences relied on the \emph{transitivity} and \emph{completeness} assumptions. The formulation of transitivity includes two aspects: strict and weak. Strict transitivity states that one choice option is strictly preferred over another, while weak transitivity allows \emph{indifference} between given choice options.
Transitivity as a modeling assumption assumes that people can give consistent ranks to given choice options. However, this assumption may be violated mainly in a sequential judgment context~\cite{debreu1960individual, bettman1998constructive} since a previous preference may be inverted later in the process and result in a rock-paper-scissors relation due to context changes and cognitive biases~\cite{tversky1974judgment}.

Moreover, completeness assumes that people always have a preference, but this does not hold for decision-making under uncertain information, where in some cases, the preference may be undecided~\cite{ok2012incomplete}. Identifying the violation of the completeness assumption is more fundamental and complex by definition, as it is harder to define from a behavior perspective.
Consequently, violating these assumptions can be framed as preference or decision uncertainty, and it is often strongly assumed that an inverted choice constitutes a noisy observation (e.g.,~\cite{gonzalez2017pbo, lin2022preference, pfannschmidt2022learning, benavoli2021choice}) without a detailed inspection from the human perspective.

\subsection{Modeling Human Feedback with Bayesian Optimization}
\label{sec:bo}
Bayesian optimization~\cite{močkus1975on} is a technique used for the global optimization of expensive-to-evaluate functions. It builds a probabilistic model of the objective function and iteratively selects the next point to evaluate based on an acquisition function that balances exploration and exploitation. The model is updated as new evaluations are made, improving its accuracy over time. Bayesian optimization is beneficial when the objective function is black-box, expensive to evaluate, and multiple local optima exist. It has been successfully applied to various optimization problems in machine learning, engineering, and science~\cite{shahriari2016bo-review}.

In addition to engineering aspects, BO can also model human feedback and enables HITL optimization~\cite{shahriari2016bo-review} utilizing, for example, biological measurements~\cite{kim2017human, ding2018human}, interaction behavior~\cite{chong2021interactive}, and measured preferential choices~\cite{brochu2007active, gonzalez2017pbo, koyama2017sequential}. It learns a \emph{posterior} from the measured human feedback and aims to search for a global maximum of an unknown function by exploration and exploitation. In conjunction with measured human feedback, two popular genres of BO are actively being researched, as shown in \autoref{tab:bo-optimizers}.
Utility-based BO~\cite{močkus1975on} requires measuring human feedback as a utility value. In contrast, preference-based BO (PBO) can propose examples using an acquisition function, ask the human to provide a choice, and infer the underlying preference iteratively.

\begin{table}[t]
\caption{Examples of different Bayesian Optimization (BO) frameworks. Types of feedback can be either utility-based or preference-based. Most of the BO frameworks do not model indifference, and none of them consider incomplete preferences. Note that 1) rating can be considered either as absolute utility or preferential ranking distance depending on the underlying preference model, 2) selection is a special case of ranking where one item is strictly preferred over other options, and the ranking order of those unselected options is not specified, which means that their preference relation may be considered incomplete.}
\centering
\begin{tabularx}{\columnwidth}{Xcccc}
\toprule
\textbf{\footnotesize Optimizer} & \textbf{\footnotesize Input Form} & \textbf{\footnotesize Feedback Type} & \textbf{\footnotesize Options} & \textbf{\footnotesize Modeled Indifference/Incompleteness} \\
\midrule
Naive BO~\cite{močkus1975on} & Rating & Utility & Pointwise & No/No \\
\midrule
2GC~\cite{brochu2007active} & Selection & Preferential & Pairwise & No/No \\
PBO~\cite{gonzalez2017pbo} & Selection & Preferential & Pairwise & No/No\\
EUBO~\cite{lin2022preference} & Selection & Preferential & Pairwise & No/No \\
4GC~\cite{brochu2010animation} & Rating & Preferential & Listwise & No/No \\
SLS~\cite{koyama2017sequential} & Selection & Preferential & Listwise & No/No \\
SPS~\cite{koyama2020sequential} & Selection & Preferential & Listwise & No/No \\
PPBO ~\cite{mikkola2020projective} & Selection & Preferential & Listwise & No/No \\
Top-k BO~\cite{nguyen2021topk} & Ranking & Preferential & Listwise & Yes/No \\
\bottomrule
\end{tabularx}
\label{tab:bo-optimizers}
\end{table}

When dealing with preference-based human feedback,
PBO as a specialized category of BO has received increasing development in recent studies~\cite{gonzalez2017pbo, mikkola2020projective}. While BO learns based on absolute utility (e.g., absolute ratings of a given option), PBO learns relative rating or preferential selection as ranking information. To avoid the violations of the transitivity described above, the recent extensions~\cite{koyama2020sequential, siivola2021preferential, benavoli2021choice} of PBO transited from using a binary pairwise comparison to using a reasonable amount of options, resulting in a listwise comparison. These extensions could largely prevent violations of transitivity and infer more information at a time because they either consider choosing a set of options as winners among all given options~\cite{koyama2020sequential, benavoli2021choice}; or provide a ranking of all given options, where options may share the same level of rank~\cite{siivola2021preferential}. Note that more ranking elements may also increase the uncertainty for users and cause them to make imperfect decisions~\cite{monarch2021hitl} due to increased workload. Thus, one should carefully consider the presented number of elements.
Although PBO has used pairwise comparisons to mitigate various issues regarding unstable human judgment, these current approaches could also violate preference assumptions due to cognitive limitations. \citet{tversky1974judgment} have widely presented how heuristic biases might influence the choice behavior. External causes can also produce a considerable amount of noisy observations in choice~\cite{kahneman2021noise}.
To overcome the violation of preference assumptions, PBO has also considered handling noisy inputs~\cite{letham2019constrained} and guarantees theoretical convergence when dealing with unstable choices. Despite all these developments in PBO, we still observe two challenges in practice. The first is that a human might change their objective during the optimization process, even when using pairwise comparisons. However, PBO assumes a fixed implicit underlying choice function that it can learn. Although PBO can handle noisy input, the other challenge is that it requires much more iterations to let the optimization converge. This is usually very costly when involving a human. We, therefore, need to design the UI carefully to mitigate these challenges and reduce errors.

\subsection{Opinion Measurement Interfaces}

\begin{figure}[t]
\centering
\includegraphics[width=\linewidth]{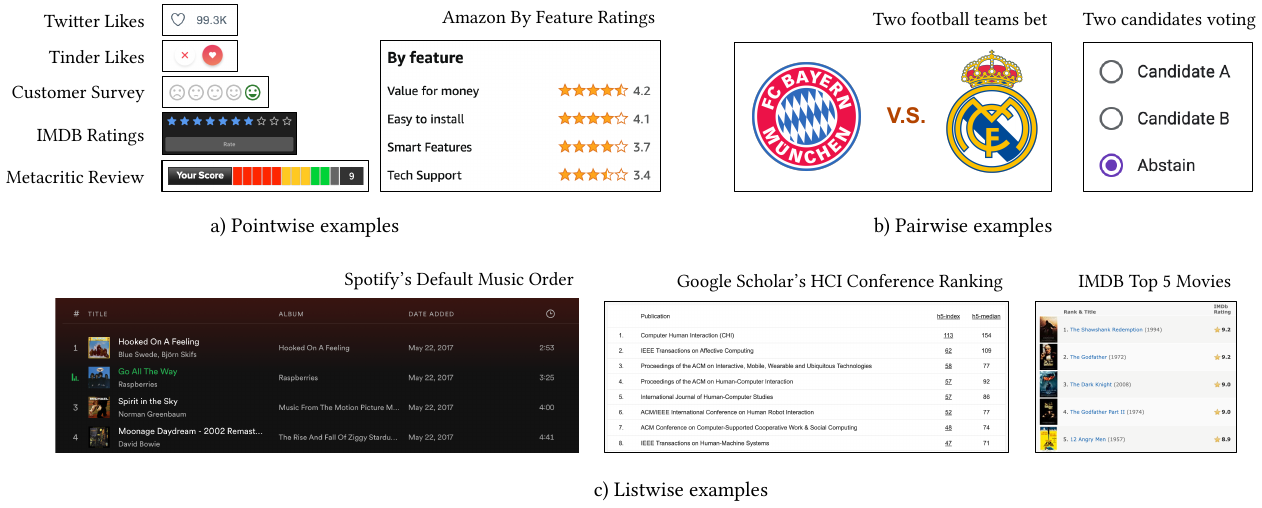}
\caption{Real-world measured opinions in different interfaces: a)~Commonly used rating scales to measure opinion regarding one item; b)~Two alternative candidates (non-)forced choice; c)~Music album song orders without specified criteria, and forced ranking of conferences and movies using a distance.}
\label{fig:examples}
\end{figure}

Conventional opinion measurement interfaces have different absolute measures, such as categorical choice and ordinal Likert-scale measures. The underlying analysis is based on Thurstone's law of comparison~\cite{thurstone1927comparelaw}. The optimal design of a pointwise measuring UI has been explored in psychology to reduce measuring error, for example, by asking repeated questions and checking the overall consistency\cite{sherer1998the},
regarding the influence of ticks~\cite{matejka2016effect}, or by allowing users to express uncertainty about their own input~\cite{greis2017input}. Previous work also discussed improving rating accuracy in the HITL context~\cite{nguyen2013rating, dudley2018imlui}.

A first systematic evaluation regarding the design space was provided by \citet{nobarany2012opinion}. They also observed that the rating interface might not be ideal for measuring opinions and discussed how to support ranking in the interface. In a way, this is similar to how PBO was designed~\cite{brochu2007active, brochu2010animation, gonzalez2017pbo}; opinion measurement interfaces that support comparison were also found beneficial here.
\citet{kalloori2018eliciting} discussed a few reasons for better-measured opinions using preference-based feedback: 1)~Explicit rating utility suffers from calibration of drift, for concepts with different types of interpretations might lead to differently rated results. 2)~Users might change their opinion from time to time due to anchoring~\cite{payne1993the}. 3)~Relative questions might potentially be faster and easier to answer.
Due to the increased complexity of feedback types, ranking UIs are rarely explored, although there is initial work in this direction~\cite{luk1994multirank, neubarth2010dnd}.
\paragraph{Interfaces used in the Real World}
Opinions are frequently measured in real life for a variety of purposes. We can generally observe three genres of these UIs: pointwise, pairwise, and listwise UIs, as shown in \autoref{fig:examples}.

\begin{figure}[t]
\centering
\includegraphics[width=\linewidth]{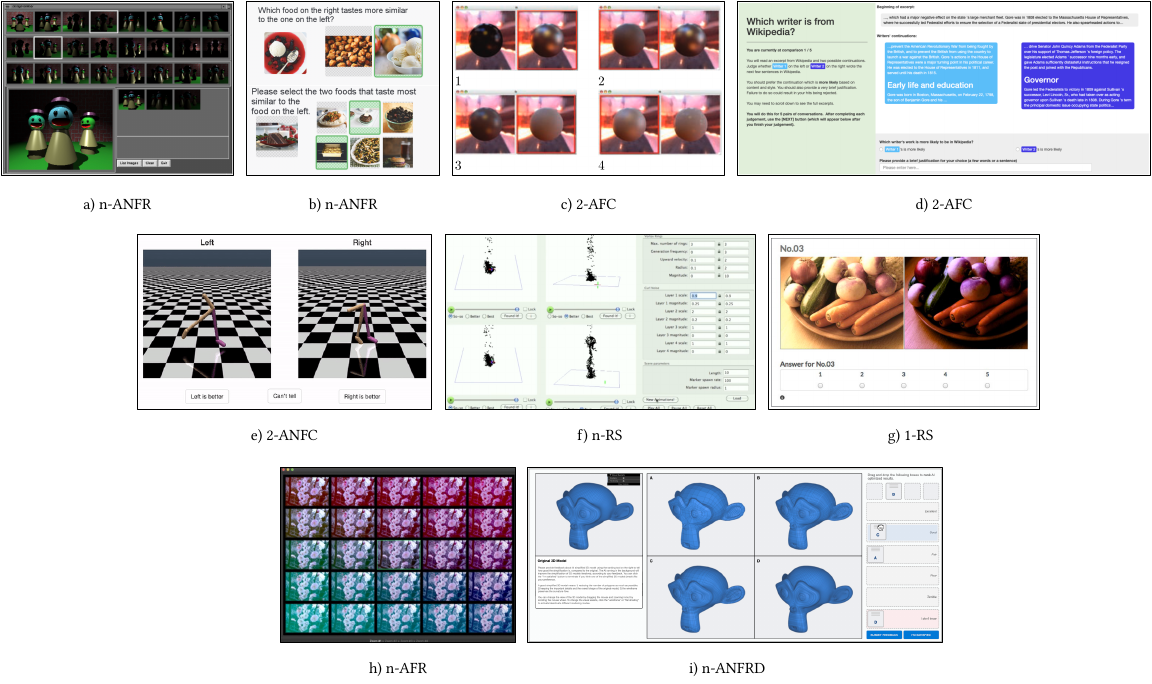}
\caption{Examples of opinion measurement interfaces being used in HITL optimizations: 
\textbf{a)} A user chooses from a list of candidates to optimize illumination~\cite{marks1997design-galleries}; 
\textbf{b)} A user is asked to choose items that are closer to a given reference~\cite{wilber2014cost}; 
\textbf{c)} A user is required to decide which of the two given options is closer to their desired material~\cite{brochu2007active}; 
\textbf{d)} A user is asked to choose which of two presented texts is more likely to be written on Wikipedia by a human~\cite{welleck2020neural}; 
\textbf{e)} A user is required to choose which trained machine's back-flip is better~\cite{christiano2017deep}, note that the interface uses a ``can't tell''; 
\textbf{f)} A user rates four presented graphics animations using a 3-step rating scale to show how close an animation is to what they are looking for~\cite{brochu2010animation}; 
\textbf{g)} A user rates which of the given paired designs is more preferred~\cite{koyama2014crowd}; 
\textbf{h)} A user must select one target out of a list of choice alternatives~\cite{koyama2020sequential}.
\textbf{i)} A user ranks the given design variations, and each rank position has an associated ordinal rating label~\cite{ou2023expertise}, and ties are allowed.
These interfaces fit in a design space and are discussed in~\autoref{sec:space1}.
}
\label{fig:hitlui}
\end{figure}

\paragraph{Interfaces used in HITL Optimization}
As shown in \autoref{fig:hitlui}, many designs for opinion measurement interfaces are used in HITL optimization. Specifically, the presented examples can be treated as a mix of pairwise and listwise comparisons. The underlying feedback type that models the UI's input form is either utility- or preference-based. Because of this diversity, we will provide a taxonomy to categorize them and determine our study design in the next section.

\section{Design Space}
\label{sec:space}
We revisit the design space of opinion measurement interfaces and extend it to include multiple design variations of the listwise approach. \autoref{tab:ui-options} provides an overview of the design space of opinion measurement interfaces. The design space formally contains these two dimensions: the number of \emph{alternative options} and the underlying \emph{feedback type}.

The number of alternative options encodes the complexity when measuring opinion. With more alternative options, we can gather more feedback from a user simultaneously. An interface is \emph{pointwise} if it only measures the opinion regarding one item.
Similarly, interfaces that measure information about two paired items are \emph{pairwise}. With more than two options, the interface is \emph{listwise}.
For example, in \autoref{fig:examples}a all interfaces are pointwise as they measure opinions regarding a single instance. Conceptually, a \emph{listwise} interface also covers interfaces that potentially allow exploring infinite options (n=$\infty$, e.g., \citet{koyama2017sequential}). However, regardless of the number of options, since a user has to select or mark a subset of preferential relations among all options, e.g.,~\citet{wilber2014cost}, a multiple-item selection can be considered a strict preference between selected and unselected ones. In contrast, the preference order among (un)selected ones remains unspecified. Note that although an interface presents paired options to measure human opinion, it could still ask humans to estimate the distance between two options, which only conveys absolute distance utility as a rating regarding one item to the reference. Furthermore, although \autoref{fig:hitlui}f presents four alternatives and asks how close they are to what the user is looking for, this is an absolute rating only concerning the user's idea of an optimum (which remains unknown to everybody else). These considerations imply that the underlying feedback type, which the framed UI presents as a usage instruction, is also crucial for opinion measurement.

The underlying type of feedback can be either \emph{utility}-based or \emph{preference}-based. Utility-based interfaces query a human to provide an absolute ordinal or cardinal value regarding an option (the interpretation of the value can differ depending on the context). Preference-based interfaces only measure relative information regarding paired options and, therefore, need further consideration to infer a latent utility.
On the one hand, since utility-based interfaces only provide numeric information directly associated with the option, the relative information between options may not be explicit. However, we can sort the utility value. On the other hand, preference-based interfaces only provide local information. A \emph{hybrid} feedback type provides a subtle design consideration to mix utility- and preference-based information. Similarly, we have two design alternatives, either strict or weak, depending on the type of preference.

\subsection{Applicability}
\label{sec:space1}
We can place the previously presented examples into this design space. \autoref{fig:examples}a contains 1-RS and n-RS: Twitter and Tinder only utilize binary responses regarding one option (like or not). The Amazon ``by feature'' rating example is n-RS but evaluates multiple dimensions using scales.
In \autoref{fig:examples}b, there are two examples of 2-AFC or 2-ANFC. 
\autoref{fig:examples}c shows three listwise interfaces: n-AFR (songs are ranked without allowing ties and without information about how they are being ranked), n-AFRD (options ranked with a distance, items with the same utility value are forced to be placed on different ranks).
\autoref{fig:hitlui}a and \autoref{fig:hitlui}b are $n$-ANFR because they require humans to rank options (permitting ties) without specifying a distance to other options,
\autoref{fig:hitlui}c and \autoref{fig:hitlui}d are $2$-AFC because they are judging two options and cannot tell indifference through the UI,
\autoref{fig:hitlui}e is $2$-ANFC because it offers ``can't tell'' to its user,
\autoref{fig:hitlui}f is $n$-RS, and \autoref{fig:hitlui}g is $1$-RS because it is judging how far one item is away from the other, and only judges one option.
\autoref{fig:hitlui}h is $n$-AFR because it selects one out of all options as top-1 ranking.
In fact, \autoref{fig:hitlui}a and \autoref{fig:hitlui}h also use a meaningful 2D layout to present the alternative choices instead of a plain list. While this may improve the efficiency and user experience in specific tasks, it does not change the general structure and, therefore, the classification of the interface.
Lastly, \autoref{fig:hitlui}i is $n$-ANFRD that asks users to rank options with a distance but also allows ties.

\begin{table}[t]
\caption{A design space of opinion measurement interfaces concerning feedback type, ordering type, and batch size. The numbers in bold are those of the interfaces implemented in our study. All of them allow users to express incomplete preferences using an ``I don't know'' checkbox.}
\centering
\begin{tabularx}{\linewidth}{Xccc}
\toprule
& \multicolumn{3}{c}{\textbf{Alternative Options ($n\in\mathbb{N}$)}} \\
\cmidrule(lr){2-4}
\textbf{Feedback Type} & \textbf{Pointwise ($n=1$)} & \textbf{Pairwise ($n=2$)} & \textbf{Listwise ($n>2$)} \\
\midrule
Utility & 1-RS (\textbf{UI1}) & 2-RS & $n$-RS (\textbf{UI3}) \\
Preferential (strict) & N/A & 2-AFC & $n$-AFR (\textbf{UI4}) \\
Preferential (weak) & N/A & 2-ANFC (\textbf{UI2}) & $n$-ANFR (\textbf{UI5}) \\
\midrule
Hybrid (strict) & N/A & N/A & $n$-AFRD \\
Hybrid (weak)   & N/A & N/A & $n$-ANFRD (\textbf{UI6}) \\
\bottomrule
\end{tabularx}
\label{tab:ui-options}
\end{table}

\section{Hypotheses}
\label{sec:hypothesis}
Using this design space, we identified six promising opinion measurement interfaces, i.e., 1-RS, 2-ANFC, $n$-RS, $n$-AFR, $n$-ANFR, $n$-ANFRD (UI1 to UI6), see \autoref{fig:teaser}. Our selection is informed by results from related work, which hints that some might be better than others. Thus, we formulate four major hypotheses concerning the performance of HITL optimization when using such an interface:

\paragraph{Hypothesis 1: Baseline}
Early work on HITL optimizations has shown great performance when using 2-ANFC~\cite{brochu2007active, brochu2007material-design, brochu2010animation}. Their results showed that 2-ANFC systems generally perform better than 1-RS and greatly support HITL systems. We use their initial findings as our baseline comparison and to validate our apparatus. Thus, we hypothesize (\textbf{H1}) that the \textit{preference-based} interface will outperform the \textit{utility-based} interface regarding performance measures (UI1 vs. UI2).

\paragraph{Hypothesis 2: Impact of Listwise Approach} If \textbf{H1} is true, this indicates that giving the user more options for comparison will improve the overall system performance. In fact, this is what findings by \citet{cao2007learning} suggest. Thus, we hypothesize that a \textit{listwise} comparison ($n > 2$), will outperform comparisons with fewer options.
As \textit{Listwise} is the most flexible in terms of feedback, we next investigate the different strategies that emerged from related work~\cite{koyama2017sequential, koyama2020sequential, mikkola2020projective}. In this work, we investigate design variations in which the human can either give preferential ratings (e.g.,~\citet{koyama2014crowd}), provide a strict ranking selection (e.g.,~\citet{koyama2020sequential}), a weak ranking selection (e.g.,~\citet{marks1997design-galleries}), or a combination of those (hybrid).
Consequently, we hypothesize (\textbf{H2a}) that a \textit{listwise} interface will outperform a \textit{pointwise} interface when using \textit{utility-based} feedback (UI1 vs. UI3).
And we hypothesize (\textbf{H2b}) that a \textit{listwise} interface will outperform a \textit{pairwise} interface when using \textit{preference-based} feedback (UI2 vs. UI5).

\paragraph{Hypothesis 3: Impact of Listwise Design Variations} Based on the initial findings by \citet{mikkola2020projective}, we additionally hypothesize \textbf{H3a} and \textbf{H3b}. In detail, we hypothesize \textbf{H3a} that a \textit{preference-based} listwise ranking interface will outperform a \textit{utility-based} listwise rating interface (cf. \cite{brochu2007material-design, brochu2007active, brochu2010animation}) while allowing for expressing indifference (UI3 vs. UI5). Further, regarding indifference, we hypothesize \textbf{H3b} that being able to express indifference in the interface (\textit{weak preference}) will be better than a forced choice (\textit{strict preference}) (UI4 vs. UI5).

\paragraph{Hypothesis 4: Impact of Hybrid Approach} Based on hypotheses \textbf{H1}-\textbf{H3} and assuming that their outcome in this work is in line with prior work, %
\citet{ou2023expertise} used a new hybrid feedback interface. For this, \textit{n-ANFRD} (UI6), we pick what we consider to be the best to support the HITL optimization, thus, getting the most information from the user. Thus, we hypothesize that the hybrid listwise interface will outperform: \textbf{H4a} the pairwise interface (UI2), \textbf{H4b} the listwise rating interface (UI3), \textbf{H4c} the listwise strict ranking interface (UI4), and \textbf{H4d} the listwise weak ranking interface (UI5).

\section{User Study}
We conducted a user study to investigate how different opinion measurement interfaces impact the overall HITL optimization performance to answer our research questions (\textbf{RQ1} and \textbf{RQ2} in \autoref{sec:intro}) and verify our four hypotheses. Here, we compare the six user interfaces (6 conditions: UI1-UI6, \autoref{tab:ui-options} and \autoref{fig:teaser}) using a between-subject design (\textsc{Interface}). We used tasks from two domains as a between-subject variable (\textsc{Domain}) to shorten the participation time and prevent a learning effect. One group did \textit{text summarization} tasks from the natural language processing domain. The other group did \textit{image color enhancement} tasks from the design domain. As dependent variables, we calculated interaction measures from usage logs and feedback from questionnaires (individual rating scales and open questions.)

\subsection{User Interfaces}
\begin{figure}[p]
\centering
\begin{subfigure}[b]{\textwidth}
\centering
\includegraphics[width=0.739\textwidth]{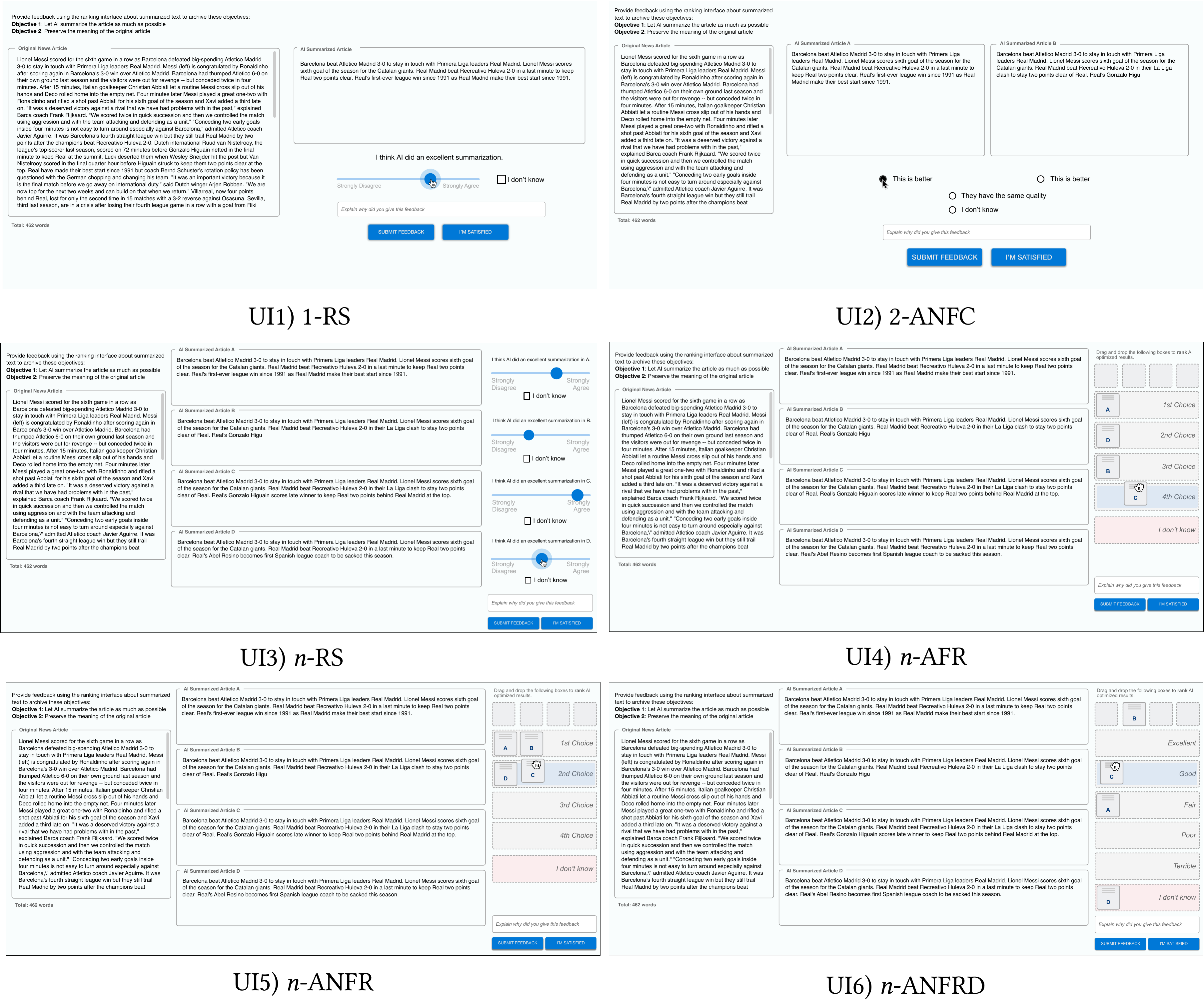}
\caption{Text Summarization}
\label{fig:ttextsumuis:text}
\end{subfigure}
\hfill\vspace{0pt}
\begin{subfigure}[b]{\textwidth}
\centering
\includegraphics[width=0.75\textwidth]{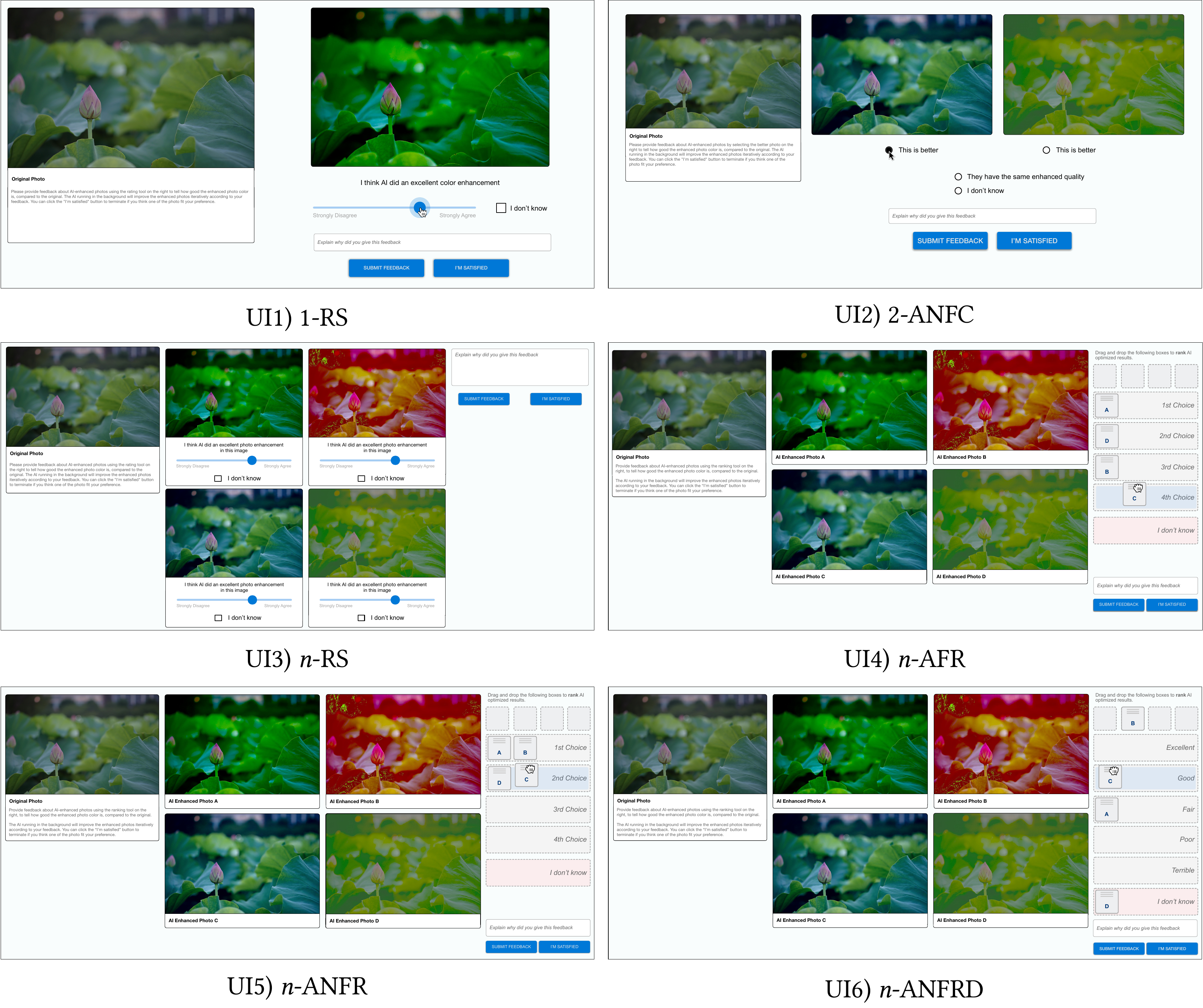}
\caption{Image Color Enhancement}
\label{fig:ttextsumuis:image}
\end{subfigure}
\caption{The twelve UIs we implemented in our studies for text summarization and image color enhancement, with (a) a news article selected from the CNN daily email dataset, and (b) a photo taken from \emph{Sequential Gallery}~\cite{koyama2020sequential}.
}
\label{fig:ttextsumuis}
\end{figure}

As shown in \autoref{fig:teaser}, we selected six generic interfaces from this design space (1-RS, 2-ANFC, $n$-RS, $n$-AFR, $n$-ANFR, $n$-ANFRD) to facilitate our study design better. All interfaces are shown in \autoref{fig:ttextsumuis} for both tasks. The following describes the behavior of the six user interfaces:
\begin{itemize}
\item[\textbf{UI1}] 1-RS uses a bipolar slider-based Likert scale, similar to $n$-RS.

\item[\textbf{UI2}] 2-ANFC uses two radio buttons for preference indication.

\item[\textbf{UI3}] n-RS uses a bipolar slider-based Likert scale to indicate the  preference for each option.

\item[\textbf{UI4/5}] $n$-AFR requires a full ranking of all options, and the interface does not allow expressing indifference in this case, whereas $n$-ANFR (UI5) can, which characterizes strict and weak preferences.

To guarantee a pure ranking behavior in the interface, $n$-AFR (UI4), and $n$-ANFR (UI5) automatically adjust the distance between options. For example, if a user dropped two boxes on the 1st and 4th choice container box. Then the 4th choice will automatically move up to the 2nd choice and inform users that their ranking is automatically aligned and normalized.

\item[\textbf{UI6}] $n$-ANFRD is the most flexible listwise ranking interface that permits users to place the option box in any given container box.
\end{itemize}
Additionally, all interfaces include either checkbox options or container boxes to allow users to express incomplete preferences.

\subsection{Task}
As many applications use HITL optimization, any domain with a set of system parameters to be determined could be used as a HITL optimization task. Examples of such domains are text summarization~\cite{simpson2020interactive, ou2023expertise}, image color enhancement~\cite{koyama2016selph, koyama2017sequential,koyama2020sequential, ou2023expertise}, melody composition~\cite{zhou2021interactive}, 3D meshing~\cite{ou2022infloop, ou2023expertise}, 3D material design~\cite{brochu2007active}, 3D animation~\cite{brochu2010animation}, and 3D illumination~\cite{marks1997design-galleries}.
In line with previous work \cite{ou2023expertise}, we selected \textit{text summarization} and \textit{image color enhancement} for the following reasons: First, text and image responses between client and server are easy to handle, lightweight, and fast. Second, the quality of text summarization and image color enhancement is a partly aesthetic judgment, for example, regarding the exact wording and color style. Therefore, it is still challenging to measure this quality objectively~\cite{bhandari2020re, ou2023expertise}, making it plausible to bring a human into the loop.

\subsection{Apparatus}
We developed the frontend UI1 to UI6 (see \autoref{fig:ttextsumuis} and Supplement Video that illustrates the use of all interfaces) using Material UI\footnote{\url{https://mui.com/}, \emph{last accessed \today}} and React DnD\footnote{\url{https://react-dnd.github.io/react-dnd/about}, \emph{last accessed \today}}.
We wrote out backend in Go %
for easier concurrency management. It serves our frontend interfaces, data collection, and communications with other dedicated computing microservices, including two \emph{domain services} and an \emph{optimizer service}. We deployed the backend on our institute infrastructure (Ubuntu 20.04, 8-Core Intel Core i9-9900K, 64GB RAM, and NVIDIA GeForce RTX 2080 Ti with 11GB of GPU memory).

\subsubsection{Domain Services: Text Summarization} For text summarization, we initially used GPT-3 by OpenAI for text summarization. However, we switched to our self-hosted language model for reproducibility and unlimited quota in the main study. We implemented the backend using Flask and handled text summarization with GPU acceleration. A pressure benchmark shows the backend can handle an average of 6 concurrent requests per second, which determines our rate control to launch parallel participation in our study. We used a pre-trained BART model\footnote{\url{https://huggingface.co/sshleifer/distilbart-cnn-12-6}, \emph{last accessed \today}} for text summarization. The model is fine-tuned for CNN daily mail with inference time 307ms, ROUGE-2 score of 21.26, and ROUGE-L score of 30.59, which outperforms a popular baseline \cite{lewis2019bart}.

We use Nucleus sampling~\cite{holtzman2020topp} as a stochastic text decoding strategy\footnote{\url{https://huggingface.co/blog/how-to-generate}, \emph{last accessed \today}} for our inference use case because it allows for a bounded parameter space (between 0 and 1) and can generate diverse human-like sentences in the inference phase. Since we designed our user task to consider the length of summarization as a decision criterion, we used a \emph{summarization ratio} parameter as a hard limit that controls the text generation length and a \emph{length penalty} parameter as a selected soft limit that encourages the model to generate shorter text. As a result, our hosted text summarization service allows 4 adjustable system parameters at every model inference stage:
1)~\emph{summarization ratio},
2)~\emph{length penalty},
3)~\emph{top-p}, and
4)~\emph{temperature}.

\subsubsection{Domain Services: Color Enhancement} 
For image color enhancement, we used a parameterized photo enhancer~\cite{koyama2016selph, koyama2017sequential, koyama2020sequential} as an image processing service. It allows 5 adjustable system parameters that can range from 0 to 1:
1)~\emph{brightness},
2)~\emph{contrast},
3)~\emph{saturation},
4)~\emph{temperature}, and
5)~\emph{tint}.

\subsubsection{Optimizer Service and Bayesian Optimizers}
\label{sec:bo-service}
We used Bayesian optimization to give users new options based on their input history. As there is no one Bayesian optimizer to cater to the needs of all six interfaces, we selected the best Bayesian optimizer for each interface. In detail, UI1 and UI3 used naive BO~\cite{mockus1978application}, UI2 used EUBO~\cite{lin2022preference}, and UI4-UI6 used the same advanced EUBO. While this could lead to performance differences, we perform benchmark comparisons to ensure they have equal performance; see \autoref{appendix:benchmark} for full details.

We used BoTorch~\cite{balandat2020botorch} to implement all selected optimizers, which reads the user responses to estimate the next optimal system parameters for exploration. The Bayesian optimizer for rating-based interfaces (UI1 and UI3) used the Gaussian process as a surrogate model with Martern 5/2 kernel. The acquisition function is an expected improvement acquisition function that models noisy inputs~\cite{letham2019constrained}.
We used EUBO~\cite{lin2022preference}, a BO variate cf. \autoref{sec:bo}, as a state-of-the-art PBO optimizer that considers noisy inputs to estimate system parameters for the pairwise interface (UI2). The acquisition function is the Analytic Expected Utility Of Best Option acquisition function~\cite{lin2022preference}.
The design of the EUBO algorithm guarantees that the current optimum is included in pairwise or listwise comparisons, and we can use the maximum posterior expected utility~\cite{chu2005preference} to understand the impact on outcome utility.

To guarantee fair comparisons and control experiment variables, instead of using different optimizers for all other listwise interfaces (i.e., UI4, UI5, and UI6), we extended EUBO to fit all listwise interfaces\footnote{See Supplemental Material for the implementation and further discussions on \url{https://github.com/pytorch/botorch/discussions/1367}, \emph{last accessed: \today}}: We first fit a Gaussian process using the user's rank data, then use the optimizer learned latent utility $f(\cdot)$ to fit another Gaussian process and infer the next batch of exploration positions. As EUBO maximizes the outcome utility difference between iterations, one can use the utility to analyze the optimization process. Moreover, in UI2 and UI4-UI6, although EUBO does not assume noiseless observations, we discarded indifference and incomplete preference input instead of generating a random preference order as 1) EUBO does not model indifference, 2) providing noisy input can decrease optimization performance and make a strong assumption about the preference model. These three Bayesian optimizers can optimize for our objectives and outperform pure random exploration in various simulated tasks, see \autoref{appendix:benchmark} for comparisons.

\begin{figure}[t]
\centering
\includegraphics[width=\linewidth]{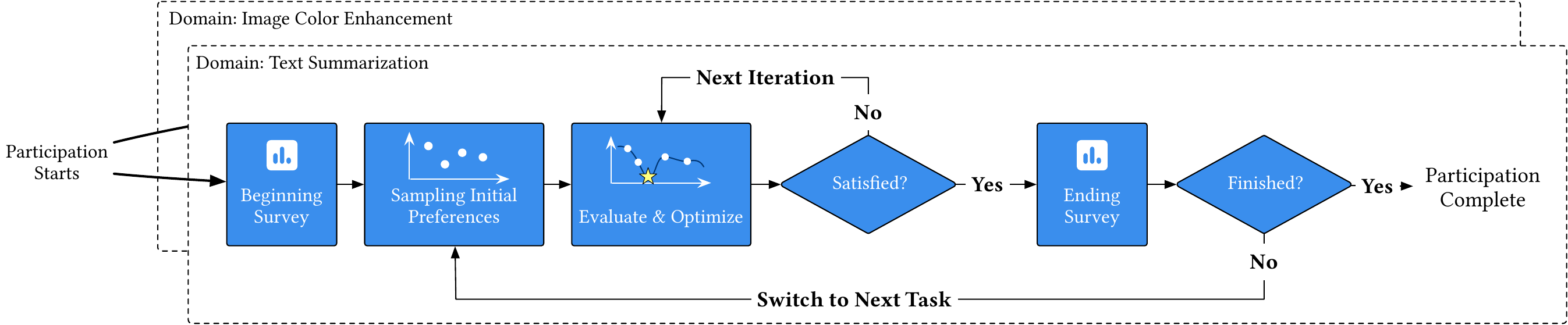}
\caption{The study procedure. Each participant completed three news articles or three photos, and each task item underwent at least 4 initial iterations for exploring the participant's preference prior.}
\label{fig:procedure}
\end{figure}

\subsection{Procedure}
The overall procedure for the participants is visualized in \autoref{fig:procedure}. 
Participants started the study with an informed consent form and answered initial demographic questions, including their age, gender, and levels of language proficiency. 

The opinion measurement interface then presents a set of evaluation options to the participants. The main task is to provide feedback, using the UI, to the AI to \emph{summarize the given news articles as much as possible while preserving the meaning of the article} or \emph{enhance the image color as much as possible according to the preferences of the participants}.
Because the optimizers need initialization samples, in the first 4 iterations, a participant evaluated outcomes produced by quasi-random Sobol sampled~\cite{rezende2014stochastic, wilson2017reparam} system parameters.
After acquiring these preference priors from participants, starting from the 5th iteration, the optimizer is used, and participants can freely terminate when satisfied with the current system outcomes. The task ends automatically after 20 iterations to limit participation time.
After termination, participants answered six questions about their satisfaction with the system outcomes and their experience using the interface to provide feedback.

Each participant completed three Latin square-shuffled 1) news articles selected from the CNN daily mail\footnote{\url{https://huggingface.co/datasets/cnn_dailymail}, \emph{last accessed \today}} dataset, which covered topics of sports, politics, and entertainment\footnote{Article IDs: ea06fd0b25cb9793397a, 35f0e33de7923036a97a, 42c027e4ff9730fbb3de}, or 2) photos\footnote{\url{https://github.com/yuki-koyama/sequential-gallery/tree/main/resources/scaled}, \emph{last accessed \today}} selected from the \emph{Sequential Gallery} by \citet{koyama2020sequential}.

\subsection{Participants}
We recruited participants worldwide via Prolific. %
Because participants had different median completion times in different experimental conditions, we paid between £3 to £9 upon completion, corresponding to an hourly wage of £9/h. Participation was under informed consent, and the study adhered to the European General Data Protection Regulation (GDPR). In total, we collected data from 635 participants from 25 countries.
To guarantee high-quality responses, we only consider participants if they: 1)~had an approval rate of 95\%, 2)~completed the study only once, 3)~answered with consistent demographics, e.g., not more than five years of age difference in the study compared to the platform registration information, and 4)~provided their response in at least a reasonable amount of time, i.e., spent longer than 3 seconds in each iteration to read the summarized text, longer than 1 second in each iteration to check the image and interact with the interface according to our pilot study observations.
Therefore, we will report our results based on 360 participants (171 female, 185 male, and 4 diverse; age $\mu=28.14, \sigma=8.39$, range 18-66), with varied self-indicated English proficiency (CEFR scale\footnote{CEFR scale: \url{https://www.coe.int/en/web/common-european-framework-reference-languages/table-1-cefr-3.3-common-reference-levels-global-scale}, \emph{last accessed \today}}: A1 0.55\%, A2 0.55\%, B1 3.89\%, B2 21.67\%, C1 43.33\%, C2 30.00\%) and varied self-indicated image color enhancement expertise (None 12.2\%, Novice 46.7\%, Intermediate 30.0\%, Experienced 10.6\%, Expert 0.6\%). We tested the six \textsc{Interfaces} with 30 participants per \textsc{Domain} (\textit{Text Summarization} and \textit{Image Color Enhancement}).

\subsection{Performance Indicators and Measurements}
\label{sec:measures}
A high-performance HITL optimization loop is considered successful from the human side when the human spends less time on providing feedback, can achieve a satisfactory result in fewer iterations, and the interface allows the human to give more precise and clear feedback. From the machine side, the underlying optimizer can learn human feedback more effectively and propose outcomes of high objective quality faster. Therefore, we measured participants' interaction details in each interface as \emph{behavior measures}. Next, we recorded participants' provided feedback and corresponding system outcomes as \emph{feedback measures} for the optimization loop. Lastly, we measured users' subjective experience with the system after each completion as \emph{questionnaire measures}.

\subsubsection{Behavior Measures}
We measured the participants' interaction in each interface using six measurements. 
1)~\emph{Decision Time per Iteration}, measured from when the evaluation options were presented to the last time when participants interacted with the interface,
2)~\emph{Decision Time per Item}, the \emph{Decision Time per Iteration} divided by the number of items shown,
3)~\emph{number of iterations} per task, 
4)~ \emph{Number of Ranking Interactions}, the number of operations to rank the given system outcomes in listwise interfaces,
5)~\emph{Number of Indifferent Preference}, number of pairs that are rated the same, and
6)~\emph{Number of Incomplete Preference} counted as the number of times the ``I don't know'' option was ticked per task.

\subsubsection{Feedback Measures}
\label{sec:feedback-measures}
To evaluate the optimization loop on the human side, we collected participants' rating and preferential ranking data in each iteration. Here, we used the normalized \emph{number of iterations} as \textsc{Progress} over the course of the HITL optimization. Moreover, we quantify human feedback using the \emph{user feedback} from direct rating  (UI1, UI3, and UI6\footnote{The ranking distance labels in UI6 are mapped to utility accordingly, e.g., 0 for \emph{Terrible} and 100 for \emph{Excellent}.}) or the optimizer \emph{learned latent utility} from preferential rankings (UI2, UI4, UI5, and UI6).
The \emph{learned latent utility} is a Laplace approximation of the posterior mean of the estimated utility values~\cite{chu2005preference}.
As EUBO maximizes the posterior expected utility~\cite{lin2022preference}, it represents the estimated user feedback when the query is chosen in the input space based on the available data and the current Gaussian process model. By comparing the posterior mean values at different iterations, we can gain insights into how the optimizer performs and whether it converges toward the user's expected solution.

On the machine side, we computed the \emph{Length} of the text, the ROUGE (\emph{ROUGE-1}, \emph{ROUGE-2}, and \emph{ROUGE-L})~\cite{lin2004rouge}, and \emph{BLEU}~\cite{papineni2002bleu} scores for the optimized text summarization to measure the objective quality of summarized texts as they correlate positively with human evaluation. \citet{ou2023expertise} used the exact same measures in their HITL text summarization task. ROUGE measures how many human-summarized words appear in the machine-generated summaries, and BLEU measures precision regarding how many words in the machine-generated words appear in the human reference summaries. 
For the image color enhancement, we computed \emph{Saturation}, \emph{Contrast}, and \emph{Brightness} using a mean of pixel-wise subtraction between source and outcome in the H, S, and V channels correspondingly. Furthermore, we computed the mean pixel-wise difference of U channels in YUV color space for \emph{Tint} changes and similarly in the V channel for \emph{Temperature} changes. These five measures are also in line with prior work by \citet{ou2023expertise}.

\emph{Note:} While the computed quality metrics are not true measurements of quality per se, together with the strategy of the optimizer to always show the current best choice, a consistent trend here can be seen as an indication of continued improvements and considered a successful optimization if at least one measured objective is being improved. Thus, we measure the semantic meaning of text summarization and only compare it to one human writer's ground truth, provided in the CNN daily mail dataset. Similarly, the color metrics only correlate to human judgment and do not measure individual color preferences. Although those metrics stay technical, they positively correlate with human judgment.

\subsubsection{Questionnaire Measures}
\label{sec:questionnaire-measures}
In line with work by \citet{ou2023expertise}, we asked six questions to obtain 100-point bipolar Likert scale responses for robust input \citep{funke2012why} without ticks to prevent biases \citep{matejka2016effect}. Q1)~\textit{I strongly agree the final result matches my expectation of the results}; Q2)~\textit{I firmly believe I can do better a result alone than this AI}; Q3)~\textit{I am strongly satisfied with the final result}; Q4)~\textit{I strongly believe the AI indeed improved the result based on my feedback.}; Q5)~\textit{I think the ``I don't know'' option helped me to express my feedback to the AI more precisely}, and Q6)~\textit{I believe I had given clear feedback to the AI}. Here, \citet{ou2023expertise} showed that Q1-Q4 have a high correlation and therefore used the aggregated value as an indicator for \emph{Satisfaction}. Q5 indicates whether expressing \emph{``I don't know''} was considered helpful, and Q6 indicates whether users think that the UI can give feedback clearly. Lastly, we obtained open-ended responses from participants with the instruction ``\textit{Describe your experience while working with this AI.}''

\section{Results}
In this section, we first present participants' behavior regarding interface interactions, including their total iterations, decision time per iteration, and how they express indifference and incomplete preferences.
Then, we show the effectiveness of the feedback optimization loop regarding the objective quality and feedback utility. Finally, we present subjective responses and comments from the questionnaire.
Since our hypotheses \textbf{H1} and \textbf{H2} in \autoref{sec:hypothesis} compare two UIs, we analyze normally distributed data using a t-test or a Mann-Whitney-Wilcoxon test in case the Shapiro-Wilk test~\cite{razali2011power} indicates that the normality is violated.
We use linear mixed models (LMMs)~\cite{bates2014fitting, kuznetsova2017lmertest} for \textbf{H3} and \textbf{H4} to report our results and to quantify the optimization loop statistically. The models include participants as a random effect and compare pairwise and listwise UIs between groups.
For non-significant independent two-sample tests, we conduct either a Bayesian independent t-test or a Bayesian Mann-Whitney test and analyze the Bayes Factor~\cite{wagenmakers2007bf, rouder2009bayesian} to quantify the likelihood of hypotheses.
The full results are available from reproducible scripts in supplemental materials: \url{https://github.com/changkun/rankui}.

\subsection{Behaviors and Interactions}

\begin{figure}[t]
\centering
\begin{subfigure}[b]{\textwidth}
\centering
\includegraphics[width=0.32\linewidth]{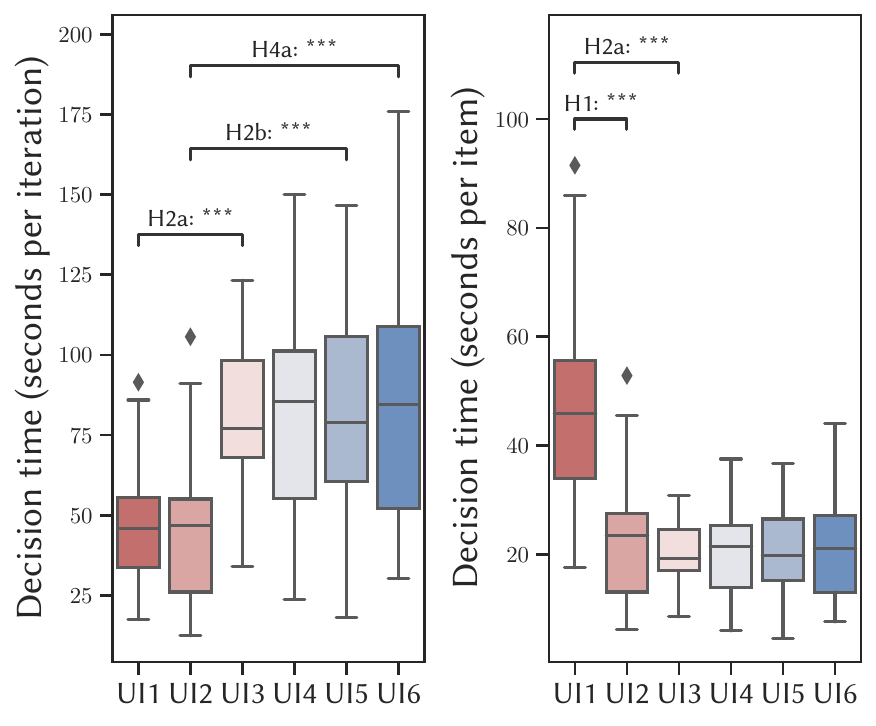}
\includegraphics[width=0.157\linewidth]{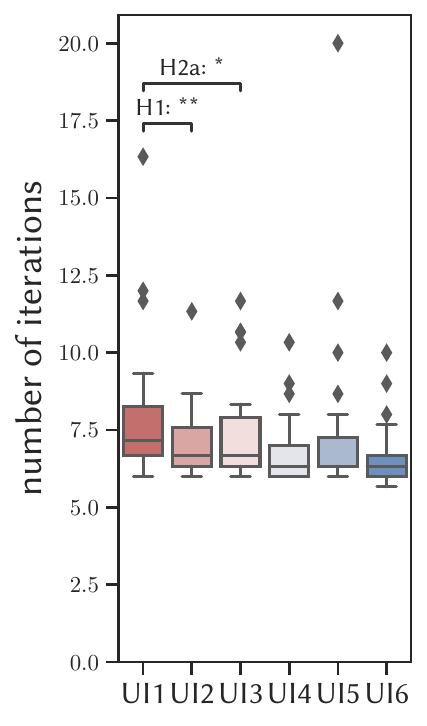}
\includegraphics[width=0.132\linewidth]{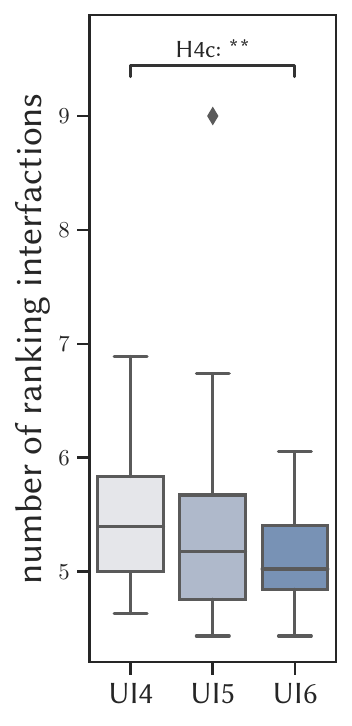}
\includegraphics[width=0.16\linewidth]{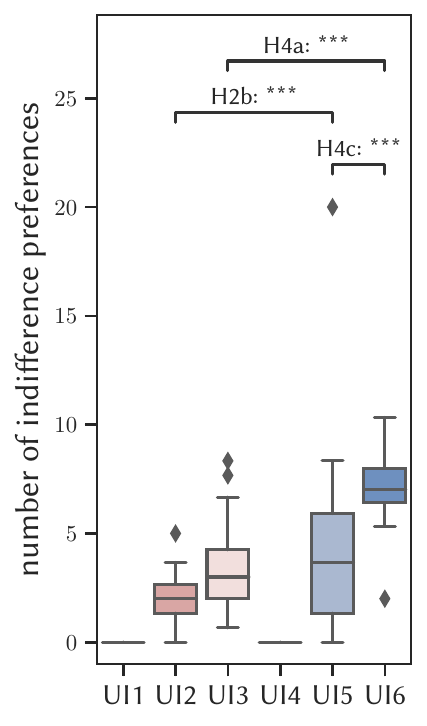}
\includegraphics[width=0.16\linewidth]{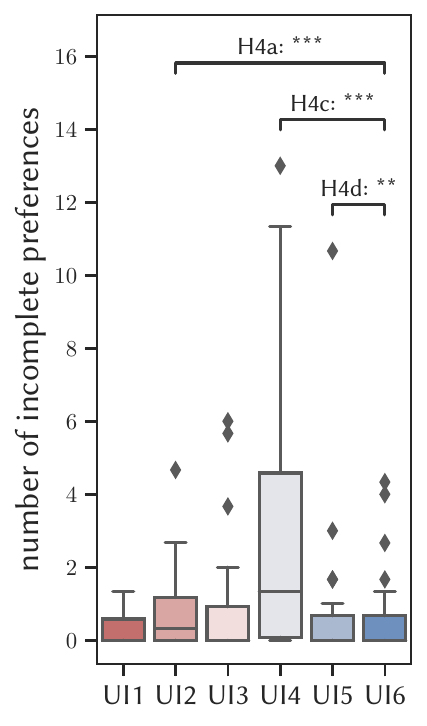}
\caption{Text Summarization}
\end{subfigure}
\hfill\vspace{0pt}
\begin{subfigure}[b]{\textwidth}
\centering
\includegraphics[width=0.32\linewidth]{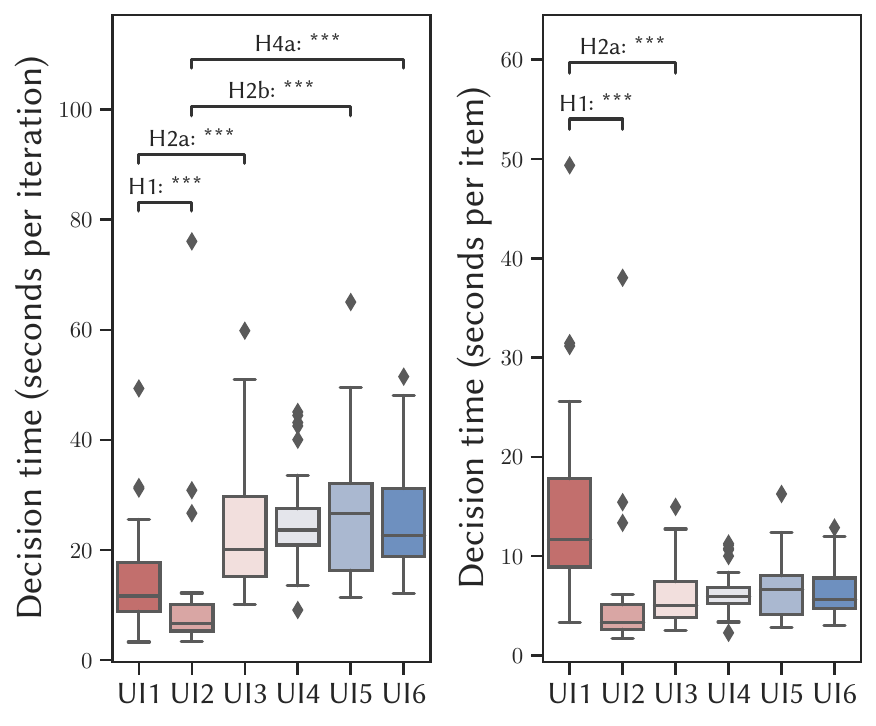}
\includegraphics[width=0.157\linewidth]{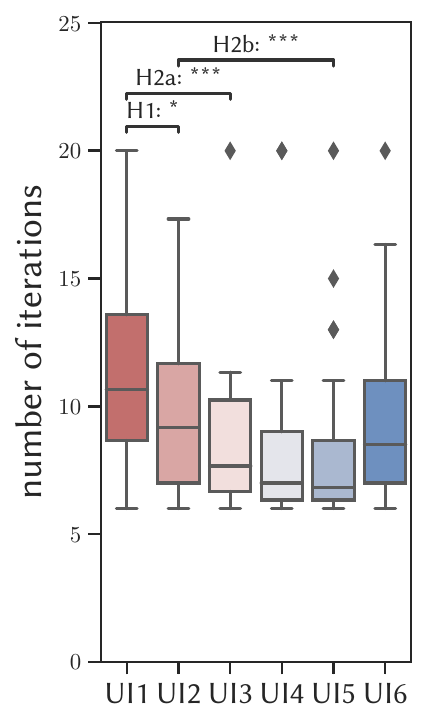}
\includegraphics[width=0.132\linewidth]{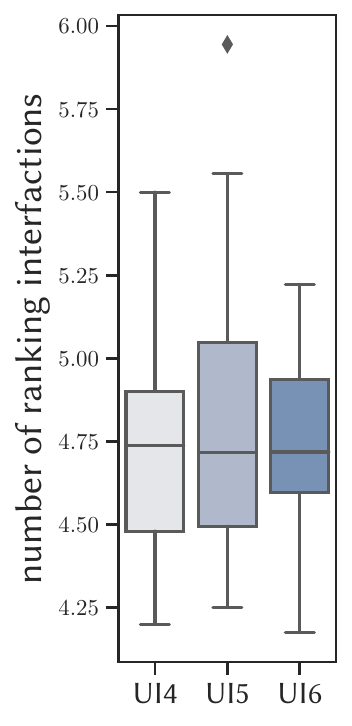}
\includegraphics[width=0.16\linewidth]{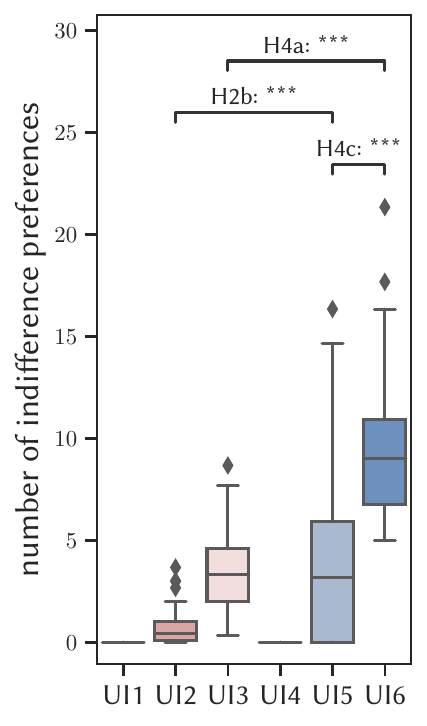}
\includegraphics[width=0.16\linewidth]{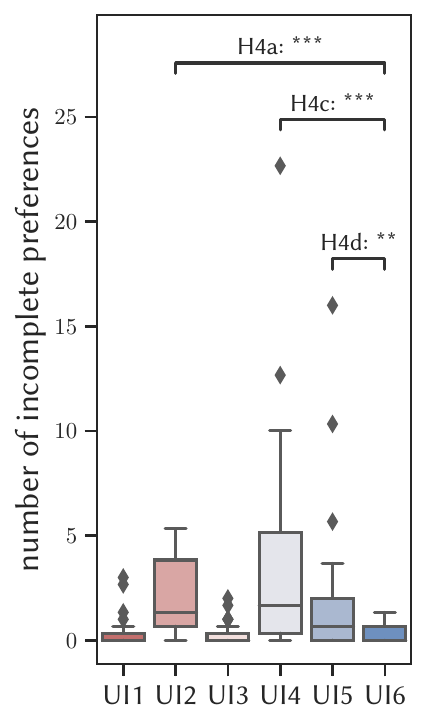}
\caption{Image Color Enhancement}
\end{subfigure}
\caption{Comparisons between opinion measurement interfaces, 
from left to right: \textbf{1)} \emph{Decision time}: Listwise approaches (UI3, UI4, UI5, UI6) significantly increased decision time per iteration. However, decision time per item in listwise approaches does not differ from the pairwise (UI2); \textbf{2)} \emph{Iterations}: Participants iterate significantly more in the pointwise (UI1) than in a pairwise (UI2) or listwise (UI3) approach; \textbf{3)} \emph{Ranking interactions}: There is a significant difference between listwise interfaces, when indifference is not permitted (UI4 vs. UI6) in the text domain, but not in the image domain; \textbf{4)} \emph{Indifference}: Participants express more indifference in listwise approaches if they can (UI5 and UI6) than in the pairwise approach (UI2.) Note that UI1 and UI4 cannot express indifference; \textbf{5)} \emph{Incomplete preferences}: Participants express incomplete preferences significantly more when they cannot express indifference in listwise approaches (UI4 vs. UI5 and UI6.)}
\label{fig:interactions}
\end{figure}

We show comparisons between all \textsc{Interfaces} in \autoref{fig:interactions} and complement them with detailed statistics across \textsc{Domains} in \autoref{tab:lmms}. Below, we will elaborate on the results regarding interaction behavior in all interfaces for both studied domains (text summarization and image color enhancement) in light of our hypotheses.

\begin{table}[b]
\vspace{-1mm}
\centering
\caption{Statistics of participants' behaviors and interactions. Results for H1, H2a, and H2b are tested using Wilcoxon (W$\times 10^5$), and LMMs are used for H3 and H4. Note that 1) $BF_{10}$ are provided for non-significant results; 2) it is not possible to express indifference in UI1 and UI4; hence H1, H2a, and H3b statistics are omitted; 3) ranking behavior is only measured for listwise interfaces (UI4-UI6: H4c and H4d), and their results are shown in \autoref{sec:rankinginteractions}.}
\label{tab:lmms}
\fontsize{4.5}{4}\selectfont
\begin{tabularx}{\linewidth}{X*{20}{d{0.2}}}
\toprule
& \multicolumn{8}{c}{\tiny \textbf{\textsc{Decision}}}
& \multicolumn{4}{c}{\tiny \textbf{\textsc{Interaction}}}
& \multicolumn{8}{c}{\tiny \textbf{\textsc{Preference}}}\\
& \multicolumn{8}{c}{\tiny \textbf{\textsc{Time}}}
& \multicolumn{4}{c}{\tiny \textbf{\textsc{Behavior}}}
& \multicolumn{8}{c}{\tiny \textbf{\textsc{Type}}}\\
\cmidrule(r){2-9}\cmidrule(lr){10-13}\cmidrule(lr){14-21}
& \multicolumn{4}{c}{\tiny per iteration} & \multicolumn{4}{c}{\tiny per item} 
& \multicolumn{4}{c}{\tiny number of interactions}
& \multicolumn{4}{c}{\tiny indifference} & \multicolumn{4}{c}{\tiny incompleteness} \\
\midrule
\multicolumn{1}{l}{\tiny \textbf{Text Sum.}}
& \multicolumn{1}{c}{W} & \multicolumn{1}{c}{r} & \multicolumn{1}{c}{BF$_{10}$} & \multicolumn{1}{c}{p}
& \multicolumn{1}{c}{W} & \multicolumn{1}{c}{r} & \multicolumn{1}{c}{BF$_{10}$} & \multicolumn{1}{c}{p}
& \multicolumn{1}{c}{W} & \multicolumn{1}{c}{r} & \multicolumn{1}{c}{BF$_{10}$} & \multicolumn{1}{c}{p}
& \multicolumn{1}{c}{W} & \multicolumn{1}{c}{r} & \multicolumn{1}{c}{BF$_{10}$} & \multicolumn{1}{c}{p}
& \multicolumn{1}{c}{W} & \multicolumn{1}{c}{r} & \multicolumn{1}{c}{BF$_{10}$} & \multicolumn{1}{c}{p} \\
\cmidrule(r){2-5}\cmidrule(lr){6-9}\cmidrule(lr){10-13}
\cmidrule(lr){14-17}\cmidrule(lr){18-21}
\multicolumn{1}{l}{\tiny \textbf{H1}:\hspace{1.5mm}UI1 vs. UI2}
& 2.140 & -.040 & .085 & .168
& 2.930 &  .310 & - & \multicolumn{1}{B{.}{.}{1,3}}{<.001} 
& .049 & .210 & - & \multicolumn{1}{B{.}{.}{1,3}}{.007} 
& - & - & - & -
& .037 & -.080 & .314 & .202 \\
\multicolumn{1}{l}{\tiny \textbf{H2a}: UI1 vs. UI3}
&1.650 & -.290 & - & \multicolumn{1}{B{.}{.}{1,3}}{<.001}
&1.360 & -.360 & - & \multicolumn{1}{B{.}{.}{1,3}}{<.001}
&.047 & .180 & - & \multicolumn{1}{B{.}{.}{1,3}}{.028}
& - & - & - & -  
& .036  & -.100 & .386 & .128 \\
\multicolumn{1}{l}{\tiny \textbf{H2b}: UI2 vs. UI5}
&3.220 & .390 & - & \multicolumn{1}{B{.}{.}{1,3}}{<.001}
&2.150 & .010 & .072 & .697 
&.043 & .050 & .188 & .506
& .030 & -.270 & - & \multicolumn{1}{B{.}{.}{1,3}}{.002} 
& .043 & .070 & .227 & .255 \\
\cmidrule(r){2-5}\cmidrule(lr){6-9}\cmidrule(lr){10-13}
\cmidrule(lr){14-17}\cmidrule(lr){18-21}
& \multicolumn{1}{c}{$\beta$} & \multicolumn{1}{c}{SE} & \multicolumn{1}{c}{t} & \multicolumn{1}{c}{p}
& \multicolumn{1}{c}{$\beta$} & \multicolumn{1}{c}{SE} & \multicolumn{1}{c}{t} & \multicolumn{1}{c}{p} 
& \multicolumn{1}{c}{$\beta$} & \multicolumn{1}{c}{SE} & \multicolumn{1}{c}{t} & \multicolumn{1}{c}{p} 
& \multicolumn{1}{c}{$\beta$} & \multicolumn{1}{c}{SE} & \multicolumn{1}{c}{t} & \multicolumn{1}{c}{p} 
& \multicolumn{1}{c}{$\beta$} & \multicolumn{1}{c}{SE} & \multicolumn{1}{c}{t} & \multicolumn{1}{c}{p} \\
\cmidrule(r){2-5}\cmidrule(lr){6-9}\cmidrule(lr){10-13}
\cmidrule(lr){14-17}\cmidrule(lr){18-21}
\multicolumn{1}{l}{\tiny \hspace{5mm}UI5 vs.}
& .015 & .001 & 13.307 & < .001 
& .059 & .004 & 13.307 & < .001
&1.978 & .046 & 42.693 & < .001
&1.075  & .154 &  6.997 & < .001
&-1.741 & .437 & -3.984 & < .001 \\
\multicolumn{1}{l}{\tiny \textbf{H3a}: \hspace{6.5mm}UI3} 
&  .001 & .002 & -.572 & .567
& -.004 & .006 & -.572 & .567
& -.001 & .065 & -.013 & .990
& -.010 & .213 & -.047 & .963
&  .312 & .566 &  .550 & .582 \\
\multicolumn{1}{l}{\tiny \textbf{H3b}: \hspace{6.5mm}UI4}
& -.001 & -.002 & -.552 & .581 
& -.003 &  .006 & -.552 & .581
& -.065 &  .066 & -.984 & .325
& - & - & - & - 
&1.831  &  .542 & 3.376 &\multicolumn{1}{B{.}{.}{1,3}}{<.001} \\
\cmidrule(r){2-5}\cmidrule(lr){6-9}\cmidrule(lr){10-13}
\cmidrule(lr){14-17}\cmidrule(lr){18-21}
\multicolumn{1}{l}{\tiny \hspace{5mm}UI6 vs.} 
& .015 & .001 & 11.021 & < .001
& .057 & .005 & 12.357 & < .001
&1.897 & .043 & 44.615 & < .001
&1.940 & .100 & 19.359 & < .001
&-1.904 &.440 & -4.329 & < .001\\
\multicolumn{1}{l}{\tiny \textbf{H4a}: \hspace{6.5mm}UI2}
& .010 & .002 & 5.020 & \multicolumn{1}{B{.}{.}{1,3}}{<.001}
&-.005 & .006 & -.736 & .462
& .042 & .059 &  .712 & .476
&-1.367 & .157 &-8.713 &\multicolumn{1}{B{.}{.}{1,3}}{<.001}
&.528 & .567 &  .932 & .351\\
\multicolumn{1}{l}{\tiny \textbf{H4b}: \hspace{6.5mm}UI3}
& -.001 & .002 & -.355 & .723
& -.002 & .006 & -.320 & .749
&  .086 & .059 & 1.463 & .143
& -.836 & .149 & -5.61 &\multicolumn{1}{B{.}{.}{1,3}}{<.001}
&  .482 & .570 &  .845 & .398\\
\multicolumn{1}{l}{\tiny \textbf{H4c}: \hspace{6.7mm}UI4}
& -.001 & .002 & -.271 & .786
& -.002 & .006 & -.281 & .779
&  .021 & .060 &  .358 & .720
& - & - & - & - 
& 1.998 & .547 & 3.653 &\multicolumn{1}{B{.}{.}{1,3}}{<.001}\\
\multicolumn{1}{l}{\tiny \textbf{H4d}: \hspace{6.5mm}UI5}
& .000 & .002 & .198  & .843
& .002 & .006 & .246  & .806
& .090 & .059 &1.534  & .125
&-.764 & .150 &-5.084 &\multicolumn{1}{B{.}{.}{1,3}}{<.001}
& .170 & .577 & .295 &  .768\\
\midrule

\multicolumn{1}{l}{\tiny \textbf{Color Enh.}}
& \multicolumn{1}{c}{W} & \multicolumn{1}{c}{r} & \multicolumn{1}{c}{BF$_{10}$} & \multicolumn{1}{c}{p}
& \multicolumn{1}{c}{W} & \multicolumn{1}{c}{r} & \multicolumn{1}{c}{BF$_{10}$} & \multicolumn{1}{c}{p}
& \multicolumn{1}{c}{W} & \multicolumn{1}{c}{r} & \multicolumn{1}{c}{BF$_{10}$} & \multicolumn{1}{c}{p}
& \multicolumn{1}{c}{W} & \multicolumn{1}{c}{r} & \multicolumn{1}{c}{BF$_{10}$} & \multicolumn{1}{c}{p}
& \multicolumn{1}{c}{W} & \multicolumn{1}{c}{r} & \multicolumn{1}{c}{BF$_{10}$} & \multicolumn{1}{c}{p} \\
\cmidrule(r){2-5}\cmidrule(lr){6-9}\cmidrule(lr){10-13}
\cmidrule(lr){14-17}\cmidrule(lr){18-21}
\multicolumn{1}{l}{\tiny \textbf{H1}:\hspace{1.5mm}UI1 vs. UI2}
& 5.480  &  .270 & - & \multicolumn{1}{B{.}{.}{1,3}}{.001}
& 7.130  &  .650 & - & \multicolumn{1}{B{.}{.}{1,3}}{.001}
& .048  &  .200 & - & \multicolumn{1}{B{.}{.}{1,3}}{.023}
& - & - & - & -
& .026  & -.360 & - & \multicolumn{1}{B{.}{.}{1,3}}{.001} \\
\multicolumn{1}{l}{\tiny \textbf{H2a}: UI1 vs. UI3}
& 1.870 & -.520 & - & \multicolumn{1}{B{.}{.}{1,3}}{.001}
& .674  & -.800 & - & \multicolumn{1}{B{.}{.}{1,3}}{.001}
& .058  &  .420 & - & \multicolumn{1}{B{.}{.}{1,3}}{<.001}
& - & - & - & -
& .040 & -.001 & .198 & .986 \\
\multicolumn{1}{l}{\tiny \textbf{H2b}: UI2 vs. UI5}
& .674  & -.800   & - & \multicolumn{1}{B{.}{.}{1,3}}{.001}
& 1.590 & -.520  & - & \multicolumn{1}{B{.}{.}{1,3}}{.001}
& .053  &  .310   & - & \multicolumn{1}{B{.}{.}{1,3}}{<.001}
& .026  & -.360   & - & \multicolumn{1}{B{.}{.}{1,3}}{<.001}
& .044  &  .100  & .252 & .203 \\
\cmidrule(r){2-5}\cmidrule(lr){6-9}\cmidrule(lr){10-13}
\cmidrule(lr){14-17}\cmidrule(lr){18-21}
& \multicolumn{1}{c}{$\beta$} & \multicolumn{1}{c}{SE} & \multicolumn{1}{c}{t} & \multicolumn{1}{c}{p}
& \multicolumn{1}{c}{$\beta$} & \multicolumn{1}{c}{SE} & \multicolumn{1}{c}{t} & \multicolumn{1}{c}{p} 
& \multicolumn{1}{c}{$\beta$} & \multicolumn{1}{c}{SE} & \multicolumn{1}{c}{t} & \multicolumn{1}{c}{p} 
& \multicolumn{1}{c}{$\beta$} & \multicolumn{1}{c}{SE} & \multicolumn{1}{c}{t} & \multicolumn{1}{c}{p} 
& \multicolumn{1}{c}{$\beta$} & \multicolumn{1}{c}{SE} & \multicolumn{1}{c}{t} & \multicolumn{1}{c}{p} \\
\cmidrule(r){2-5}\cmidrule(lr){6-9}\cmidrule(lr){10-13}
\cmidrule(lr){14-17}\cmidrule(lr){18-21}
\multicolumn{1}{l}{\tiny \hspace{5mm}UI5 vs.}
& .046  & .004 &  12.170 & <.001
& .059  & .004 & 13.307 & < .001
&1.978  & .046 & 42.693 & < .001
&1.075  & .154 &  6.997 & < .001
&-1.741 & .437 & -3.984 & < .001 \\
\multicolumn{1}{l}{\tiny \textbf{H3a}: \hspace{6.5mm}UI3} 
&  .005  & .005 & .994   & .320 
&  .021  & .021 & .994   & .320
&  .045  & .086 & .520   & .603
&  .424  & .307 & 1.381  & .167
& -1.726 & .526 & -3.282 &  \multicolumn{1}{B{.}{.}{1,3}}{.001}\\
\multicolumn{1}{l}{\tiny \textbf{H3b}: \hspace{6.5mm}UI4}
& -.001 & .005 &  -.100 & .920
& -.002 & .021 &  -.100 & .920
&  .007 & .087 &   .077 & .938
& - & - & - & -
&  .759 & .460 &  1.651 & .099\\
\cmidrule(r){2-5}\cmidrule(lr){6-9}\cmidrule(lr){10-13}
\cmidrule(lr){14-17}\cmidrule(lr){18-21}
\multicolumn{1}{l}{\tiny \hspace{5mm}UI6 vs.} 
&  .046  &  .001 & 45.100 & <.001 
&  .183  &  .017 & 10.791 & <.001
&  2.200 &  .060 & 36.631 & <.001
&  2.197 &  .154 & 14.304 & <.001
& -1.682 &  .343 & -4.907 & <.001\\
\multicolumn{1}{l}{\tiny \textbf{H4a}: \hspace{6.5mm}UI2}
&  .095  & .001 &  87.911 & \multicolumn{1}{B{.}{.}{1,3}}{<.001}
&  .112  & .024 &   4.572 & \multicolumn{1}{B{.}{.}{1,3}}{<.001}
&  .050  & .085 &    .586 &  .558
& -2.740 & .258 & -10.612 & \multicolumn{1}{B{.}{.}{1,3}}{<.001}
&  1.978 & .428 &   4.624 & \multicolumn{1}{B{.}{.}{1,3}}{<.001}\\
\multicolumn{1}{l}{\tiny \textbf{H4b}: \hspace{6.5mm}UI3}
&  .005 & .001 &  4.984 & \multicolumn{1}{B{.}{.}{1,3}}{<.001}
&  .022 & .024 &   .901 & .367
& -.089 & .085 & -1.038 & .299
& - & - & - & -
& -.326 & .482 & -.676  & .499 
\\
\multicolumn{1}{l}{\tiny \textbf{H4c}: \hspace{6.7mm}UI4}
& -.034 &  .001 & -.319 & .750
& -.002 &  .024 & -.077 & .939
& -.127 &  .086 & -4.926 & \multicolumn{1}{B{.}{.}{1,3}}{<.001}
&  2.0345 &  .4273 &  4.762 & \multicolumn{1}{B{.}{.}{1,3}}{<.001}
\\
\multicolumn{1}{l}{\tiny \textbf{H4d}: \hspace{6.5mm}UI5}
&  .000  &  .001 &   .044 & .965
&  .001  &  .024 &   .022 & .983
& -.134  &  .086 & -1.557 & .119
& -1.405 &  .234 & -6.005 & \multicolumn{1}{B{.}{.}{1,3}}{<.001}
&  1.308 &  .436 &  3.001 & \multicolumn{1}{B{.}{.}{1,3}}{.003}\\
\bottomrule
\end{tabularx}
\end{table}

\subsubsection{Decision Time per Iteration}
For \textbf{H1}, for decision time per iteration, we found no significant difference between UI1 and UI2 in text summarization, but UI1 caused a significant increase in decision time compared to UI2 in image color enhancement, see \autoref{fig:interactions} and \autoref{tab:lmms}. Regarding average decision time per item, there is a significant difference between UI1 and UI2. \emph{This means that providing comparisons enables faster decisions.}
For \textbf{H2}, UI3 results in a significantly longer decision time per iteration than UI1, and UI5 has a significantly longer decision time per iteration than UI2, which is true for both domains. This indicates that \emph{users need more decision time to judge more given options}.

\subsubsection{Decision Time per Item}
However,  when checking decision time per item, UI1 takes significantly more time than UI3 in both domains, while we found no significant difference between UI2 and UI5. This indicates that \emph{the listwise approach did not increase decision time per item.}
Finally, we fitted general linear mixed models for \textbf{H3} and \textbf{H4} using the Gamma family with an inverse link~\cite{lo2015to}. None showed significant differences in UI3-UI6 regarding decision time, except \textbf{H4a}: It compares UI2 to UI6, and the effect of UI2 is statistically significant and positive regarding decision time per iteration. \emph{Although the pairwise interface enables faster total decision time when there are more options, it does not make per-item decisions significantly faster than the listwise approach.}

\subsubsection{Interaction Behavior – Number of Iterations}
For the total iterations, we found that participants explored significantly more with UI1 than with UI2 (\textbf{H1}) and UI3 (\textbf{H2a}) in both domains, see \autoref{fig:interactions} and \autoref{tab:lmms}. However, we found no significant differences between UI2 and UI3 (\textbf{H2b}). We fitted Poisson mixed models for \textbf{H3} and \textbf{H4}, and in total iterations, there were no significant differences for most of the comparisons between pairwise and listwise interfaces (UI2-UI6) in both domains. \emph{While the pointwise interface (UI1) resulted in significantly more iterations than the other interfaces, there were no significant differences across domains between pairwise and listwise interfaces (UI2-UI6).}

\subsubsection{Interaction Behavior – Number of Ranking Interactions}
\label{sec:rankinginteractions}
We recorded the ranking interaction from UI4 to UI6 for \textbf{H4}, see \autoref{fig:interactions}.
In text summarization, a fitted Poisson mixed model showed that compared to the total ranking interactions in UI6 ($\beta$=1.627, SE=0.023, t=69.873, p<.001), the effect for UI4 is statistically significant and positive ($\beta$=0.065, SE=0.033, t=1.986, p=.047). For UI5, the effect is statistically non-significant
and positive ($\beta$=0.037, SE=0.032, t=1.156, p=.248). However, in image color enhancement, compared to the total ranking interactions in UI6 ($\beta$=4.700, SE=0.064, t=73.761, p < .001), the effects for UI4 ($\beta$=-0.001, SE=0.092, t=-0.008, p=.994) and UI5 ($\beta$=-0.001, SE=.092, t=-.008, p=.479) are both statistically non-significant.
In sum, we found that \emph{with the same BO optimizer under the hood, participants tend to produce fewer ranking interactions with UI6 than with other listwise interfaces (UI4 and UI5).}

\subsubsection{Preference Type – Expressing Indifference}
Following our study design, UI1 and UI4 cannot express indifference for indifference preference. Thus, we do not evaluate \textbf{H2a} and \textbf{H3b} here. We found that UI5 prompted participants to express significantly more indifference preference than UI2 (\textbf{H2b}) for preference-based feedback; see \autoref{fig:interactions} and \autoref{tab:lmms}. However, we found no significant difference between UI3 and UI5 for expressing indifference preference.
We fitted Poisson mixed models for \textbf{H4a}, \textbf{H4b}, and \textbf{H4d}, comparing them to UI6, see \autoref{fig:interactions} and \autoref{tab:lmms}. The effect of UI3 is statistically significant and negative compared to UI6. The effect of UI5 is statistically significant and negative. These results apply to both domains we studied. Hence, \emph{compared to all other UIs, participants express indifference significantly more often when using UI6}. 

\subsubsection{Preference Type – Expressing Incompleteness}
We first inspected the Wilcoxon signed rank test for participants who moved their sliders and used the ``I don't know'' checkbox in UI1 and UI4 because these interfaces cannot express indifference preference.
We found no significant difference between adjusting the slider to the midpoint and clicking ``I don't know'' (W = 239.50, p = .219; r = -0.24, $CI_{95\%}=[-0.55, 0.13], BF_{10}=0.638$).
\emph{This result verifies the prior work regarding the indifference between incomplete preference and midpoint rating~\cite{armstrong1987mid} in a bipolar Likert scale and validates our apparatus.}
In text summarization, for \textbf{H1}, participants with UI2 did not express significantly more incomplete preferences than with UI1,
similar when comparing UI1 vs. UI3 (\textbf{H2a}), and UI2 vs. UI5 (\textbf{H2b}), see \autoref{fig:interactions} and \autoref{tab:lmms}.
For \textbf{H3}, a fitted Poisson mixed model shows when compared to UI5, UI3 is statistically non-significant
and positive, but the effect of UI4 is statistically significant and
positive.
Another fitted Poisson mixed model for \textbf{H4} shows that, compared to UI6, only UI4 is statistically significant and positive. Similarly, in image color enhancement, results showed that UI6 is expressed significantly less than UI2, UI4, and UI5. Therefore, \emph{compared to all other UIs, participants tend to express fewer incomplete preferences when using UI6.}

\subsection{The Optimization Loop}
The overall optimization loop produces three types of data: 1) the optimized system outcomes, 2) user feedback (either absolute ratings or preferential rankings), and 3) an optimizer-inferred (latent) utility function.
To analyze the optimization loop, we selected a similar approach that analyzes the stationarity of rating responses~\cite{ou2022infloop} and the optimizer predictions~\cite{ou2023expertise}: in a successful optimization loop, the system outcomes should align and come closer to user expectations over iterations based on user feedback. Otherwise, the system outcomes should go back and forth or further away from user expectations.
As a result, to compare how the optimization loops work differently in different UIs, we should expect to observe: 1) The objective quality (see~\autoref{sec:feedback-measures} for the selected metrics) of optimized system outcomes progress significantly in at least one objective. 2) User feedback should give higher utility responses progressively. 3) Optimizer learned latent utility (see ~\autoref{sec:bo-service}) increases progressively.

\subsubsection{Objective Quality of System Outcomes}
The goal of all \textsc{Interfaces} is to reduce the text length while preserving the meaning of the text or to enhance the image color towards participants' preferences depending on the \textsc{Domain}.

\begin{figure}[t]
\centering
    \begin{subfigure}[b]{\textwidth}
        \centering
        \includegraphics[width=\linewidth]{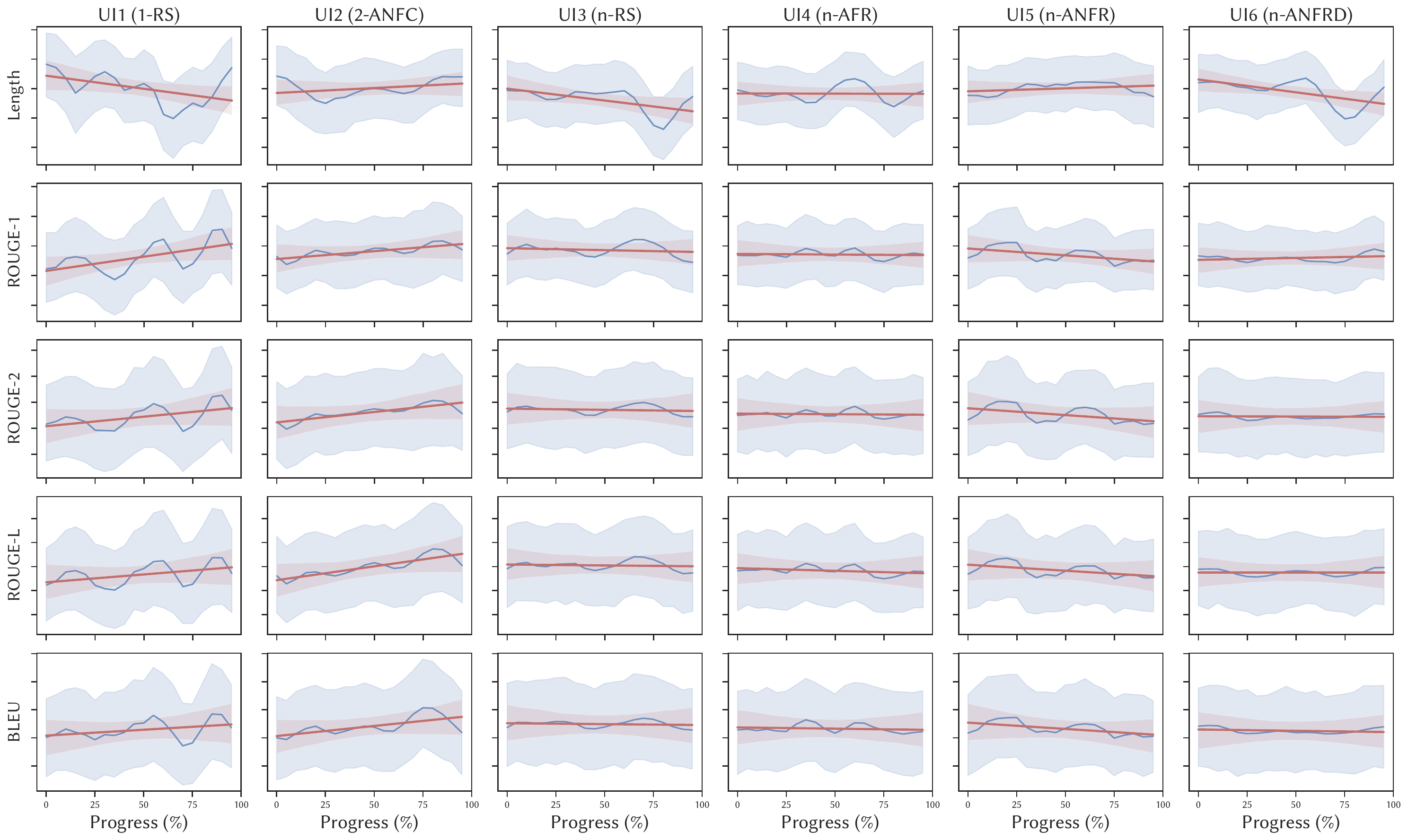}
        \caption{Text Summarization}
        \label{fig:outcomes:text}
    \end{subfigure}
 
    \begin{subfigure}[b]{\textwidth}
        \includegraphics[width=\linewidth]{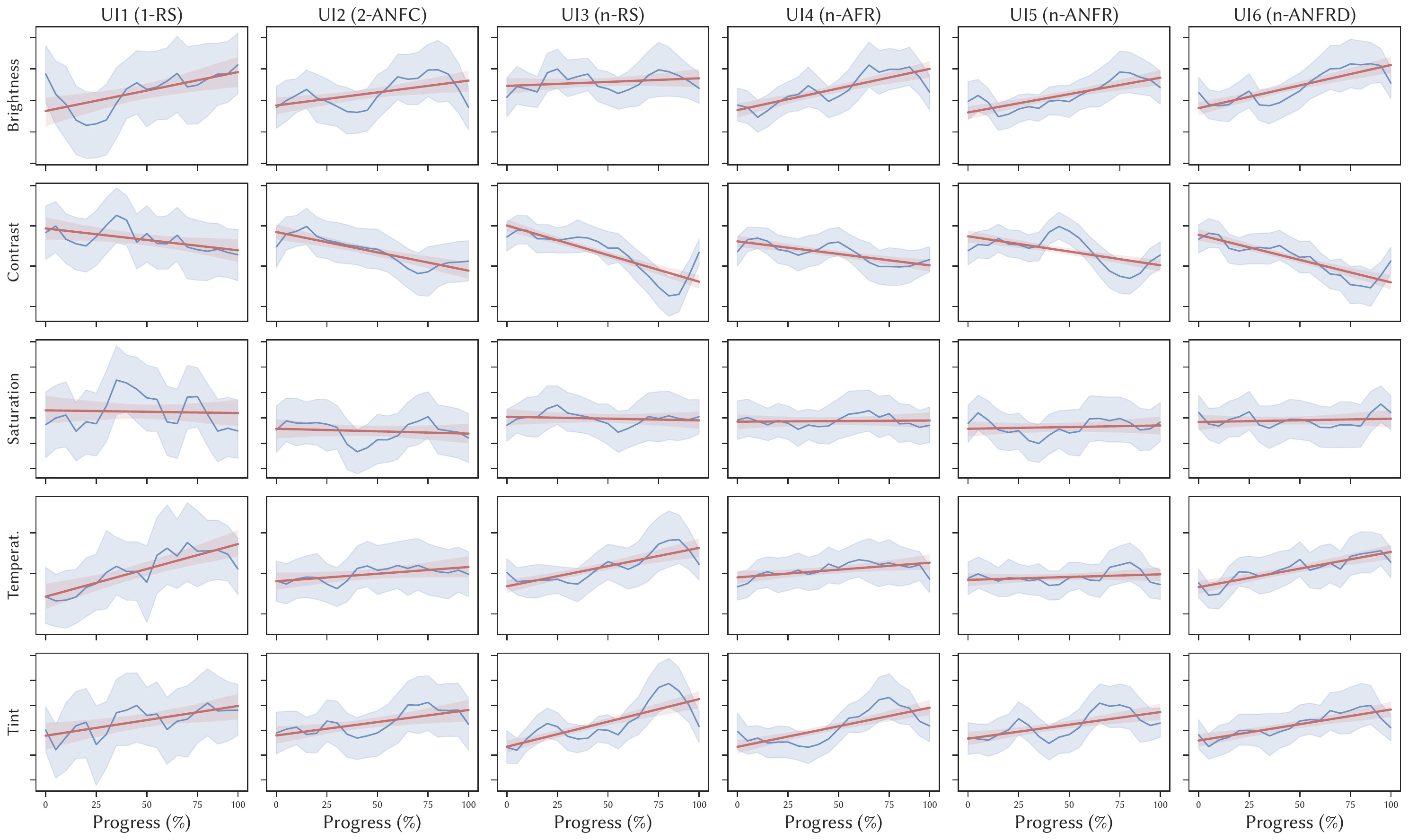}
        \caption{Image Color Enhancement}
        \label{fig:outcomes:image}
    \end{subfigure}
\caption{Comparing objective quality between interfaces. The blue line is the average utility. The red line is a linear regression of the blue line, with each individual slope indicated in red.}
\label{fig:outcomes}
\end{figure}

In summary, our results of the objective quality validate that \emph{all interfaces can optimize objective metrics with a consistent trend}. In more detail, for the text summarization domain, we measured objective quality using \emph{Length}, ROUGE (\emph{ROUGE-1}, \emph{ROUGE-2}, and \emph{ROUGE-L}) and \emph{BLEU}, which compare the system outcome throughout the optimization loop. We show the progress of the HITL optimization over time for the measurements in \autoref{fig:outcomes:text}, and all statistical results are consolidated in \autoref{tab:bo-behavior}.
Regarding summarized text \emph{Length}, UI6 outperforms UI2 and UI4, and the effect of \textsc{Progress} is statistically significant.
Moreover, UI6 outperforms UI3 with respect to \emph{ROUGE-1}, and the effect of \textsc{Progress} is statistically significant for \emph{ROUGE-1}.
For the image enhancement domain, we computed how five different metrics changed with \textsc{Progress} and looked for consistent trends. Namely we investigate \emph{Brightness}, \emph{Contrast}, \emph{Saturation}, \emph{Temperature}, and \emph{Tint}. We show the progress of the HITL optimization over time for the measurements in \autoref{fig:outcomes:image}, and all statistical results are consolidated in \autoref{tab:bo-behavior}. Here, UI6 outperforms UI3 in terms of \emph{Brightness}. Moreover, the effect of \textsc{Progress} is statistically significant for all color metrics except \emph{Saturation}.

\begin{table}[b]
\centering
\caption{Statistics of the optimization loop. The main effect of iteration is evaluated for all measured metrics in text summarization and image color enhancement. Similarly, the main effect \textsc{Interface} is evaluated by the comparison between UI6 to all other interfaces. User input utilities are only compared between rating interfaces (UI1/UI3.) Learned latent utilities are only compared between preferential interfaces (UI2/UI4/UI5.) Results indicate UI6 outperforms all other interfaces in utility, and the progress over iterations showed that the optimization loop has a consistent trend. See also \autoref{fig:outcomes}, \autoref{fig:utility_user}, and \autoref{fig:utility_latent} for visualizations.}
\label{tab:bo-behavior}
\fontsize{5.5}{5.5}\selectfont
\begin{tabularx}{\linewidth}{X*{18}{d{0.2}}}
\toprule
& \multicolumn{3}{c}{\textbf{\textsc{}}}
& \multicolumn{15}{c}{\scriptsize \textbf{\textsc{Opinion Measurement Interfaces} (UI6 vs.)}} \\
\cmidrule(l){5-19}
& \multicolumn{3}{c}{\scriptsize \textbf{\textsc{Progress}}}
& \multicolumn{3}{c}{UI1}
& \multicolumn{3}{c}{UI2}
& \multicolumn{3}{c}{UI3}
& \multicolumn{3}{c}{UI4}
& \multicolumn{3}{c}{UI5} \\
\cmidrule(r){2-4}\cmidrule(lr){5-7}\cmidrule(lr){8-10}
\cmidrule(lr){11-13}\cmidrule(lr){14-16}\cmidrule(l){17-19}
& \multicolumn{1}{c}{$\beta$} & \multicolumn{1}{c}{SE} & \multicolumn{1}{c}{p} 
& \multicolumn{1}{c}{$\beta$} & \multicolumn{1}{c}{SE} & \multicolumn{1}{c}{p} 
& \multicolumn{1}{c}{$\beta$} & \multicolumn{1}{c}{SE} & \multicolumn{1}{c}{p}
& \multicolumn{1}{c}{$\beta$} & \multicolumn{1}{c}{SE} & \multicolumn{1}{c}{p}
& \multicolumn{1}{c}{$\beta$} & \multicolumn{1}{c}{SE} & \multicolumn{1}{c}{p}
& \multicolumn{1}{c}{$\beta$} & \multicolumn{1}{c}{SE} & \multicolumn{1}{c}{p} \\
\midrule

\multicolumn{19}{l}{\textbf{Text Summerization}}\\
\multicolumn{1}{l}{User Feedback}
& .005 & .001 & \multicolumn{1}{B{.}{.}{1,3}}{<.001}
& -.078 & .031 & \multicolumn{1}{B{.}{.}{1,3}}{.012}
& \multicolumn{1}{c}{–} & \multicolumn{1}{c}{–} & \multicolumn{1}{c}{–}
& -.042 & .031 & .169
& \multicolumn{1}{c}{–} & \multicolumn{1}{c}{–} & \multicolumn{1}{c}{–}
& \multicolumn{1}{c}{–} & \multicolumn{1}{c}{–} & \multicolumn{1}{c}{–} \\
\multicolumn{1}{l}{Learned Latent Utility}
& .011 & .001 & \multicolumn{1}{B{.}{.}{1,3}}{<.001}
& \multicolumn{1}{c}{–} & \multicolumn{1}{c}{–} & \multicolumn{1}{c}{–}
& .229 & .028 & \multicolumn{1}{B{.}{.}{1,3}}{<.001}
& \multicolumn{1}{c}{–} & \multicolumn{1}{c}{–} & \multicolumn{1}{c}{–}
& -.007 & .028 & .780
& -.024 & .028 & .390 \\
\multicolumn{1}{l}{Length}
& -.087 & .020 & \multicolumn{1}{B{.}{.}{1,3}}{<.001}
&  .260 & .427 & .544
& -.914 & .427 & \multicolumn{1}{B{.}{.}{1,3}}{.033}
& -.596 & .427 & .164
& -.949 & .427 & \multicolumn{1}{B{.}{.}{1,3}}{.027}
& -.803 & .427 & .061 \\
\multicolumn{1}{l}{ROUGE-1}
&  .000 & .000 & .495
& -.007 & .004 & \multicolumn{1}{B{.}{.}{1,3}}{.070}
&  .000 & .004 & .897
&  .009 & .004 & \multicolumn{1}{B{.}{.}{1,3}}{.038}
&  .004 & .004 & .333
&  .008 & .004 & .063\\
\multicolumn{1}{l}{ROUGE-2}
& -.000 &  .000 & .905
& -.008 &  .004 & .065
& -.005 &  .004 & .256
&  .007 &  .004 & .114
&  .002 &  .004 & .625
&  .006 &  .004 & .138 \\
\multicolumn{1}{l}{ROUGE-L} 
& -.000 &  .000 & .999
& -.008 &  .005 & .093
& -.006 &  .005 & .189
&  .007 &  .005 & .135
&  .004 &  .005 & .453
&  .007 &  .005 & .177\\
\multicolumn{1}{l}{BLEU}
& -.000 &  .000 &   .553
& -.004 &  .004  &  .218
& -.005 &  .004  &  .186
&  .005 &  .004  &  .160
&  .001 &  .004  &  .675
&  .005 &  .004  &  .165\\
\midrule
\multicolumn{19}{l}{\textbf{Image Color Enhancement}}\\
\multicolumn{1}{l}{User Feedback}
& .012 & .001 & \multicolumn{1}{B{.}{.}{1,3}}{<.001}
& -.123 & .028 & \multicolumn{1}{B{.}{.}{1,3}}{<.001}
& \multicolumn{1}{c}{–} & \multicolumn{1}{c}{–} & \multicolumn{1}{c}{–}
& -.183 & .028 & \multicolumn{1}{B{.}{.}{1,3}}{<.001}
& \multicolumn{1}{c}{–} & \multicolumn{1}{c}{–} & \multicolumn{1}{c}{–}
& \multicolumn{1}{c}{–} & \multicolumn{1}{c}{–} & \multicolumn{1}{c}{–}\\
\multicolumn{1}{l}{Learned Latent Utility}
&  .003 & .001 & \multicolumn{1}{B{.}{.}{1,3}}{.001}
& \multicolumn{1}{c}{–} & \multicolumn{1}{c}{–} & \multicolumn{1}{c}{–}
&  -.170 & .028 & \multicolumn{1}{B{.}{.}{1,3}}{<.001}
& \multicolumn{1}{c}{–} & \multicolumn{1}{c}{–} & \multicolumn{1}{c}{–}
& -.061 & .028 & \multicolumn{1}{B{.}{.}{1,3}}{.032}
& -.062 & .028 & \multicolumn{1}{B{.}{.}{1,3}}{.028}\\
\multicolumn{1}{l}{Brightness}
&  .721  &  .118 & \multicolumn{1}{B{.}{.}{1,3}}{<.001}
& -.896  & 3.012 &  .766
&  1.036 & 2.985 &  .729
&  7.059 & 2.985 & \multicolumn{1}{B{.}{.}{1,3}}{.019}
& -.616  & 2.959 &  .835
& -1.355 & 2.959 &  .647\\
\multicolumn{1}{l}{Contrast}
& -1.245 &  .158 & \multicolumn{1}{B{.}{.}{1,3}}{<.001}
&  3.154 & 4.310 & .465
&  1.344 & 4.271 & .753
&  4.663 & 4.271 & .276
& -3.247 & 4.235 & .444
& -.734  & 4.235 & .862\\
\multicolumn{1}{l}{Saturation}
&   .034 &  .069 & .619
&  2.296 & 1.962 & .243
& -1.503 & 1.945 & .440
&  1.036 & 1.945 & .595
&   .067 & 1.928 & .972
& -1.326 & 1.928 & .492\\
\multicolumn{1}{l}{Temperature}
&   .229 &  .046 &  \multicolumn{1}{B{.}{.}{1,3}}{<.001}
& -1.165 & 1.206 &   .335
&   .804 & 1.195 &   .502
&   .111 & 1.195 &   .926
&  1.210 & 1.185 &   .308
&   .901 & 1.185 &   .448\\
\multicolumn{1}{l}{Tint}
&  .164 &  .038 & \multicolumn{1}{B{.}{.}{1,3}}{<.001}
&  .485 & 1.104 &  .661
&  .433 & 1.094 &  .692
& -.624 & 1.094 &  .569
& -.636 & 1.085 &  .558
&  .182 & 1.085 &  .867\\
\bottomrule
\end{tabularx}
\end{table}

\subsubsection{User Feedback and Optimizer Learned Utility of Preferences}
\label{sec:utility}
When using UI1 and UI3, we asked users to provide absolute ratings of system-generated solutions (for utility-based optimizers), which we consider a \emph{user feedback}. In contrast, in UI2, UI4, and UI5, users provide ranking responses (for preference-based optimizers) that give local preference relations between system outcomes. Based on this, the Bayesian optimizers then infer the \emph{learned latent utility}. Lastly, in UI6, as it mixes absolute ratings and rankings, we can analyze both \emph{user feedback} and the \emph{learned latent utility}. We specify the construction of the measure in \autoref{sec:feedback-measures}. Thus, we can use either the \emph{user feedback} or the \emph{learned latent utility} to compare the optimization progress between UIs. 

\begin{figure}[t]
\centering
\begin{subfigure}[b]{0.49\textwidth}
\centering
\includegraphics[width=\linewidth]{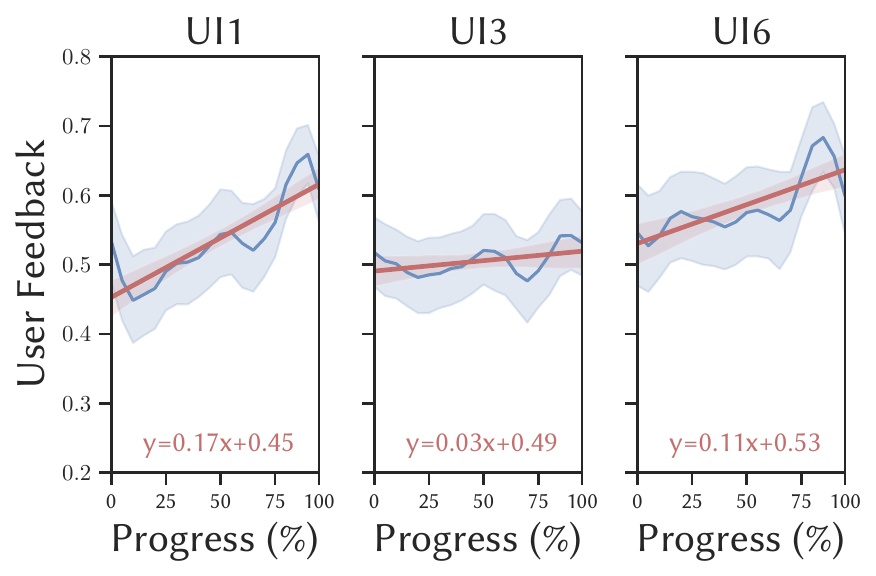}
\caption{Text Summarization}
\end{subfigure}
\hfill
\begin{subfigure}[b]{0.49\textwidth}
\centering
\includegraphics[width=\linewidth]{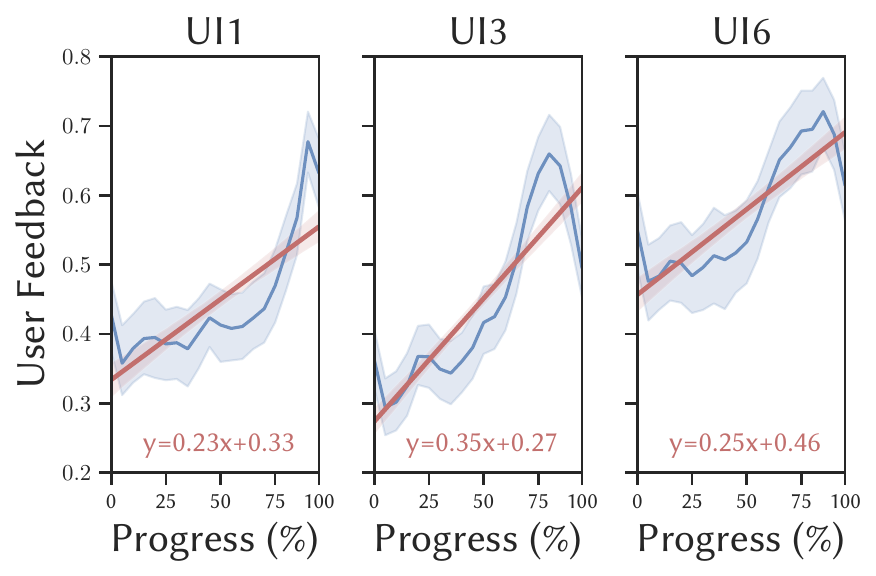}
\caption{Image Color Enhancement}
\end{subfigure}
\caption{The \emph{user feedback} (user input rating utility through the UIs directly) throughout the optimization loop. The blue line is the average \emph{user feedback}, and the red line is a linear regression of the blue line. An increasing \emph{user feedback} trends indicate the optimization goes more toward the users' preferences. Thus, increasing their feedback over time. Here, UI6 outperforms UI1 and UI3 in both domains after completing the optimization.}
\label{fig:utility_user}
\end{figure}

\begin{figure}[t]
\centering
\begin{subfigure}[b]{0.49\textwidth}
\includegraphics[width=\linewidth]{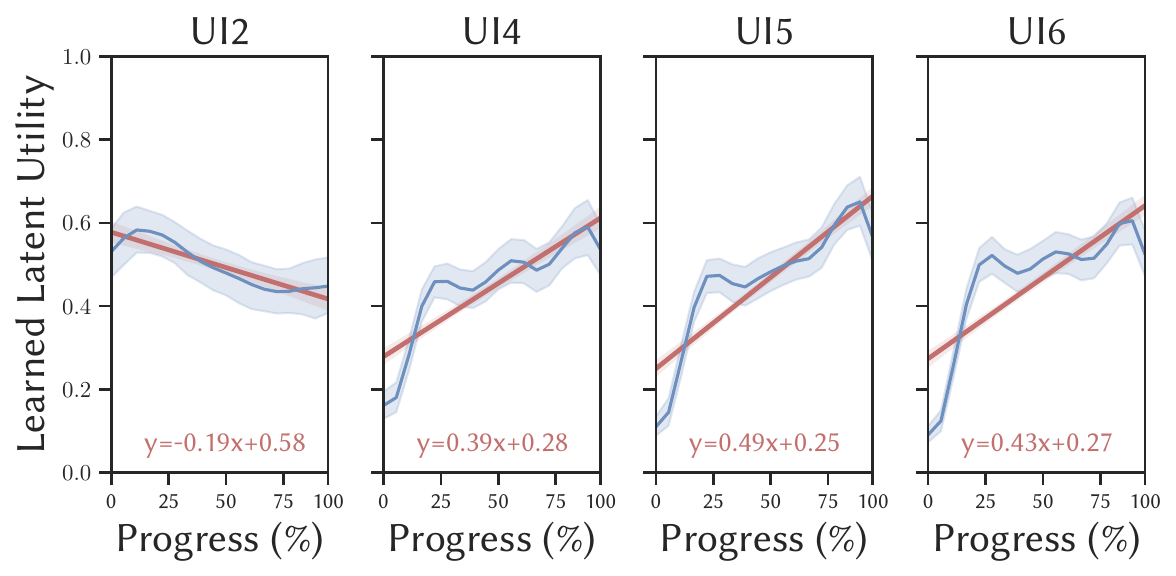}
\caption{Text Summarization}
\end{subfigure}
\hfill
\begin{subfigure}[b]{0.49\textwidth}
\includegraphics[width=\linewidth]{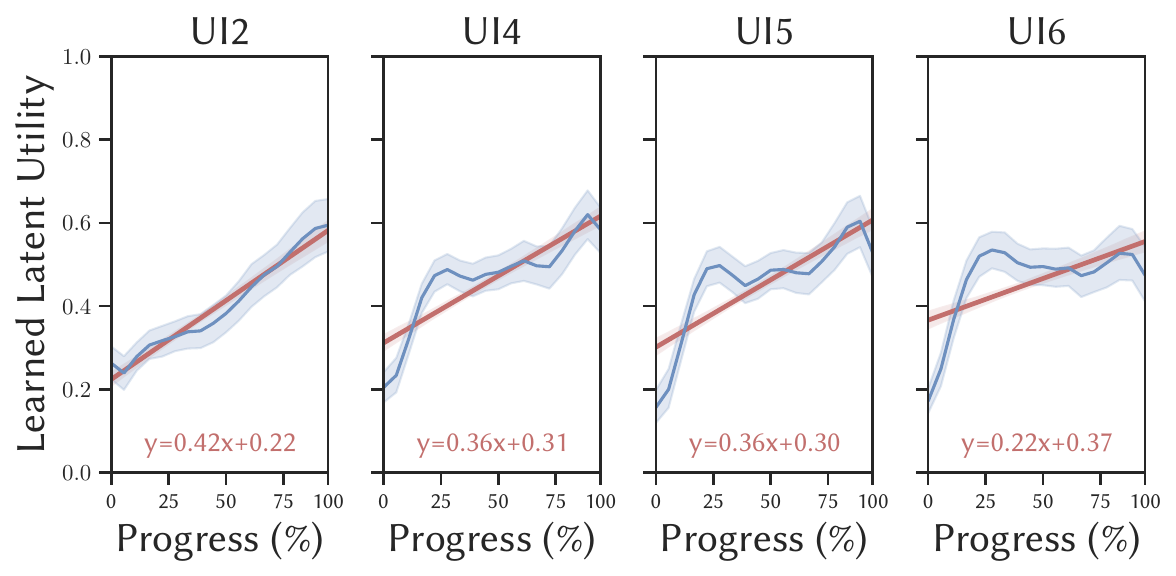}
\caption{Image Color Enhancement}
\end{subfigure}
\caption{The optimizer \emph{learned latent utility} (inferred ranking utility through preferential Bayesian optimization) throughout the optimization loop. The blue line is the average utility, and the red line is a second-order polynomial regression of the blue line. An increasing utility (either direct or latent) trend indicates the optimized text is more toward users' preferences. In latent utility comparisons of the two domains, UI6 outperforms UI2, UI4, and UI5.}
\label{fig:utility_latent}
\end{figure}

In total, we use four LMMs to compare the different UIs using either \emph{user feedback} or \emph{lerned latent utility} as dependent variables. Therefore, we fitted four LMMs regarding \emph{user feedback} and \emph{lerned latent utility} independently from the \textsc{Domain}. We used \textsc{Interface} and \textsc{Progress} as fixed main effects to compare the performance between UIs; see \autoref{fig:utility_user}, \autoref{fig:utility_latent}, and \autoref{tab:bo-behavior}.
First, we investigate the \emph{user feedback} in both domains. 
For the text summarization domain, the LMM results showed the effect of iteration is statistically significant and positive, and UI6 outperforms UI1 and UI3; see ``User feedback'' in \autoref{tab:bo-behavior}.
For the image color enhancement domain, the results show that UI6 significantly outperforms the other UIs.
Second, we investigate the \emph{learned latent utility} in both domains. 
The results of the LMM for the text summarization domain showed the effect of iteration is statistically non-significant and negative. However, UI6 is significantly better than UI2, UI4, and UI5.
For the image color enhancement domain, UI6 significantly outperforms the other UIs.

In summary, UI6 significantly outperforms the other UIs regarding optimizer \emph{lerned latent utility} and \emph{user feedback}. These results show that \emph{through HITL optimizations, UI6 produced significantly higher utility than all other interfaces (UI1-UI5) in both domains}.

\subsection{Questionnaire Results}

\begin{table}[t]
\centering
\caption{Statistics of ending questionnaire answers. Results for H1, H2a, and H2b are tested using Wilcoxon (W$\times 10^5$), and LMMs are used for H3 and H4. Note that $BF_{10}$ are provided for non-significant results.}
\label{tab:qs}
\fontsize{7}{10}\selectfont
\begin{tabularx}{\linewidth}{Xd{2.3}d{1.3}d{2.3}d{1.3}d{2.3}d{1.3}d{1.3}d{1.3}d{2.3}d{1.3}d{2.3}d{1.3}}
\toprule
& \multicolumn{4}{c}{\textbf{Satisfaction (Q1-Q4)}}
& \multicolumn{4}{c}{\textbf{Incompleteness Indication (Q5)}}
& \multicolumn{4}{c}{\textbf{Feedback Clarity (Q6)}}\\
\midrule
\multicolumn{13}{l}{\textbf{Text Summarization}}\\
& \multicolumn{1}{c}{W} & \multicolumn{1}{c}{r} & \multicolumn{1}{c}{BF$_{10}$} & \multicolumn{1}{c}{p}
& \multicolumn{1}{c}{W} & \multicolumn{1}{c}{r} & \multicolumn{1}{c}{BF$_{10}$} & \multicolumn{1}{c}{p}
& \multicolumn{1}{c}{W} & \multicolumn{1}{c}{r} & \multicolumn{1}{c}{BF$_{10}$} & \multicolumn{1}{c}{p}\\
\cmidrule(r){2-5}\cmidrule(lr){6-9}\cmidrule(lr){10-13}
\multicolumn{1}{l}{\textbf{H1}:\hspace{2.1mm}UI1 vs. UI2}
& .028 & -.300 & \multicolumn{1}{c}{–}    & \multicolumn{1}{B{.}{.}{2,3}}{<.001}
& .039 & -.050 & .221 & .578
& .039 & -.050 & .196 & .588 \\
\multicolumn{1}{l}{\textbf{H2a}: UI1 vs. UI3}
& .155 & -.040 & .274 & .511 
& .135 & -.160 & \multicolumn{1}{c}{–} & \multicolumn{1}{B{.}{.}{1,3}}{.008}
& .160 & -.008 & .189 & .902 \\
\multicolumn{1}{l}{\textbf{H2b}: UI2 vs. UI5}
& .053 & .310 & \multicolumn{1}{c}{–} & \multicolumn{1}{B{.}{.}{1,3}}{<.001} 
& .041 & .003 & .174 & .970 
& .047 & .150 & .498 & .076 \\
\cmidrule(r){2-5}\cmidrule(lr){6-9}\cmidrule(lr){10-13}
& \multicolumn{1}{c}{$\beta$} & \multicolumn{1}{c}{SE} & \multicolumn{1}{c}{t} & \multicolumn{1}{c}{p}
& \multicolumn{1}{c}{$\beta$} & \multicolumn{1}{c}{SE} & \multicolumn{1}{c}{t} & \multicolumn{1}{c}{p} 
& \multicolumn{1}{c}{$\beta$} & \multicolumn{1}{c}{SE} & \multicolumn{1}{c}{t} & \multicolumn{1}{c}{p}\\
\cmidrule(r){2-5}\cmidrule(lr){6-9}\cmidrule(lr){10-13}
\multicolumn{1}{l}{\hspace{5mm}UI5 vs.}
& 59.306 & 2.694 & 22.015 & <.001 & 35.722 & 4.484 & 7.967 & <.001 & 71.222 & 3.254 & 21.886 & <.001\\
\multicolumn{1}{l}{\textbf{H3a}: \hspace{6.5mm}UI3} 
& -3.483 & 3.810 & -.914 & .363 & 1.022 & 6.341 & .161 & .872 &  1.744 & 4.602 &  .379 & .706\\
\multicolumn{1}{l}{\textbf{H3b}: \hspace{6.5mm}UI4}
& -1.692 & 3.810 & -.444 & .658 & 8.144 & 6.341 & 1.284 & .202 &  2.200 & 4.602 & .478 & .634\\
\cmidrule(r){2-5}\cmidrule(lr){6-9}\cmidrule(lr){10-13}
\multicolumn{1}{l}{\hspace{5mm}UI6 vs.} 
& 64.669 & 2.740 & 23.601 & <.001 & 38.956 & 4.335 &  8.986 & <.001 & 75.211 & 3.111 & 24.177 & <.001\\
\multicolumn{1}{l}{\textbf{H4a}: \hspace{6.5mm}UI2}
&  4.675 & 3.875 &  1.206 & .230 & -3.233 & 6.131 & -.527 & .599 & .878 & 4.399 & .200 & .842\\
\multicolumn{1}{l}{\textbf{H4b}: \hspace{6.5mm}UI3}
& -8.847 & 3.875 & -2.283 & .024 & -2.211 & 6.131 & -.361 & .719 & -2.244 & 4.399 & -.510 & .611\\
\multicolumn{1}{l}{\textbf{H4c}: \hspace{6.7mm}UI4}
& -7.056 & 3.875 & -1.821 & .071 & 4.911 & 6.131 & .801 & .424 & -1.789 & 4.399 & -.407 & .685\\
\multicolumn{1}{l}{\textbf{H4d}: \hspace{6.5mm}UI5}
& -5.364 & 3.875 & -1.384 & .168 & -3.233 & 6.131 & -0.527 &  .599 & -3.989 & 4.399 & -.907 & .366\\
\midrule
\multicolumn{13}{l}{\textbf{Image Color Enhancement}}\\
& \multicolumn{1}{c}{W} & \multicolumn{1}{c}{r} & \multicolumn{1}{c}{BF$_{10}$} & \multicolumn{1}{c}{p}
& \multicolumn{1}{c}{W} & \multicolumn{1}{c}{r} & \multicolumn{1}{c}{BF$_{10}$} & \multicolumn{1}{c}{p}
& \multicolumn{1}{c}{W} & \multicolumn{1}{c}{r} & \multicolumn{1}{c}{BF$_{10}$} & \multicolumn{1}{c}{p}\\
\cmidrule(r){2-5}\cmidrule(lr){6-9}\cmidrule(lr){10-13}
\multicolumn{1}{l}{\textbf{H1}:\hspace{2.1mm}UI1 vs. UI2}
& .037, & -.10 & .251 & .256
& .025, & -.37 & \multicolumn{1}{c}{–} & \multicolumn{1}{B{.}{.}{1,3}}{< .001}
& .052, &  .28 & \multicolumn{1}{c}{–} & \multicolumn{1}{B{.}{.}{1,3}}{.001} \\
\multicolumn{1}{l}{\textbf{H2a}: UI1 vs. UI3}
& .035 & -.14 & .634 & .098
& .030 & -.26 & \multicolumn{1}{c}{–} & \multicolumn{1}{B{.}{.}{1,3}}{.002}
& .038 & -.07 & .185 & .412\\
\multicolumn{1}{l}{\textbf{H2b}: UI2 vs. UI5}
& .033 & -.19 & \multicolumn{1}{c}{–} & \multicolumn{1}{B{.}{.}{1,3}}{.029}
& .049 &  .22 & \multicolumn{1}{c}{–} & \multicolumn{1}{B{.}{.}{1,3}}{.011}
& .028 & -.32 & \multicolumn{1}{c}{–} & \multicolumn{1}{B{.}{.}{1,3}}{<.001}\\
\cmidrule(r){2-5}\cmidrule(lr){6-9}\cmidrule(lr){10-13}
& \multicolumn{1}{c}{$\beta$} & \multicolumn{1}{c}{SE} & \multicolumn{1}{c}{t} & \multicolumn{1}{c}{p}
& \multicolumn{1}{c}{$\beta$} & \multicolumn{1}{c}{SE} & \multicolumn{1}{c}{t} & \multicolumn{1}{c}{p} 
& \multicolumn{1}{c}{$\beta$} & \multicolumn{1}{c}{SE} & \multicolumn{1}{c}{t} & \multicolumn{1}{c}{p}\\
\cmidrule(r){2-5}\cmidrule(lr){6-9}\cmidrule(lr){10-13}
\multicolumn{1}{l}{\hspace{5mm}UI5 vs.}
&  64.344 & 2.760  & 23.312 & <.001 &  36.189 & 4.042  &  8.954 & <.001 & 78.822 & 2.799 & 28.158 & <.001 \\
\multicolumn{1}{l}{\textbf{H3a}: \hspace{6.5mm}UI3} 
&  -3.597 & 3.903  & -.922 &  .359 &   5.144 & 5.716  &  .900 &  .371 &  .678 & 3.959 &  .171 &  .864 \\
\multicolumn{1}{l}{\textbf{H3b}: \hspace{6.5mm}UI4}
&  -2.722 & 3.903  & -.697 &  .487 &   6.756 & 5.716  &  1.182 &  .240 & -2.078 & 3.959 & -.525 &  .601 \\
\cmidrule(r){2-5}\cmidrule(lr){6-9}\cmidrule(lr){10-13}
\multicolumn{1}{l}{\hspace{5mm}UI6 vs.} 
& 60.061 & 2.786 & 21.558 & <.001 & 21.900 & 3.912 & 5.599 & <.001 &  82.800 & 2.723 & 30.408 & < .001 \\
\multicolumn{1}{l}{\textbf{H4a}: \hspace{6.5mm}UI2}
& -1.894 & 3.940 & -0.481 &  .631 & 24.644 & 5.532 & 4.455 & \multicolumn{1}{B{.}{.}{1,3}}{<.001} & -13.933 & 3.851 & -3.618 & \multicolumn{1}{B{.}{.}{1,3}}{< .001} \\
\multicolumn{1}{l}{\textbf{H4b}: \hspace{6.5mm}UI3}
&   .686 & 3.940 &  .174 &  .862 & 19.433 & 5.532 & 3.513 &  \multicolumn{1}{B{.}{.}{1,3}}{.001} &  -3.300 & 3.851 & -0.857 &   .393 \\
\multicolumn{1}{l}{\textbf{H4c}: \hspace{6.7mm}UI4}
&  1.561 & 3.940 &  .396 &  .693 & 21.044 & 5.532 & 3.804 & \multicolumn{1}{B{.}{.}{1,3}}{<.001} &  -6.056 & 3.851 & -1.573 &   .118 \\
\multicolumn{1}{l}{\textbf{H4d}: \hspace{6.5mm}UI5}
&  4.283 & 3.940 &  1.087 &  .279 & 14.289 & 5.532 & 2.583 &  \multicolumn{1}{B{.}{.}{1,3}}{.011} &  -3.978 & 3.851 & -1.033 &   .303 \\
\bottomrule
\end{tabularx}
\end{table}

\begin{figure}[t]
\centering
\begin{subfigure}[b]{0.49\textwidth}
\includegraphics[width=\linewidth]{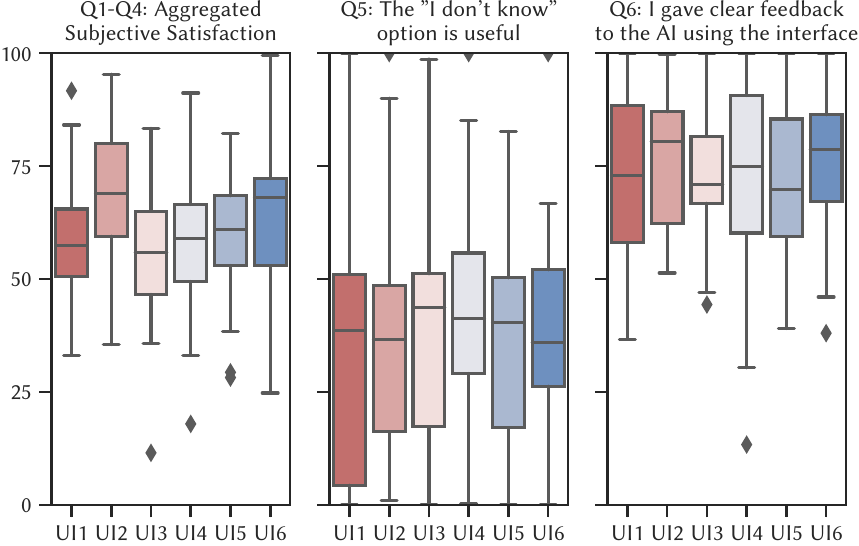}
\caption{Text Summarization}
\end{subfigure}
\hfill
\begin{subfigure}[b]{0.49\textwidth}
\includegraphics[width=\linewidth]{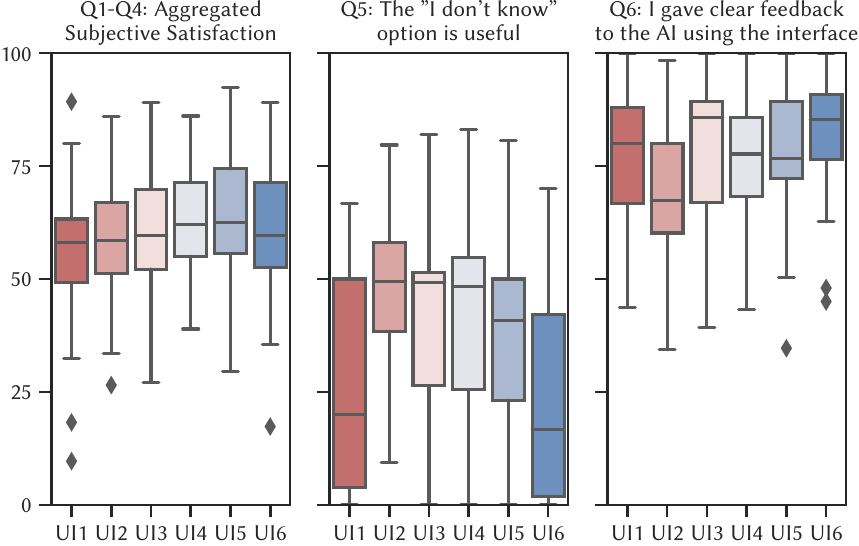}
\caption{Image Color Enhancement}
\end{subfigure}
\caption{Questionnaire results. Participants answered six questions per task and were collected using a bipolar slider-based Likert scale from text summarization and image color enhancement tasks. 100 means ``Strongly agree``, and 0 means ``Strongly disagree''. Participants also provided open-ended responses, which are quantified to sentiment ratings. The results indicate significant differences between the pointwise (UI1) and pairwise (UI2) approaches, and the hybrid approach (UI6) outperforms other listwise approaches (UI3, UI4, and UI5) but shares similar satisfaction and sentiment compared to the pairwise interface (UI2). When using UI6, participants considered expressing incomplete preference using ``I don't know'' as the least useful.}
\label{fig:questionnaires}
\end{figure}

We asked participants six questions (Q1-Q6) to verify their satisfaction (see~\autoref{sec:questionnaire-measures}), and the results are shown in \autoref{fig:questionnaires}.
Among these questions, in line with~\citet{ou2023expertise} are intended to measure subjective satisfaction (fairly high Cronbach's $\alpha$=0.792, $CI_{95\%}$=[0.771, 0.812]). Hence, we aggregate these questions to a combined \emph{Satisfaction} indicator.
Overall, all interfaces in both \textsc{Domains} received, on average, positive ratings for satisfaction (>50).

In text summarization, for \textbf{H1}, participants did not report significantly higher satisfaction in UI2 (M=63.75, SD=18.66) compared to UI1 (M=57.02, SD=20.32).
For \textbf{H2a}, also UI1 compared to UI3(M=59.01, SD=18.45) showed no significant difference.
Finally, for \textbf{H2b}, the results show a significant difference when comparing UI2 to UI5 (M=73.67, SD=20.39).
For \textbf{H3} and \textbf{H4}, we found no significant differences between all interfaces (UI2-UI6) in terms of satisfaction.
Similarly, we found allowing expressing incompleteness preference in UI6 is significantly less helpful compared to UI2-UI5 (all p<.050) but not when compared to UI1 in the image color enhancement domain.
For Q6, we found that participants felt they gave more clear feedback when using UI6 compared to UI2 ($\beta = -13.933$, $SE = 3.851$, $t = -3.618$, $p<.001$) - \textbf{H4a} in image color enhancement domain.

\emph{These questionnaire results show that participants were significantly more satisfied (Q1-Q4) with listwise interfaces than with a pairwise interface. They reported that UI6 was clearer in expressing their preference and hence expressed fewer incomplete preferences (Q5) and could give more clear instructions (Q6).}

\section{Discussion}
Overall, we found significant differences in that all other interfaces outperform the pointwise interfaces in both studied domains. Still, there are subtle differences between the list- and pair-wise user interfaces. In this section, we structure the results according to our initial hypotheses and summarize possible implications and trade-offs when determining the use of an interface for HITL optimization.

\subsection{Validating Hypotheses}
\paragraph{H1: Baseline Apparatus Verification}
Based on our results, the 2-ANFC pairwise interface (UI2) outperforms the 1-RS pointwise interface (UI1) because it has a significantly lower decision time per item and fewer iterations. The 2-ANFC also helped participants to express indifference and their incomplete preference for better system parameter exploration and exploitation.
Participants using 2-ANFC also reported significantly higher satisfaction compared to 1-RS. This result verifies that our apparatus is valid and aligns with discussions in prior work~\cite{brochu2007active, brochu2007material-design, brochu2010animation}.

\paragraph{H2: The Listwise Approach}
In terms of UI interactions, our results showed that listwise interfaces ($n$-RS/$n$-ANFR, UI3/UI5) provide the following advantages (compared to 1-RS/2-ANFC, UI1/UI2):
1)~they allow users to evaluate more options without increasing decision time and iterations;
2)~users express more indifferent preferences and less incomplete preferences;
3)~they allow implicit comparisons among all given options.
However, the direct use of the listwise approach (i.e., using more sliders to rate more items than 1-RS or rank multiple items without distance in $n$-ANFR) did not significantly improve the performance of the optimization loop.

\paragraph{H3: Design Variations of The Listwise Approach}
When comparing design variations regarding querying different feedback types for ranking (\textbf{H3a}: $n$-RS vs. $n$-ANFR, i.e., UI3 vs. UI5), allowing expressing indifference and incomplete preferences (\textbf{H3b}: $n$-AFR vs. $n$-ANFR, i.e., UI4 vs. UI5) in listwise approach,
we did not find enough evidence to show a difference in decision time, involved iterations, and the performance of the optimization loop. The subjectively reported satisfaction also did not differ significantly from each other.
However, $n$-AFR expresses significantly more incomplete preferences compared to $n$-ANFR.

\paragraph{H4: The Hybrid Approach}
We found that the hybrid approach ($n$-ANFRD, UI6) did not differ from 2-ANFC (\textbf{H4a}), $n$-RS (\textbf{H4b}), $n$-AFR (\textbf{H4c}), $n$-ANFR (\textbf{H4d}) in terms of decision time per item and involved iterations, and subjective satisfaction.
However, $n$-ANFR lets participants express indifference significantly more than other interfaces, resulting in fewer ranking interactions. When checking the performance of the optimization loop, $n$-ANFRD showed significantly higher user input utility throughout the loop and higher latent preferent utility compared to pairwise and other listwise interfaces.

\subsection{Interpretation and Implications}

While we initially assumed using a listwise approach that increased the number of parallel comparisons could improve the overall performance, the results did not support this assumption, particularly for acquired decision time. Based on the validation of \textbf{H3}, neither $n$-RS (UI3, utility-based) nor $n$-ANFR (UI5, preference-based) show a significant advantage and only increased the overall decision time in the loop compared to 2-ANFC (UI2, preference-based).
This might be because neither $n$-RS nor $n$-ANFR provided enough support for participants to reduce the overall noisy observations, and there is room for improvement when using a BO optimizer specifically designed for listwise UIs. In $n$-RS, despite multiple given alternatives, participants might not notice the relative position between different sliders; in $n$-ANFR, the UI did not demonstrate enough absolute cues to participants on how pure ranking can influence their input, and a pure local ranking may not give participants information on a global overview.

Instead, the $n$-ANFRD (UI6) hybrids collect utility-based and preference-based feedback by providing rating scales in the regular listwise ranking UI. Similar to other listwise UIs, based on the validation of \textbf{H4}, although $n$-ANFRD increased total decision time, our results show the general advantage of preference elicitation (expressing more indifference, fewer ranking interactions, fewer incomplete preferences, and better performance in the optimization loop).
This could be interpreted as the UI starting to offer participants implicit recall support for their past ratings if their ranking positions are improving. In addition, $n$-ANFRD resulted in fewer ranking interactions and made it clearer for participants to express a ranking decision compared to other purely listwise ranking interfaces.
Furthermore, our results regarding designing UIs to support indifference and incomplete preferences offer implications for future modeling algorithms. In Bayesian optimization, indifference preference behavior could be an indicator to inform the acquisition function to avoid exploiting these areas; incomplete preferences might be another indicator to balance exploration and exploitation.
Depending on the context, one might design the algorithm to a)~avoid exploring a space where people had an incomplete preference to reduce input uncertainty, or alternatively b)~explore it even more for creativity support and then integrate the measured preferences with more interaction behavior, such as interface features and task completion time~\cite{dudley2019crowdsourcing}, to better support the optimization process.

\subsection{Design Trade-Offs}
As we validated in our hypotheses, there are many subtle differences between pairwise and listwise interfaces when permitting users to express indifference and incomplete preferences. This gives us evidence of how they could impact HITL optimization performance. Below, we discuss potential trade-offs when using these interfaces.
\subsubsection{Guideline 1: Pointwise Approach}
The pointwise slider-based interface has proven useful in many disciplines, such as psychology, subjective scale measures, etc. It is useful to identify social agreement among populations and measure the direct utility given by a human. The technique for processing these types of data is also mature. But it is precarious and problematic when directly applying it to individual sequential measurements. Future HITL optimization systems or any sequential measurement should generally avoid using this approach.
\subsubsection{Guideline 2: Pairwise Approach}
A pairwise interface only requires a human to judge based on the comparison of two given items. Since there is no measured utility and it only provides relative information about two items, the modeling of the latent utility function could suffer from a change of criteria and intransitive preference when the anchoring point between two sessions is not chosen properly.
When an iteration receives an incomplete preference, the underlying optimizer could not receive much useful modeling information.
However, when a design objective is to optimize for fast decisions, the pairwise approach (2-ANFC) is a winner among alternative listwise designs.
\subsubsection{Guideline 3: Listwise Approach}
Listwise interfaces establish a comparison by increasing the number of alternatives.
Similar to the pairwise interface, a listwise interface without measuring ranking distance ($n$-AFR and $n$-ANFR in our study) requires a properly sampled anchoring option between iterations to establish a ranking function.
Otherwise, there is no direct association between iterations. Instead, a listwise interface that allows expressing ranking distance ($n$-RS and $n$-ANFRD in our study) conveys an implicit association between different iterations regardless of whether the human provides local or global judgments it is more suited to use in a HITL optimization context.
Furthermore, $n$-ANFRD is better than $n$-RS because a concentrated participant may not notice the relative slider position between parallel judging options in $n$-RS, but $n$-ANFRD explicitly visualizes the ranking in the UI for users, which helps users more clearly express their opinion in a sequential task.
As design suggestions, when there is a limited measuring budget, e.g., one can only query humans for a very limited number of opinions, applying listwise interfaces is considered better than pairwise because it provides more measurement of their opinion and does not decrease the optimization performance.
Furthermore, allowing users to express the ranking distance ($n$-ANFRD) is the winner of all listwise variations ($n$-RS, $n$-AFR, and $n$-ANFR).
\subsubsection{Guideline 4: Measuring Indifference and Incomplete Preferences}
When measuring more accurate user opinions, allowing them to express indifference and incomplete preferences is crucial. For the pointwise interface, these two different types of preference were implicitly ignored and treated as being indicated by centering the slider. This interpretation of a central slider position may be true with proper task wording. Still, it may be easily misused and produce more noisy input when a user does not know how to express their opinion with the interface. Therefore, in the HITL optimization scenario, it is better to explicitly design these two functionalities to support users in expressing their indifference and incomplete preferences. This information could also inform optimizers to avoid exploring these areas. This is particularly important when the evaluation budget is very limited in the HITL optimization context.

\subsection{Limitations and Future Work}
Our study used a text summarization task and an image color enhancement task to explore the impact of interfaces in HITL optimization. While these domains are common and make it easy to run a study online at scale, we did not verify our results on more domains. Although we did not find a large difference between the text and image domains, the same study paradigm could be evaluated in other domains, such as the 3D domain, where visual decisions are harder to make due to the complexity of geometry, which may result in different observed behaviors. This might be interesting for future investigation.

Since our optimization context is highly dependent on humans and modeled human feedback as objective functions, the system performance highly depends on the quality of human input. The responses collected online may be influenced by participants’ environments, such as slow decision time due to external distractions. Moreover, for the preferential settings, depending on the underlying optimizer (e.g., \citet{brochu2007active}), the decision time might be different if the system repeatedly returns the current optimum.

Our study evaluated opinion measurement interfaces, not optimizers. However, differently designed listwise interfaces could benefit from dedicated Bayesian optimizers. Without a specialized BO optimizer, our suggested $n$-ANFRD interface supports the user to express a ranking distance and still outperforms other interfaces in various aspects, such as fewer ranking interactions, more clearly stated indifference, and rarely expressed incomplete preferences. We also did not look into a detailed inspection of the cause of indifference and incomplete preference. This leaves a large potential for 1) future HCI research to study factors, such as involved expertise, and 2) future BO algorithmic research to model the influence of different optimizers on the same interface, the interaction effects between interfaces and optimizers, and more realistic modeling of indifference and incomplete preference.

Lastly, we focused on the design space for aggregated preferences and used Bayesian optimizers that assumed one latent preference function. However, this could be limited by the human preference aggregation process, where multiple criteria may not be expressed simultaneously. Future work can extend the design space to multi-criteria opinion measurement interfaces and further exploit multi-objective Bayesian optimization methods~\cite{benavoli2021choice, daulton2020differentiable}. Similarly, our discussed taxonomy and design space only considered structured measures. Hence, we did not explore UIs that utilize unstructured input, such as text prompts~\cite{dang2022prompt} and human-sketched edits~\cite{chong2021interactive} of generative models.

\section{Conclusion}
In this paper, we revisited the design space of opinion measurement interfaces in the HITL optimization context. Concretely, we aimed to determine the best-suited interface to query human judgments in sequential evaluation.
We found subtle differences between pairwise opinion measurement interfaces and listwise ones: In particular, listwise interfaces helped users express more indifference preferences over incomplete preferences than the pairwise interface. Moreover, the hybrid interface that mixes the utility-based and preferential ranking-based feedback to counterbalance the limitations of each approach could better emphasize the relative relation of options for users and achieve better HITL optimization performance. As winners, both $n$-ANFRD and 2-ANFC have unique advantages. The design decision between them is a trade-off of decision time, the precision of measured human feedback, and overall optimization loop performance.

Looking ahead, the exploration of interface designs for HITL optimization does not end here: Our considered design space only covers structured human inputs bound to suitable optimizers, i.e., utility and preferential data.
With the increasing developments in large language models, it would also be interesting and challenging to explore, quantify, and evaluate human-in-the-loop interfaces that allow unstructured inputs such as using a text prompt~\cite{dang2022prompt} to alter media data using diffusion models~\cite{rombach2022stablediffusion}. Our results here provide a starting point in this direction.

\section*{Open Science}
\label{sec:open-science}
We encourage readers to reproduce and extend our results. Our developed user interfaces, optimizers, collected datasets, and analysis scripts are open source and made available at \url{https://github.com/changkun/rankui}.

\begin{acks}
We thank
Robin Welsch for discussions of multilevel modeling,
Fiona Draxler for feedback on measuring language proficiency,
Eyke Hüllermeier and Clemens Damke for discussions of ranking analysis,
Jerry Lin for the discussion of pairwise Gaussian processes,
Hai Dang for the discussion of large language models,
Francesco Chiossi for the discussion of modeling reaction time distributions,
and Yifei Zhan for suggestions on the ranking interface implementation.
Work on this project is also partly funded by the Bavarian State Ministry of Science and the Arts and coordinated by the Bavarian Research Institute for Digital Transformation (bidt).
\end{acks}

\bibliographystyle{ACM-Reference-Format}
\bibliography{ref}


\begin{thebibliography}{76}


\ifx \showCODEN    \undefined \def \showCODEN     #1{\unskip}     \fi
\ifx \showDOI      \undefined \def \showDOI       #1{#1}\fi
\ifx \showISBNx    \undefined \def \showISBNx     #1{\unskip}     \fi
\ifx \showISBNxiii \undefined \def \showISBNxiii  #1{\unskip}     \fi
\ifx \showISSN     \undefined \def \showISSN      #1{\unskip}     \fi
\ifx \showLCCN     \undefined \def \showLCCN      #1{\unskip}     \fi
\ifx \shownote     \undefined \def \shownote      #1{#1}          \fi
\ifx \showarticletitle \undefined \def \showarticletitle #1{#1}   \fi
\ifx \showURL      \undefined \def \showURL       {\relax}        \fi
\providecommand\bibfield[2]{#2}
\providecommand\bibinfo[2]{#2}
\providecommand\natexlab[1]{#1}
\providecommand\showeprint[2][]{arXiv:#2}

\bibitem[Armstrong(1987)]%
        {armstrong1987mid}
\bibfield{author}{\bibinfo{person}{Robert~L. Armstrong}.}
  \bibinfo{year}{1987}\natexlab{}.
\newblock \showarticletitle{The Midpoint on a Five-Point Likert-Type Scale}.
\newblock \bibinfo{journal}{\emph{Perceptual and Motor Skills}}
  \bibinfo{volume}{64}, \bibinfo{number}{2} (\bibinfo{date}{Apr}
  \bibinfo{year}{1987}), \bibinfo{pages}{359–362}.
\newblock
\showISSN{0031-5125}
\urldef\tempurl%
\url{https://doi.org/10.2466/pms.1987.64.2.359}
\showDOI{\tempurl}


\bibitem[Balandat et~al\mbox{.}(2020)]%
        {balandat2020botorch}
\bibfield{author}{\bibinfo{person}{Maximilian Balandat}, \bibinfo{person}{Brian
  Karrer}, \bibinfo{person}{Daniel Jiang}, \bibinfo{person}{Samuel Daulton},
  \bibinfo{person}{Ben Letham}, \bibinfo{person}{Andrew~G Wilson}, {and}
  \bibinfo{person}{Eytan Bakshy}.} \bibinfo{year}{2020}\natexlab{}.
\newblock \showarticletitle{BoTorch: A Framework for Efficient Monte-Carlo
  Bayesian Optimization}. In \bibinfo{booktitle}{\emph{Advances in Neural
  Information Processing Systems}},
  \bibfield{editor}{\bibinfo{person}{H.~Larochelle},
  \bibinfo{person}{M.~Ranzato}, \bibinfo{person}{R.~Hadsell},
  \bibinfo{person}{M.F. Balcan}, {and} \bibinfo{person}{H.~Lin}} (Eds.),
  Vol.~\bibinfo{volume}{33}. \bibinfo{publisher}{Curran Associates, Inc.},
  \bibinfo{address}{Virtual}, \bibinfo{pages}{21524--21538}.
\newblock
\urldef\tempurl%
\url{https://proceedings.neurips.cc/paper/2020/file/f5b1b89d98b7286673128a5fb112cb9a-Paper.pdf}
\showURL{%
\tempurl}


\bibitem[Bates et~al\mbox{.}(2014)]%
        {bates2014fitting}
\bibfield{author}{\bibinfo{person}{Douglas Bates}, \bibinfo{person}{Martin
  Mächler}, \bibinfo{person}{Ben Bolker}, {and} \bibinfo{person}{Steve
  Walker}.} \bibinfo{year}{2014}\natexlab{}.
\newblock \bibinfo{title}{Fitting Linear Mixed-Effects Models using lme4}.
\newblock
\newblock
\urldef\tempurl%
\url{https://doi.org/10.48550/arXiv.1406.5823}
\showDOI{\tempurl}


\bibitem[Benavoli et~al\mbox{.}(2021)]%
        {benavoli2021choice}
\bibfield{author}{\bibinfo{person}{Alessio Benavoli}, \bibinfo{person}{Dario
  Azzimonti}, {and} \bibinfo{person}{Dario Piga}.}
  \bibinfo{year}{2021}\natexlab{}.
\newblock \bibinfo{title}{Choice functions based multi-objective Bayesian
  optimisation}.
\newblock
\newblock
\urldef\tempurl%
\url{http://arxiv.org/abs/2110.08217}
\showURL{%
\tempurl}


\bibitem[Bettman et~al\mbox{.}(1998)]%
        {bettman1998constructive}
\bibfield{author}{\bibinfo{person}{James~R Bettman},
  \bibinfo{person}{Mary~Frances Luce}, {and} \bibinfo{person}{John~W Payne}.}
  \bibinfo{year}{1998}\natexlab{}.
\newblock \showarticletitle{Constructive consumer choice processes}.
\newblock \bibinfo{journal}{\emph{Journal of consumer research}}
  \bibinfo{volume}{25}, \bibinfo{number}{3} (\bibinfo{year}{1998}),
  \bibinfo{pages}{187--217}.
\newblock


\bibitem[Bhandari et~al\mbox{.}(2020)]%
        {bhandari2020re}
\bibfield{author}{\bibinfo{person}{Manik Bhandari},
  \bibinfo{person}{Pranav~Narayan Gour}, \bibinfo{person}{Atabak Ashfaq},
  \bibinfo{person}{Pengfei Liu}, {and} \bibinfo{person}{Graham Neubig}.}
  \bibinfo{year}{2020}\natexlab{}.
\newblock \showarticletitle{Re-evaluating Evaluation in Text Summarization}. In
  \bibinfo{booktitle}{\emph{Proceedings of the 2020 Conference on Empirical
  Methods in Natural Language Processing (EMNLP)}}.
  \bibinfo{publisher}{Association for Computational Linguistics},
  \bibinfo{address}{Online}, \bibinfo{pages}{9347--9359}.
\newblock
\urldef\tempurl%
\url{https://doi.org/10.18653/v1/2020.emnlp-main.751}
\showDOI{\tempurl}


\bibitem[Brochu et~al\mbox{.}(2010)]%
        {brochu2010animation}
\bibfield{author}{\bibinfo{person}{Eric Brochu}, \bibinfo{person}{Tyson
  Brochu}, {and} \bibinfo{person}{Nando de Freitas}.}
  \bibinfo{year}{2010}\natexlab{}.
\newblock \showarticletitle{A Bayesian Interactive Optimization Approach to
  Procedural Animation Design}. In \bibinfo{booktitle}{\emph{Proceedings of the
  2010 ACM SIGGRAPH/Eurographics Symposium on Computer Animation}} (Madrid,
  Spain) \emph{(\bibinfo{series}{SCA '10})}. \bibinfo{publisher}{Eurographics
  Association}, \bibinfo{address}{Goslar, DEU}, \bibinfo{pages}{103–112}.
\newblock


\bibitem[Brochu et~al\mbox{.}(2007a)]%
        {brochu2007active}
\bibfield{author}{\bibinfo{person}{Eric Brochu}, \bibinfo{person}{Nando~de
  Freitas}, {and} \bibinfo{person}{Abhijeet Ghosh}.}
  \bibinfo{year}{2007}\natexlab{a}.
\newblock \showarticletitle{Active Preference Learning with Discrete Choice
  Data}. In \bibinfo{booktitle}{\emph{Proceedings of the 20th International
  Conference on Neural Information Processing Systems}} (Vancouver, British
  Columbia, Canada) \emph{(\bibinfo{series}{NIPS'07})}.
  \bibinfo{publisher}{Curran Associates Inc.}, \bibinfo{address}{Red Hook, NY,
  USA}, \bibinfo{pages}{409–416}.
\newblock
\showISBNx{9781605603520}


\bibitem[Brochu et~al\mbox{.}(2007b)]%
        {brochu2007material-design}
\bibfield{author}{\bibinfo{person}{Eric Brochu}, \bibinfo{person}{Abhijeet
  Ghosh}, {and} \bibinfo{person}{Nando de Freitas}.}
  \bibinfo{year}{2007}\natexlab{b}.
\newblock \showarticletitle{Preference galleries for material design}. In
  \bibinfo{booktitle}{\emph{{ACM} {SIGGRAPH} 2007 posters}}
  \emph{(\bibinfo{series}{{SIGGRAPH} '07})}. \bibinfo{publisher}{ACM Press},
  \bibinfo{address}{New York, NY, USA}, \bibinfo{pages}{105--es}.
\newblock
\showISBNx{978-1-4503-1828-0}
\urldef\tempurl%
\url{https://doi.org/10.1145/1280720.1280834}
\showURL{%
\tempurl}


\bibitem[Cao et~al\mbox{.}(2007)]%
        {cao2007learning}
\bibfield{author}{\bibinfo{person}{Zhe Cao}, \bibinfo{person}{Tao Qin},
  \bibinfo{person}{Tie-Yan Liu}, \bibinfo{person}{Ming-Feng Tsai}, {and}
  \bibinfo{person}{Hang Li}.} \bibinfo{year}{2007}\natexlab{}.
\newblock \showarticletitle{Learning to rank: from pairwise approach to
  listwise approach}. In \bibinfo{booktitle}{\emph{Proceedings of the 24th
  international conference on Machine learning}} \emph{(\bibinfo{series}{ICML
  ’07})}. \bibinfo{publisher}{Association for Computing Machinery},
  \bibinfo{address}{New York, NY, USA}, \bibinfo{pages}{129–136}.
\newblock
\showISBNx{978-1-59593-793-3}
\urldef\tempurl%
\url{https://doi.org/10.1145/1273496.1273513}
\showDOI{\tempurl}


\bibitem[Chan et~al\mbox{.}(2022)]%
        {chan2022chi}
\bibfield{author}{\bibinfo{person}{Liwei Chan}, \bibinfo{person}{Yi-Chi Liao},
  \bibinfo{person}{George~B Mo}, \bibinfo{person}{John~J Dudley},
  \bibinfo{person}{Chun-Lien Cheng}, \bibinfo{person}{Per~Ola Kristensson},
  {and} \bibinfo{person}{Antti Oulasvirta}.} \bibinfo{year}{2022}\natexlab{}.
\newblock \showarticletitle{Investigating Positive and Negative Qualities of
  Human-in-the-Loop Optimization for Designing Interaction Techniques}. In
  \bibinfo{booktitle}{\emph{Proceedings of the 2022 CHI Conference on Human
  Factors in Computing Systems}} (New Orleans, LA, USA)
  \emph{(\bibinfo{series}{CHI '22})}. \bibinfo{publisher}{Association for
  Computing Machinery}, \bibinfo{address}{New York, NY, USA}, Article
  \bibinfo{articleno}{112}, \bibinfo{numpages}{14}~pages.
\newblock
\showISBNx{9781450391573}
\urldef\tempurl%
\url{https://doi.org/10.1145/3491102.3501850}
\showDOI{\tempurl}


\bibitem[Chong et~al\mbox{.}(2021)]%
        {chong2021interactive}
\bibfield{author}{\bibinfo{person}{Toby Chong}, \bibinfo{person}{I-Chao Shen},
  \bibinfo{person}{Issei Sato}, {and} \bibinfo{person}{Takeo Igarashi}.}
  \bibinfo{year}{2021}\natexlab{}.
\newblock \showarticletitle{Interactive Optimization of Generative Image
  Modelling using Sequential Subspace Search and Content-based Guidance}.
\newblock \bibinfo{journal}{\emph{Computer Graphics Forum}}
  \bibinfo{volume}{40}, \bibinfo{number}{1} (\bibinfo{year}{2021}),
  \bibinfo{pages}{279--292}.
\newblock
\urldef\tempurl%
\url{https://doi.org/10.1111/cgf.14188}
\showDOI{\tempurl}


\bibitem[Christiano et~al\mbox{.}(2017)]%
        {christiano2017deep}
\bibfield{author}{\bibinfo{person}{Paul~F. Christiano}, \bibinfo{person}{Jan
  Leike}, \bibinfo{person}{Tom~B. Brown}, \bibinfo{person}{Miljan Martic},
  \bibinfo{person}{Shane Legg}, {and} \bibinfo{person}{Dario Amodei}.}
  \bibinfo{year}{2017}\natexlab{}.
\newblock \showarticletitle{Deep Reinforcement Learning from Human
  Preferences}. In \bibinfo{booktitle}{\emph{Proceedings of the 31st
  International Conference on Neural Information Processing Systems}} (Long
  Beach, California, USA) \emph{(\bibinfo{series}{NIPS'17})}.
  \bibinfo{publisher}{Curran Associates Inc.}, \bibinfo{address}{Red Hook, NY,
  USA}, \bibinfo{pages}{4302–4310}.
\newblock
\showISBNx{9781510860964}


\bibitem[Chu and Ghahramani(2005)]%
        {chu2005preference}
\bibfield{author}{\bibinfo{person}{Wei Chu} {and} \bibinfo{person}{Zoubin
  Ghahramani}.} \bibinfo{year}{2005}\natexlab{}.
\newblock \showarticletitle{Preference Learning with Gaussian Processes}. In
  \bibinfo{booktitle}{\emph{Proceedings of the 22nd International Conference on
  Machine Learning}} (Bonn, Germany) \emph{(\bibinfo{series}{ICML '05})}.
  \bibinfo{publisher}{Association for Computing Machinery},
  \bibinfo{address}{New York, NY, USA}, \bibinfo{pages}{137–144}.
\newblock
\showISBNx{1595931805}
\urldef\tempurl%
\url{https://doi.org/10.1145/1102351.1102369}
\showDOI{\tempurl}


\bibitem[Dang et~al\mbox{.}(2022)]%
        {dang2022prompt}
\bibfield{author}{\bibinfo{person}{Hai Dang}, \bibinfo{person}{Lukas Mecke},
  \bibinfo{person}{Florian Lehmann}, \bibinfo{person}{Sven Goller}, {and}
  \bibinfo{person}{Daniel Buschek}.} \bibinfo{year}{2022}\natexlab{}.
\newblock \bibinfo{title}{How to Prompt? Opportunities and Challenges of Zero-
  and Few-Shot Learning for Human-AI Interaction in Creative Applications of
  Generative Models}.
\newblock
\newblock
\urldef\tempurl%
\url{https://doi.org/10.48550/arXiv.2209.01390}
\showDOI{\tempurl}


\bibitem[Daulton et~al\mbox{.}(2020)]%
        {daulton2020differentiable}
\bibfield{author}{\bibinfo{person}{Samuel Daulton}, \bibinfo{person}{Maximilian
  Balandat}, {and} \bibinfo{person}{Eytan Bakshy}.}
  \bibinfo{year}{2020}\natexlab{}.
\newblock \showarticletitle{Differentiable Expected Hypervolume Improvement for
  Parallel Multi-Objective Bayesian Optimization}. In
  \bibinfo{booktitle}{\emph{Advances in Neural Information Processing
  Systems}}, Vol.~\bibinfo{volume}{33}. \bibinfo{publisher}{Curran Associates,
  Inc.}, \bibinfo{address}{Virtual Conference}, \bibinfo{pages}{9851–9864}.
\newblock
\urldef\tempurl%
\url{https://proceedings.neurips.cc/paper/2020/hash/6fec24eac8f18ed793f5eaad3dd7977c-Abstract.html}
\showURL{%
\tempurl}


\bibitem[Debreu(1960)]%
        {debreu1960individual}
\bibfield{author}{\bibinfo{person}{Gerard Debreu}.}
  \bibinfo{year}{1960}\natexlab{}.
\newblock \bibinfo{title}{Individual choice behavior: A theoretical analysis}.
\newblock
\newblock


\bibitem[Ding et~al\mbox{.}(2018)]%
        {ding2018human}
\bibfield{author}{\bibinfo{person}{Ye Ding}, \bibinfo{person}{Myunghee Kim},
  \bibinfo{person}{Scott Kuindersma}, {and} \bibinfo{person}{Conor~J. Walsh}.}
  \bibinfo{year}{2018}\natexlab{}.
\newblock \showarticletitle{Human-in-the-loop optimization of hip assistance
  with a soft exosuit during walking}.
\newblock \bibinfo{journal}{\emph{Science Robotics}} \bibinfo{volume}{3},
  \bibinfo{number}{15} (\bibinfo{year}{2018}), \bibinfo{pages}{eaar5438}.
\newblock
\urldef\tempurl%
\url{https://doi.org/10.1126/scirobotics.aar5438}
\showDOI{\tempurl}


\bibitem[Dreier(1996)]%
        {dreier1996rational}
\bibfield{author}{\bibinfo{person}{James Dreier}.}
  \bibinfo{year}{1996}\natexlab{}.
\newblock \showarticletitle{Rational preference: Decision theory as a theory of
  practical rationality}.
\newblock \bibinfo{journal}{\emph{Theory and Decision}} \bibinfo{volume}{40},
  \bibinfo{number}{3} (\bibinfo{date}{May} \bibinfo{year}{1996}),
  \bibinfo{pages}{249–276}.
\newblock
\showISSN{1573-7187}
\urldef\tempurl%
\url{https://doi.org/10.1007/BF00134210}
\showDOI{\tempurl}


\bibitem[Dudley et~al\mbox{.}(2019)]%
        {dudley2019crowdsourcing}
\bibfield{author}{\bibinfo{person}{John~J. Dudley}, \bibinfo{person}{Jason~T.
  Jacques}, {and} \bibinfo{person}{Per~Ola Kristensson}.}
  \bibinfo{year}{2019}\natexlab{}.
\newblock \showarticletitle{Crowdsourcing Interface Feature Design with
  Bayesian Optimization}. In \bibinfo{booktitle}{\emph{Proceedings of the 2019
  CHI Conference on Human Factors in Computing Systems}} (Glasgow, Scotland Uk)
  \emph{(\bibinfo{series}{CHI '19})}. \bibinfo{publisher}{Association for
  Computing Machinery}, \bibinfo{address}{New York, NY, USA},
  \bibinfo{pages}{1–12}.
\newblock
\showISBNx{9781450359702}
\urldef\tempurl%
\url{https://doi.org/10.1145/3290605.3300482}
\showDOI{\tempurl}


\bibitem[Dudley and Kristensson(2018)]%
        {dudley2018imlui}
\bibfield{author}{\bibinfo{person}{John~J. Dudley} {and}
  \bibinfo{person}{Per~Ola Kristensson}.} \bibinfo{year}{2018}\natexlab{}.
\newblock \showarticletitle{A Review of User Interface Design for Interactive
  Machine Learning}.
\newblock \bibinfo{journal}{\emph{ACM Transactions on Interactive Intelligent
  Systems}} \bibinfo{volume}{8}, \bibinfo{number}{2} (\bibinfo{date}{Jun}
  \bibinfo{year}{2018}), \bibinfo{pages}{1–37}.
\newblock
\showISSN{2160-6455, 2160-6463}
\urldef\tempurl%
\url{https://doi.org/10.1145/3185517}
\showDOI{\tempurl}


\bibitem[Eagly and Chaiken(1993)]%
        {eagly1993the}
\bibfield{author}{\bibinfo{person}{Alice~H. Eagly} {and}
  \bibinfo{person}{Shelly Chaiken}.} \bibinfo{year}{1993}\natexlab{}.
\newblock \bibinfo{booktitle}{\emph{The psychology of attitudes}}.
\newblock \bibinfo{publisher}{Harcourt Brace Jovanovich College Publishers},
  \bibinfo{address}{Orlando, FL, US}. xxii, 794 pages.
\newblock
\showISBNx{978-0-15-500097-1}


\bibitem[Funke and Reips(2012)]%
        {funke2012why}
\bibfield{author}{\bibinfo{person}{Frederik Funke} {and}
  \bibinfo{person}{Ulf-Dietrich Reips}.} \bibinfo{year}{2012}\natexlab{}.
\newblock \showarticletitle{Why Semantic Differentials in Web-Based Research
  Should Be Made from Visual Analogue Scales and Not from 5-Point Scales}.
\newblock \bibinfo{journal}{\emph{Field Methods}} \bibinfo{volume}{24},
  \bibinfo{number}{3} (\bibinfo{year}{2012}), \bibinfo{pages}{310--327}.
\newblock
\urldef\tempurl%
\url{https://doi.org/10.1177/1525822X12444061}
\showDOI{\tempurl}


\bibitem[Gonz\'{a}lez et~al\mbox{.}(2017)]%
        {gonzalez2017pbo}
\bibfield{author}{\bibinfo{person}{Javier Gonz\'{a}lez},
  \bibinfo{person}{Zhenwen Dai}, \bibinfo{person}{Andreas Damianou}, {and}
  \bibinfo{person}{Neil~D. Lawrence}.} \bibinfo{year}{2017}\natexlab{}.
\newblock \showarticletitle{Preferential Bayesian Optimization}. In
  \bibinfo{booktitle}{\emph{Proceedings of the 34th International Conference on
  Machine Learning - Volume 70}} (Sydney, NSW, Australia)
  \emph{(\bibinfo{series}{ICML'17})}. \bibinfo{publisher}{JMLR.org},
  \bibinfo{address}{Australia}, \bibinfo{pages}{1282–1291}.
\newblock
\urldef\tempurl%
\url{https://doi.org/10.5555/3305381.3305514}
\showURL{%
\tempurl}


\bibitem[Greis et~al\mbox{.}(2017)]%
        {greis2017input}
\bibfield{author}{\bibinfo{person}{Miriam Greis}, \bibinfo{person}{Hendrik
  Schuff}, \bibinfo{person}{Marius Kleiner}, \bibinfo{person}{Niels Henze},
  {and} \bibinfo{person}{Albrecht Schmidt}.} \bibinfo{year}{2017}\natexlab{}.
\newblock \showarticletitle{Input Controls for Entering Uncertain Data:
  Probability Distribution Sliders}.
\newblock \bibinfo{journal}{\emph{Proc. ACM Hum.-Comput. Interact.}}
  \bibinfo{volume}{1}, \bibinfo{number}{EICS}, Article \bibinfo{articleno}{3}
  (\bibinfo{date}{jun} \bibinfo{year}{2017}), \bibinfo{numpages}{17}~pages.
\newblock
\urldef\tempurl%
\url{https://doi.org/10.1145/3095805}
\showDOI{\tempurl}


\bibitem[Holtzman et~al\mbox{.}(2020)]%
        {holtzman2020topp}
\bibfield{author}{\bibinfo{person}{Ari Holtzman}, \bibinfo{person}{Jan Buys},
  \bibinfo{person}{Li Du}, \bibinfo{person}{Maxwell Forbes}, {and}
  \bibinfo{person}{Yejin Choi}.} \bibinfo{year}{2020}\natexlab{}.
\newblock \bibinfo{title}{The Curious Case of Neural Text Degeneration}.
\newblock
\newblock
\urldef\tempurl%
\url{http://arxiv.org/abs/1904.09751}
\showURL{%
\tempurl}


\bibitem[Kahneman et~al\mbox{.}(2021)]%
        {kahneman2021noise}
\bibfield{author}{\bibinfo{person}{Daniel Kahneman}, \bibinfo{person}{Olivier
  Sibony}, {and} \bibinfo{person}{Cass~R. Sunstein}.}
  \bibinfo{year}{2021}\natexlab{}.
\newblock \bibinfo{booktitle}{\emph{Noise}}.
\newblock \bibinfo{publisher}{HarperCollins UK}, \bibinfo{address}{UK}.
\newblock
\showISBNx{978-0-00-830901-5}


\bibitem[Kalloori et~al\mbox{.}(2018)]%
        {kalloori2018eliciting}
\bibfield{author}{\bibinfo{person}{Saikishore Kalloori},
  \bibinfo{person}{Francesco Ricci}, {and} \bibinfo{person}{Rosella Gennari}.}
  \bibinfo{year}{2018}\natexlab{}.
\newblock \showarticletitle{Eliciting pairwise preferences in recommender
  systems}. In \bibinfo{booktitle}{\emph{Proceedings of the 12th ACM Conference
  on Recommender Systems}}. \bibinfo{publisher}{ACM},
  \bibinfo{address}{Vancouver British Columbia Canada},
  \bibinfo{pages}{329–337}.
\newblock
\showISBNx{978-1-4503-5901-6}
\urldef\tempurl%
\url{https://doi.org/10.1145/3240323.3240364}
\showDOI{\tempurl}


\bibitem[Kim et~al\mbox{.}(2017)]%
        {kim2017human}
\bibfield{author}{\bibinfo{person}{Myunghee Kim}, \bibinfo{person}{Ye Ding},
  \bibinfo{person}{Philippe Malcolm}, \bibinfo{person}{Jozefien Speeckaert},
  \bibinfo{person}{Christoper~J. Siviy}, \bibinfo{person}{Conor~J. Walsh},
  {and} \bibinfo{person}{Scott Kuindersma}.} \bibinfo{year}{2017}\natexlab{}.
\newblock \showarticletitle{Human-in-the-loop Bayesian optimization of wearable
  device parameters}.
\newblock \bibinfo{journal}{\emph{PLOS ONE}} \bibinfo{volume}{12},
  \bibinfo{number}{9} (\bibinfo{date}{09} \bibinfo{year}{2017}),
  \bibinfo{pages}{1--15}.
\newblock
\urldef\tempurl%
\url{https://doi.org/10.1371/journal.pone.0184054}
\showDOI{\tempurl}


\bibitem[Koyama et~al\mbox{.}(2014)]%
        {koyama2014crowd}
\bibfield{author}{\bibinfo{person}{Yuki Koyama}, \bibinfo{person}{Daisuke
  Sakamoto}, {and} \bibinfo{person}{Takeo Igarashi}.}
  \bibinfo{year}{2014}\natexlab{}.
\newblock \showarticletitle{Crowd-powered parameter analysis for visual design
  exploration}. In \bibinfo{booktitle}{\emph{Proceedings of the 27th annual
  {ACM} symposium on {User} interface software and technology}}
  \emph{(\bibinfo{series}{{UIST} '14})}. \bibinfo{publisher}{ACM Press},
  \bibinfo{address}{New York, NY, USA}, \bibinfo{pages}{65--74}.
\newblock
\showISBNx{978-1-4503-3069-5}
\urldef\tempurl%
\url{https://doi.org/10.1145/2642918.2647386}
\showURL{%
\tempurl}


\bibitem[Koyama et~al\mbox{.}(2016)]%
        {koyama2016selph}
\bibfield{author}{\bibinfo{person}{Yuki Koyama}, \bibinfo{person}{Daisuke
  Sakamoto}, {and} \bibinfo{person}{Takeo Igarashi}.}
  \bibinfo{year}{2016}\natexlab{}.
\newblock \showarticletitle{SelPh: Progressive Learning and Support of Manual
  Photo Color Enhancement}. In \bibinfo{booktitle}{\emph{Proceedings of the
  2016 CHI Conference on Human Factors in Computing Systems}}.
  \bibinfo{publisher}{ACM}, \bibinfo{address}{San Jose California USA},
  \bibinfo{pages}{2520–2532}.
\newblock
\showISBNx{978-1-4503-3362-7}
\urldef\tempurl%
\url{https://doi.org/10.1145/2858036.2858111}
\showDOI{\tempurl}


\bibitem[Koyama et~al\mbox{.}(2020)]%
        {koyama2020sequential}
\bibfield{author}{\bibinfo{person}{Yuki Koyama}, \bibinfo{person}{Issei Sato},
  {and} \bibinfo{person}{Masataka Goto}.} \bibinfo{year}{2020}\natexlab{}.
\newblock \showarticletitle{Sequential gallery for interactive visual design
  optimization}.
\newblock \bibinfo{journal}{\emph{ACM Transactions on Graphics}}
  \bibinfo{volume}{39}, \bibinfo{number}{4} (\bibinfo{date}{Jul}
  \bibinfo{year}{2020}), \bibinfo{pages}{88:88:1--88:88:12}.
\newblock
\showISSN{0730-0301}
\urldef\tempurl%
\url{https://doi.org/10.1145/3386569.3392444}
\showDOI{\tempurl}


\bibitem[Koyama et~al\mbox{.}(2017)]%
        {koyama2017sequential}
\bibfield{author}{\bibinfo{person}{Yuki Koyama}, \bibinfo{person}{Issei Sato},
  \bibinfo{person}{Daisuke Sakamoto}, {and} \bibinfo{person}{Takeo Igarashi}.}
  \bibinfo{year}{2017}\natexlab{}.
\newblock \showarticletitle{Sequential line search for efficient visual design
  optimization by crowds}.
\newblock \bibinfo{journal}{\emph{ACM Transactions on Graphics}}
  \bibinfo{volume}{36}, \bibinfo{number}{4} (\bibinfo{date}{Jul}
  \bibinfo{year}{2017}), \bibinfo{pages}{48:1--48:11}.
\newblock
\showISSN{0730-0301}
\urldef\tempurl%
\url{https://doi.org/10.1145/3072959.3073598}
\showDOI{\tempurl}


\bibitem[Kushner(1964)]%
        {kushner1964bo}
\bibfield{author}{\bibinfo{person}{H.~J. Kushner}.}
  \bibinfo{year}{1964}\natexlab{}.
\newblock \showarticletitle{A New Method of Locating the Maximum Point of an
  Arbitrary Multipeak Curve in the Presence of Noise}.
\newblock \bibinfo{journal}{\emph{Journal of Basic Engineering}}
  \bibinfo{volume}{86}, \bibinfo{number}{1} (\bibinfo{date}{Mar}
  \bibinfo{year}{1964}), \bibinfo{pages}{97–106}.
\newblock
\showISSN{0021-9223}
\urldef\tempurl%
\url{https://doi.org/10.1115/1.3653121}
\showDOI{\tempurl}


\bibitem[Kuznetsova et~al\mbox{.}(2017)]%
        {kuznetsova2017lmertest}
\bibfield{author}{\bibinfo{person}{Alexandra Kuznetsova},
  \bibinfo{person}{Per~B. Brockhoff}, {and} \bibinfo{person}{Rune H.~B.
  Christensen}.} \bibinfo{year}{2017}\natexlab{}.
\newblock \showarticletitle{{lmerTest} Package: Tests in Linear Mixed Effects
  Models}.
\newblock \bibinfo{journal}{\emph{Journal of Statistical Software}}
  \bibinfo{volume}{82}, \bibinfo{number}{13} (\bibinfo{year}{2017}),
  \bibinfo{pages}{1--26}.
\newblock
\urldef\tempurl%
\url{https://doi.org/10.18637/jss.v082.i13}
\showDOI{\tempurl}


\bibitem[Letham et~al\mbox{.}(2019)]%
        {letham2019constrained}
\bibfield{author}{\bibinfo{person}{Benjamin Letham}, \bibinfo{person}{Brian
  Karrer}, \bibinfo{person}{Guilherme Ottoni}, {and} \bibinfo{person}{Eytan
  Bakshy}.} \bibinfo{year}{2019}\natexlab{}.
\newblock \showarticletitle{Constrained Bayesian Optimization with Noisy
  Experiments}.
\newblock \bibinfo{journal}{\emph{Bayesian Analysis}} \bibinfo{volume}{14},
  \bibinfo{number}{2} (\bibinfo{date}{Jun} \bibinfo{year}{2019}),
  \bibinfo{pages}{495–519}.
\newblock
\showISSN{1936-0975, 1931-6690}
\urldef\tempurl%
\url{https://doi.org/10.1214/18-BA1110}
\showDOI{\tempurl}


\bibitem[Lewis et~al\mbox{.}(2019)]%
        {lewis2019bart}
\bibfield{author}{\bibinfo{person}{Mike Lewis}, \bibinfo{person}{Yinhan Liu},
  \bibinfo{person}{Naman Goyal}, \bibinfo{person}{Marjan Ghazvininejad},
  \bibinfo{person}{Abdelrahman Mohamed}, \bibinfo{person}{Omer Levy},
  \bibinfo{person}{Veselin Stoyanov}, {and} \bibinfo{person}{Luke
  Zettlemoyer}.} \bibinfo{year}{2019}\natexlab{}.
\newblock \showarticletitle{{BART:} Denoising Sequence-to-Sequence Pre-training
  for Natural Language Generation, Translation, and Comprehension}.
\newblock \bibinfo{journal}{\emph{CoRR}}  \bibinfo{volume}{abs/1910.13461}
  (\bibinfo{year}{2019}), \bibinfo{numpages}{1 - 10}~pages.
\newblock
\showeprint[arXiv]{1910.13461}
\urldef\tempurl%
\url{http://arxiv.org/abs/1910.13461}
\showURL{%
\tempurl}


\bibitem[Lin(2004)]%
        {lin2004rouge}
\bibfield{author}{\bibinfo{person}{Chin-Yew Lin}.}
  \bibinfo{year}{2004}\natexlab{}.
\newblock \showarticletitle{ROUGE: A Package for Automatic Evaluation of
  Summaries}. In \bibinfo{booktitle}{\emph{Text Summarization Branches Out}}.
  \bibinfo{publisher}{Association for Computational Linguistics},
  \bibinfo{address}{Barcelona, Spain}, \bibinfo{pages}{74–81}.
\newblock
\urldef\tempurl%
\url{https://aclanthology.org/W04-1013}
\showURL{%
\tempurl}


\bibitem[Lin et~al\mbox{.}(2022)]%
        {lin2022preference}
\bibfield{author}{\bibinfo{person}{Jerry~Zhiyuan Lin}, \bibinfo{person}{Raul
  Astudillo}, \bibinfo{person}{Peter Frazier}, {and} \bibinfo{person}{Eytan
  Bakshy}.} \bibinfo{year}{2022}\natexlab{}.
\newblock \showarticletitle{Preference Exploration for Efficient Bayesian
  Optimization with Multiple Outcomes}. In
  \bibinfo{booktitle}{\emph{Proceedings of The 25th International Conference on
  Artificial Intelligence and Statistics}} \emph{(\bibinfo{series}{Proceedings
  of Machine Learning Research}, Vol.~\bibinfo{volume}{151})},
  \bibfield{editor}{\bibinfo{person}{Gustau Camps-Valls},
  \bibinfo{person}{Francisco J.~R. Ruiz}, {and} \bibinfo{person}{Isabel
  Valera}} (Eds.). \bibinfo{publisher}{PMLR}, \bibinfo{address}{Virtual
  Conference}, \bibinfo{pages}{4235--4258}.
\newblock
\urldef\tempurl%
\url{https://proceedings.mlr.press/v151/jerry-lin22a.html}
\showURL{%
\tempurl}


\bibitem[Lo and Andrews(2015)]%
        {lo2015to}
\bibfield{author}{\bibinfo{person}{Steson Lo} {and} \bibinfo{person}{Sally
  Andrews}.} \bibinfo{year}{2015}\natexlab{}.
\newblock \showarticletitle{To transform or not to transform: Using generalized
  linear mixed models to analyse reaction time data}.
\newblock \bibinfo{journal}{\emph{Frontiers in psychology}}
  \bibinfo{volume}{6} (\bibinfo{year}{2015}), \bibinfo{pages}{1171}.
\newblock


\bibitem[Luk(1994)]%
        {luk1994multirank}
\bibfield{author}{\bibinfo{person}{Wai-Lan Luk}.}
  \bibinfo{year}{1994}\natexlab{}.
\newblock \emph{\bibinfo{title}{Multi-user interface for group ranking: a
  user-centered approach}}.
\newblock \bibinfo{thesistype}{Ph.\,D. Dissertation}.
  \bibinfo{school}{University of British Columbia}.
\newblock
\urldef\tempurl%
\url{https://doi.org/10.14288/1.0087518}
\showDOI{\tempurl}


\bibitem[Marks et~al\mbox{.}(1997)]%
        {marks1997design-galleries}
\bibfield{author}{\bibinfo{person}{Joe Marks}, \bibinfo{person}{Brad Andalman},
  \bibinfo{person}{Paul~A. Beardsley}, \bibinfo{person}{William Freeman},
  \bibinfo{person}{Sarah Gibson}, \bibinfo{person}{Jessica Hodgins},
  \bibinfo{person}{T. Kang}, \bibinfo{person}{Brian Mirtich},
  \bibinfo{person}{Hanspeter Pfister}, \bibinfo{person}{Wheeler Ruml},
  \bibinfo{person}{Kathy Ryall}, \bibinfo{person}{Joshua Seims}, {and}
  \bibinfo{person}{Stuart Shieber}.} \bibinfo{year}{1997}\natexlab{}.
\newblock \showarticletitle{{Design Galleries: A General Approach to Setting
  Parameters for Computer Graphics and Animation}}. In
  \bibinfo{booktitle}{\emph{Proceedings of the 24th annual conference on
  {Computer} graphics and interactive techniques}}
  \emph{(\bibinfo{series}{{SIGGRAPH} '97})}. \bibinfo{publisher}{ACM
  Press/Addison-Wesley Publishing Co.}, \bibinfo{address}{USA},
  \bibinfo{pages}{389--400}.
\newblock
\showISBNx{978-0-89791-896-1}
\urldef\tempurl%
\url{https://doi.org/10.1145/258734.258887}
\showURL{%
\tempurl}


\bibitem[Matejka et~al\mbox{.}(2016)]%
        {matejka2016effect}
\bibfield{author}{\bibinfo{person}{Justin Matejka}, \bibinfo{person}{Michael
  Glueck}, \bibinfo{person}{Tovi Grossman}, {and} \bibinfo{person}{George
  Fitzmaurice}.} \bibinfo{year}{2016}\natexlab{}.
\newblock \showarticletitle{The Effect of Visual Appearance on the Performance
  of Continuous Sliders and Visual Analogue Scales}. In
  \bibinfo{booktitle}{\emph{Proceedings of the 2016 CHI Conference on Human
  Factors in Computing Systems}} (San Jose, California, USA)
  \emph{(\bibinfo{series}{CHI '16})}. \bibinfo{publisher}{Association for
  Computing Machinery}, \bibinfo{address}{New York, NY, USA},
  \bibinfo{pages}{5421–5432}.
\newblock
\showISBNx{9781450333627}
\urldef\tempurl%
\url{https://doi.org/10.1145/2858036.2858063}
\showDOI{\tempurl}


\bibitem[Mikkola et~al\mbox{.}(2020)]%
        {mikkola2020projective}
\bibfield{author}{\bibinfo{person}{Petrus Mikkola}, \bibinfo{person}{Milica
  Todorovi{\'c}}, \bibinfo{person}{Jari J{\"a}rvi}, \bibinfo{person}{Patrick
  Rinke}, {and} \bibinfo{person}{Samuel Kaski}.}
  \bibinfo{year}{2020}\natexlab{}.
\newblock \showarticletitle{Projective Preferential Bayesian Optimization}. In
  \bibinfo{booktitle}{\emph{International Conference on Machine Learning}}.
  PMLR, \bibinfo{publisher}{MLResearchPress}, \bibinfo{address}{Online},
  \bibinfo{pages}{6884--6892}.
\newblock


\bibitem[Mockus et~al\mbox{.}(1978)]%
        {mockus1978application}
\bibfield{author}{\bibinfo{person}{Jonas Mockus}, \bibinfo{person}{Vytautas
  Tiesis}, {and} \bibinfo{person}{Antanas Zilinskas}.}
  \bibinfo{year}{1978}\natexlab{}.
\newblock \showarticletitle{The application of Bayesian methods for seeking the
  extremum}.
\newblock \bibinfo{journal}{\emph{Towards global optimization}}
  \bibinfo{volume}{2}, \bibinfo{number}{117-129} (\bibinfo{year}{1978}),
  \bibinfo{pages}{2}.
\newblock


\bibitem[Monarch(2021)]%
        {monarch2021hitl}
\bibfield{author}{\bibinfo{person}{Robert~M. Monarch}.}
  \bibinfo{year}{2021}\natexlab{}.
\newblock \bibinfo{booktitle}{\emph{Human-in-the-Loop Machine Learning: Active
  Learning and Annotation for Human-centered AI}}.
\newblock \bibinfo{publisher}{Simon and Schuster}, \bibinfo{address}{USA}.
\newblock
\showISBNx{978-1-61729-674-1}


\bibitem[Močkus(1975)]%
        {močkus1975on}
\bibfield{author}{\bibinfo{person}{J. Močkus}.}
  \bibinfo{year}{1975}\natexlab{}.
\newblock \bibinfo{booktitle}{\emph{On Bayesian Methods for Seeking the
  Extremum}}.
\newblock \bibinfo{publisher}{Springer}, \bibinfo{address}{Berlin, Heidelberg},
  \bibinfo{pages}{400–404}.
\newblock
\showISBNx{978-3-662-38527-2}
\urldef\tempurl%
\url{https://doi.org/10.1007/978-3-662-38527-2_55}
\showDOI{\tempurl}


\bibitem[Neubarth(2010)]%
        {neubarth2010dnd}
\bibfield{author}{\bibinfo{person}{Wolfgang Neubarth}.}
  \bibinfo{year}{2010}\natexlab{}.
\newblock \bibinfo{booktitle}{\emph{Drag \& drop: A flexible method for moving
  objects, implementing rankings, and a wide range of other applications}}.
\newblock \bibinfo{publisher}{American Psychological Association},
  \bibinfo{address}{Washington, DC, US}, \bibinfo{pages}{63–74}.
\newblock
\showISBNx{978-1-4338-0695-7}
\urldef\tempurl%
\url{https://doi.org/10.1037/12076-005}
\showDOI{\tempurl}


\bibitem[Nguyen et~al\mbox{.}(2021)]%
        {nguyen2021topk}
\bibfield{author}{\bibinfo{person}{Quoc~Phong Nguyen},
  \bibinfo{person}{Sebastian Tay}, \bibinfo{person}{Bryan Kian~Hsiang Low},
  {and} \bibinfo{person}{Patrick Jaillet}.} \bibinfo{year}{2021}\natexlab{}.
\newblock \showarticletitle{Top-k Ranking Bayesian Optimization}.
\newblock \bibinfo{journal}{\emph{Proceedings of the AAAI Conference on
  Artificial Intelligence}} \bibinfo{volume}{35}, \bibinfo{number}{1010}
  (\bibinfo{date}{May} \bibinfo{year}{2021}), \bibinfo{pages}{9135–9143}.
\newblock
\showISSN{2374-3468}


\bibitem[Nguyen et~al\mbox{.}(2013)]%
        {nguyen2013rating}
\bibfield{author}{\bibinfo{person}{Tien~T. Nguyen}, \bibinfo{person}{Daniel
  Kluver}, \bibinfo{person}{Ting-Yu Wang}, \bibinfo{person}{Pik-Mai Hui},
  \bibinfo{person}{Michael~D. Ekstrand}, \bibinfo{person}{Martijn~C.
  Willemsen}, {and} \bibinfo{person}{John Riedl}.}
  \bibinfo{year}{2013}\natexlab{}.
\newblock \showarticletitle{Rating support interfaces to improve user
  experience and recommender accuracy}. In
  \bibinfo{booktitle}{\emph{Proceedings of the 7th ACM conference on
  Recommender systems}} \emph{(\bibinfo{series}{RecSys ’13})}.
  \bibinfo{publisher}{Association for Computing Machinery},
  \bibinfo{address}{New York, NY, USA}, \bibinfo{pages}{149–156}.
\newblock
\showISBNx{978-1-4503-2409-0}
\urldef\tempurl%
\url{https://doi.org/10.1145/2507157.2507188}
\showDOI{\tempurl}


\bibitem[Nobarany et~al\mbox{.}(2012)]%
        {nobarany2012opinion}
\bibfield{author}{\bibinfo{person}{Syavash Nobarany}, \bibinfo{person}{Louise
  Oram}, \bibinfo{person}{Vasanth~Kumar Rajendran}, \bibinfo{person}{Chi-Hsiang
  Chen}, \bibinfo{person}{Joanna McGrenere}, {and} \bibinfo{person}{Tamara
  Munzner}.} \bibinfo{year}{2012}\natexlab{}.
\newblock \showarticletitle{The design space of opinion measurement interfaces:
  exploring recall support for rating and ranking}. In
  \bibinfo{booktitle}{\emph{Proceedings of the SIGCHI Conference on Human
  Factors in Computing Systems}} \emph{(\bibinfo{series}{CHI ’12})}.
  \bibinfo{publisher}{Association for Computing Machinery},
  \bibinfo{address}{New York, NY, USA}, \bibinfo{pages}{2035–2044}.
\newblock
\showISBNx{978-1-4503-1015-4}
\urldef\tempurl%
\url{https://doi.org/10.1145/2207676.2208351}
\showDOI{\tempurl}


\bibitem[Ok et~al\mbox{.}(2012)]%
        {ok2012incomplete}
\bibfield{author}{\bibinfo{person}{Efe~A. Ok}, \bibinfo{person}{Pietro
  Ortoleva}, {and} \bibinfo{person}{Gil Riella}.}
  \bibinfo{year}{2012}\natexlab{}.
\newblock \showarticletitle{Incomplete Preferences Under Uncertainty:
  Indecisiveness in Beliefs versus Tastes}.
\newblock \bibinfo{journal}{\emph{Econometrica}} \bibinfo{volume}{80},
  \bibinfo{number}{4} (\bibinfo{year}{2012}), \bibinfo{pages}{1791–1808}.
\newblock
\showISSN{1468-0262}
\urldef\tempurl%
\url{https://doi.org/10.3982/ECTA8040}
\showDOI{\tempurl}


\bibitem[Ou et~al\mbox{.}(2022)]%
        {ou2022infloop}
\bibfield{author}{\bibinfo{person}{Changkun Ou}, \bibinfo{person}{Daniel
  Buschek}, \bibinfo{person}{Sven Mayer}, {and} \bibinfo{person}{Andreas
  Butz}.} \bibinfo{year}{2022}\natexlab{}.
\newblock \showarticletitle{The Human in the Infinite Loop: A Case Study on
  Revealing and Explaining Human-AI Interaction Loop Failures}. In
  \bibinfo{booktitle}{\emph{Proceedings of Mensch Und Computer 2022}}
  (2022-01-01) \emph{(\bibinfo{series}{MuC\'22})}.
  \bibinfo{publisher}{Association for Computing Machinery},
  \bibinfo{address}{Darmstadt, Germany}, \bibinfo{numpages}{16}~pages.
\newblock
\urldef\tempurl%
\url{https://doi.org/10.1145/3543758.3543761}
\showDOI{\tempurl}


\bibitem[Ou et~al\mbox{.}(2023)]%
        {ou2023expertise}
\bibfield{author}{\bibinfo{person}{Changkun Ou}, \bibinfo{person}{Sven Mayer},
  {and} \bibinfo{person}{Andreas~Martin Butz}.}
  \bibinfo{year}{2023}\natexlab{}.
\newblock \showarticletitle{The Impact of Expertise in the Loop for Exploring
  Machine Rationality}. In \bibinfo{booktitle}{\emph{Proceedings of the 28th
  International Conference on Intelligent User Interfaces}} (Sydney, NSW,
  Australia) \emph{(\bibinfo{series}{IUI '23})}.
  \bibinfo{publisher}{Association for Computing Machinery},
  \bibinfo{address}{New York, NY, USA}, \bibinfo{pages}{307–321}.
\newblock
\showISBNx{9798400701061}
\urldef\tempurl%
\url{https://doi.org/10.1145/3581641.3584040}
\showDOI{\tempurl}


\bibitem[Overgaard(2006)]%
        {overgaard2006the}
\bibfield{author}{\bibinfo{person}{Søren Overgaard}.}
  \bibinfo{year}{2006}\natexlab{}.
\newblock \showarticletitle{The Problem of Other Minds: Wittgenstein’s
  Phenomenological Perspective}.
\newblock \bibinfo{journal}{\emph{Phenomenology and the Cognitive Sciences}}
  \bibinfo{volume}{5}, \bibinfo{number}{1} (\bibinfo{date}{Mar}
  \bibinfo{year}{2006}), \bibinfo{pages}{53–73}.
\newblock
\showISSN{1572-8676}
\urldef\tempurl%
\url{https://doi.org/10.1007/s11097-005-9014-7}
\showDOI{\tempurl}


\bibitem[Papineni et~al\mbox{.}(2002)]%
        {papineni2002bleu}
\bibfield{author}{\bibinfo{person}{Kishore Papineni}, \bibinfo{person}{Salim
  Roukos}, \bibinfo{person}{Todd Ward}, {and} \bibinfo{person}{Wei-Jing Zhu}.}
  \bibinfo{year}{2002}\natexlab{}.
\newblock \showarticletitle{Bleu: a Method for Automatic Evaluation of Machine
  Translation}. In \bibinfo{booktitle}{\emph{Proceedings of the 40th Annual
  Meeting of the Association for Computational Linguistics}}.
  \bibinfo{publisher}{Association for Computational Linguistics},
  \bibinfo{address}{Philadelphia, Pennsylvania, USA},
  \bibinfo{pages}{311–318}.
\newblock
\urldef\tempurl%
\url{https://doi.org/10.3115/1073083.1073135}
\showDOI{\tempurl}


\bibitem[Payne et~al\mbox{.}(1993)]%
        {payne1993the}
\bibfield{author}{\bibinfo{person}{John~W Payne}, \bibinfo{person}{John~William
  Payne}, \bibinfo{person}{James~R Bettman}, {and} \bibinfo{person}{Eric~J
  Johnson}.} \bibinfo{year}{1993}\natexlab{}.
\newblock \bibinfo{booktitle}{\emph{The adaptive decision maker}}.
\newblock \bibinfo{publisher}{Cambridge university press},
  \bibinfo{address}{Cambridge, UK}.
\newblock


\bibitem[Pfannschmidt et~al\mbox{.}(2022)]%
        {pfannschmidt2022learning}
\bibfield{author}{\bibinfo{person}{Karlson Pfannschmidt},
  \bibinfo{person}{Pritha Gupta}, \bibinfo{person}{Björn Haddenhorst}, {and}
  \bibinfo{person}{Eyke Hüllermeier}.} \bibinfo{year}{2022}\natexlab{}.
\newblock \showarticletitle{Learning context-dependent choice functions}.
\newblock \bibinfo{journal}{\emph{International Journal of Approximate
  Reasoning}}  \bibinfo{volume}{140} (\bibinfo{date}{Jan}
  \bibinfo{year}{2022}), \bibinfo{pages}{116–155}.
\newblock
\showISSN{0888-613X}
\urldef\tempurl%
\url{https://doi.org/10.1016/j.ijar.2021.10.002}
\showDOI{\tempurl}


\bibitem[Razali et~al\mbox{.}(2011)]%
        {razali2011power}
\bibfield{author}{\bibinfo{person}{Nornadiah~Mohd Razali},
  \bibinfo{person}{Yap~Bee Wah}, {et~al\mbox{.}}}
  \bibinfo{year}{2011}\natexlab{}.
\newblock \showarticletitle{Power comparisons of shapiro-wilk,
  kolmogorov-smirnov, lilliefors and anderson-darling tests}.
\newblock \bibinfo{journal}{\emph{Journal of statistical modeling and
  analytics}} \bibinfo{volume}{2}, \bibinfo{number}{1} (\bibinfo{year}{2011}),
  \bibinfo{pages}{21--33}.
\newblock


\bibitem[Rezende et~al\mbox{.}(2014)]%
        {rezende2014stochastic}
\bibfield{author}{\bibinfo{person}{Danilo~Jimenez Rezende},
  \bibinfo{person}{Shakir Mohamed}, {and} \bibinfo{person}{Daan Wierstra}.}
  \bibinfo{year}{2014}\natexlab{}.
\newblock \showarticletitle{Stochastic Backpropagation and Approximate
  Inference in Deep Generative Models}. In
  \bibinfo{booktitle}{\emph{Proceedings of the 31st International Conference on
  Machine Learning}}. \bibinfo{publisher}{PMLR}, \bibinfo{address}{Beijing,
  China}, \bibinfo{pages}{1278–1286}.
\newblock
\showISSN{1938-7228}
\urldef\tempurl%
\url{https://proceedings.mlr.press/v32/rezende14.html}
\showURL{%
\tempurl}


\bibitem[Rombach et~al\mbox{.}(2022)]%
        {rombach2022stablediffusion}
\bibfield{author}{\bibinfo{person}{Robin Rombach}, \bibinfo{person}{Andreas
  Blattmann}, \bibinfo{person}{Dominik Lorenz}, \bibinfo{person}{Patrick
  Esser}, {and} \bibinfo{person}{Bj\"orn Ommer}.}
  \bibinfo{year}{2022}\natexlab{}.
\newblock \showarticletitle{High-Resolution Image Synthesis With Latent
  Diffusion Models}. In \bibinfo{booktitle}{\emph{Proceedings of the IEEE/CVF
  Conference on Computer Vision and Pattern Recognition (CVPR)}}.
  \bibinfo{publisher}{IEEE}, \bibinfo{address}{New Orleans, USA},
  \bibinfo{pages}{10684--10695}.
\newblock
\urldef\tempurl%
\url{https://doi.org/10.48550/arXiv.2112.10752}
\showDOI{\tempurl}


\bibitem[Rouder et~al\mbox{.}(2009)]%
        {rouder2009bayesian}
\bibfield{author}{\bibinfo{person}{Jeffrey~N. Rouder}, \bibinfo{person}{Paul~L.
  Speckman}, \bibinfo{person}{Dongchu Sun}, \bibinfo{person}{Richard~D. Morey},
  {and} \bibinfo{person}{Geoffrey Iverson}.} \bibinfo{year}{2009}\natexlab{}.
\newblock \showarticletitle{Bayesian t tests for accepting and rejecting the
  null hypothesis}.
\newblock \bibinfo{journal}{\emph{Psychonomic Bulletin \& Review}}
  \bibinfo{volume}{16}, \bibinfo{number}{2} (\bibinfo{date}{Apr}
  \bibinfo{year}{2009}), \bibinfo{pages}{225–237}.
\newblock
\showISSN{1531-5320}
\urldef\tempurl%
\url{https://doi.org/10.3758/PBR.16.2.225}
\showDOI{\tempurl}


\bibitem[Russell(2019)]%
        {russell2019human}
\bibfield{author}{\bibinfo{person}{Stuart Russell}.}
  \bibinfo{year}{2019}\natexlab{}.
\newblock \bibinfo{booktitle}{\emph{Human Compatible: AI and the Problem of
  Control}}.
\newblock \bibinfo{publisher}{Penguin UK}, \bibinfo{address}{UK}.
\newblock
\showISBNx{978-0-241-33524-6}


\bibitem[Shahriari et~al\mbox{.}(2016)]%
        {shahriari2016bo-review}
\bibfield{author}{\bibinfo{person}{B. Shahriari}, \bibinfo{person}{K. Swersky},
  \bibinfo{person}{Z. Wang}, \bibinfo{person}{R.~P. Adams}, {and}
  \bibinfo{person}{N.~de Freitas}.} \bibinfo{year}{2016}\natexlab{}.
\newblock \showarticletitle{Taking the {Human} {Out} of the {Loop}: {A}
  {Review} of {Bayesian} {Optimization}}.
\newblock \bibinfo{journal}{\emph{Proc. IEEE}} \bibinfo{volume}{104},
  \bibinfo{number}{1} (\bibinfo{date}{Jan.} \bibinfo{year}{2016}),
  \bibinfo{pages}{148--175}.
\newblock
\showISSN{1558-2256}
\urldef\tempurl%
\url{https://doi.org/10.1109/JPROC.2015.2494218}
\showURL{%
\tempurl}


\bibitem[Sherer et~al\mbox{.}(1998)]%
        {sherer1998the}
\bibfield{author}{\bibinfo{person}{Mark Sherer}, \bibinfo{person}{Paula
  Bergloff}, \bibinfo{person}{Corwin Boake}, \bibinfo{person}{Walter High~Jr},
  {and} \bibinfo{person}{Ellen Levin}.} \bibinfo{year}{1998}\natexlab{}.
\newblock \showarticletitle{The Awareness Questionnaire: factor structure and
  internal consistency}.
\newblock \bibinfo{journal}{\emph{Brain Injury}} \bibinfo{volume}{12},
  \bibinfo{number}{1} (\bibinfo{date}{Jan} \bibinfo{year}{1998}),
  \bibinfo{pages}{63–68}.
\newblock
\showISSN{0269-9052}
\urldef\tempurl%
\url{https://doi.org/10.1080/026990598122863}
\showDOI{\tempurl}


\bibitem[Siivola et~al\mbox{.}(2021)]%
        {siivola2021preferential}
\bibfield{author}{\bibinfo{person}{Eero Siivola}, \bibinfo{person}{Akash~Kumar
  Dhaka}, \bibinfo{person}{Michael~Riis Andersen}, \bibinfo{person}{Javier
  González}, \bibinfo{person}{Pablo~García Moreno}, {and}
  \bibinfo{person}{Aki Vehtari}.} \bibinfo{year}{2021}\natexlab{}.
\newblock \showarticletitle{Preferential Batch Bayesian Optimization}. In
  \bibinfo{booktitle}{\emph{2021 IEEE 31st International Workshop on Machine
  Learning for Signal Processing (MLSP)}}. \bibinfo{publisher}{IEEE},
  \bibinfo{address}{Gold Coast, Queensland, Australia}, \bibinfo{pages}{1–6}.
\newblock
\showISSN{1551-2541}
\urldef\tempurl%
\url{https://doi.org/10.1109/MLSP52302.2021.9596494}
\showDOI{\tempurl}


\bibitem[Simpson et~al\mbox{.}(2020)]%
        {simpson2020interactive}
\bibfield{author}{\bibinfo{person}{Edwin Simpson}, \bibinfo{person}{Yang Gao},
  {and} \bibinfo{person}{Iryna Gurevych}.} \bibinfo{year}{2020}\natexlab{}.
\newblock \showarticletitle{Interactive Text Ranking with Bayesian
  Optimization: A Case Study on Community QA and Summarization}.
\newblock \bibinfo{journal}{\emph{Transactions of the Association for
  Computational Linguistics}}  \bibinfo{volume}{8} (\bibinfo{date}{Dec}
  \bibinfo{year}{2020}), \bibinfo{pages}{759–775}.
\newblock
\showISSN{2307-387X}
\urldef\tempurl%
\url{https://doi.org/10.1162/tacl_a_00344}
\showDOI{\tempurl}


\bibitem[Skinner(2011)]%
        {skinner2011about}
\bibfield{author}{\bibinfo{person}{B.~F. Skinner}.}
  \bibinfo{year}{2011}\natexlab{}.
\newblock \bibinfo{booktitle}{\emph{About Behaviorism}}.
\newblock \bibinfo{publisher}{Knopf Doubleday Publishing Group},
  \bibinfo{address}{New York, NY, USA}.
\newblock
\showISBNx{978-0-307-79784-1}


\bibitem[Thurstone(1927)]%
        {thurstone1927comparelaw}
\bibfield{author}{\bibinfo{person}{L.~L. Thurstone}.}
  \bibinfo{year}{1927}\natexlab{}.
\newblock \showarticletitle{A law of comparative judgment}.
\newblock \bibinfo{journal}{\emph{Psychological Review}} \bibinfo{volume}{34},
  \bibinfo{number}{4} (\bibinfo{year}{1927}), \bibinfo{pages}{273--286}.
\newblock
\showISSN{1939-1471(Electronic),0033-295X(Print)}
\urldef\tempurl%
\url{https://doi.org/10.1037/h0070288}
\showURL{%
\tempurl}


\bibitem[Tversky and Kahneman(1974)]%
        {tversky1974judgment}
\bibfield{author}{\bibinfo{person}{Amos Tversky} {and} \bibinfo{person}{Daniel
  Kahneman}.} \bibinfo{year}{1974}\natexlab{}.
\newblock \showarticletitle{Judgment under Uncertainty: Heuristics and Biases}.
\newblock \bibinfo{journal}{\emph{Science}} \bibinfo{volume}{185},
  \bibinfo{number}{4157} (\bibinfo{date}{Sep} \bibinfo{year}{1974}),
  \bibinfo{pages}{1124–1131}.
\newblock
\showISSN{0036-8075, 1095-9203}
\urldef\tempurl%
\url{https://doi.org/10.1126/science.185.4157.1124}
\showDOI{\tempurl}


\bibitem[Wagenmakers(2007)]%
        {wagenmakers2007bf}
\bibfield{author}{\bibinfo{person}{Eric-Jan Wagenmakers}.}
  \bibinfo{year}{2007}\natexlab{}.
\newblock \showarticletitle{A practical solution to the pervasive problems ofp
  values}.
\newblock \bibinfo{journal}{\emph{Psychonomic Bulletin \& Review}}
  \bibinfo{volume}{14}, \bibinfo{number}{5} (\bibinfo{date}{Oct}
  \bibinfo{year}{2007}), \bibinfo{pages}{779–804}.
\newblock
\showISSN{1531-5320}
\urldef\tempurl%
\url{https://doi.org/10.3758/BF03194105}
\showDOI{\tempurl}


\bibitem[Welleck et~al\mbox{.}(2020)]%
        {welleck2020neural}
\bibfield{author}{\bibinfo{person}{Sean Welleck}, \bibinfo{person}{Ilia
  Kulikov}, \bibinfo{person}{Stephen Roller}, \bibinfo{person}{Emily Dinan},
  \bibinfo{person}{Kyunghyun Cho}, {and} \bibinfo{person}{Jason Weston}.}
  \bibinfo{year}{2020}\natexlab{}.
\newblock \showarticletitle{Neural Text Generation With Unlikelihood Training}.
  In \bibinfo{booktitle}{\emph{8th International Conference on Learning
  Representations, {ICLR} 2020, Addis Ababa, Ethiopia, April 26-30, 2020}}.
  \bibinfo{publisher}{OpenReview.net}, \bibinfo{address}{Virtual Conference},
  \bibinfo{numpages}{18}~pages.
\newblock
\urldef\tempurl%
\url{https://openreview.net/forum?id=SJeYe0NtvH}
\showURL{%
\tempurl}


\bibitem[Wilber et~al\mbox{.}(2014)]%
        {wilber2014cost}
\bibfield{author}{\bibinfo{person}{Michael Wilber}, \bibinfo{person}{Iljung
  Kwak}, {and} \bibinfo{person}{Serge Belongie}.}
  \bibinfo{year}{2014}\natexlab{}.
\newblock \showarticletitle{Cost-Effective HITs for Relative Similarity
  Comparisons}.
\newblock \bibinfo{journal}{\emph{Proceedings of the AAAI Conference on Human
  Computation and Crowdsourcing}} \bibinfo{volume}{2}, \bibinfo{number}{1}
  (\bibinfo{date}{Sep.} \bibinfo{year}{2014}), \bibinfo{pages}{227--233}.
\newblock
\urldef\tempurl%
\url{https://doi.org/10.1609/hcomp.v2i1.13152}
\showDOI{\tempurl}


\bibitem[Wilson et~al\mbox{.}(2017)]%
        {wilson2017reparam}
\bibfield{author}{\bibinfo{person}{James~T. Wilson}, \bibinfo{person}{Riccardo
  Moriconi}, \bibinfo{person}{Frank Hutter}, {and} \bibinfo{person}{Marc~Peter
  Deisenroth}.} \bibinfo{year}{2017}\natexlab{}.
\newblock \bibinfo{title}{The reparameterization trick for acquisition
  functions}.
\newblock
\newblock
\urldef\tempurl%
\url{https://doi.org/10.48550/arXiv.1712.00424}
\showDOI{\tempurl}


\bibitem[Yue et~al\mbox{.}(2021)]%
        {koyama2021ui}
\bibfield{author}{\bibinfo{person}{Yonghao Yue}, \bibinfo{person}{Yuki Koyama},
  \bibinfo{person}{Issei Sato}, {and} \bibinfo{person}{Takeo Igarashi}.}
  \bibinfo{year}{2021}\natexlab{}.
\newblock \showarticletitle{User interfaces for high-dimensional design
  problems: from theories to implementations}. In \bibinfo{booktitle}{\emph{ACM
  SIGGRAPH 2021 Courses}} \emph{(\bibinfo{series}{SIGGRAPH ’21})}.
  \bibinfo{publisher}{Association for Computing Machinery},
  \bibinfo{address}{New York, NY, USA}, \bibinfo{pages}{1–34}.
\newblock
\showISBNx{978-1-4503-8361-5}
\urldef\tempurl%
\url{https://doi.org/10.1145/3450508.3464551}
\showDOI{\tempurl}


\bibitem[Zhou et~al\mbox{.}(2021)]%
        {zhou2021interactive}
\bibfield{author}{\bibinfo{person}{Yijun Zhou}, \bibinfo{person}{Yuki Koyama},
  \bibinfo{person}{Masataka Goto}, {and} \bibinfo{person}{Takeo Igarashi}.}
  \bibinfo{year}{2021}\natexlab{}.
\newblock \showarticletitle{Interactive Exploration-Exploitation Balancing for
  Generative Melody Composition}. In \bibinfo{booktitle}{\emph{26th
  International Conference on Intelligent User Interfaces}}
  \emph{(\bibinfo{series}{IUI ’21})}. \bibinfo{publisher}{Association for
  Computing Machinery}, \bibinfo{address}{New York, NY, USA},
  \bibinfo{pages}{43–47}.
\newblock
\showISBNx{978-1-4503-8017-1}
\urldef\tempurl%
\url{https://doi.org/10.1145/3397481.3450663}
\showDOI{\tempurl}


\end{thebibliography}

\appendix
\section{Benchmark Validations for Implemented Bayesian Optimizers}
\label{appendix:benchmark}
\begin{figure}[H]
\centering
\includegraphics[width=\textwidth]{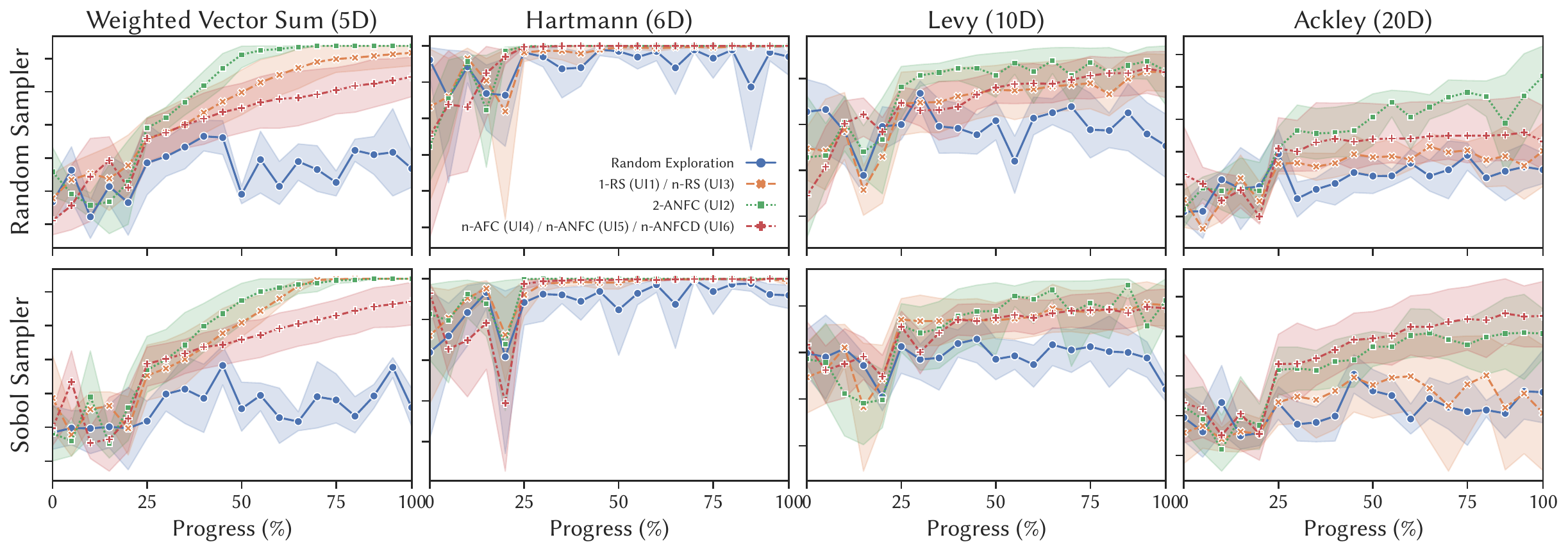}
\caption{The performance benchmark for validating our implemented Bayesian optimizers (for different UIs) compares to a random exploration (baseline). Higher y-axis utility values mean better exploitation performance. The upper row shows exploitation performance using a random sampler for the initial samples. The bottom row shows exploitation performance using a Sobol sampler. Each column shows the performance of a different synthetic function. The results validate our implemented BOs and outperform random exploration.}
\label{fig:listwise-bo}
\end{figure}
As described in \autoref{sec:bo-service}, for utility-based UIs ($1$-RS/$n$-RS, UI1/UI3), we implemented a utility-based BO; for preference-based UIs, we use EUBO~\cite{lin2022preference} that is designed for pairwise comparisons ($2$-ANFC, UI2); but also for all listwise interfaces ($n$-AFC/$n$-ANFC/$n$-ANFCD, UI4/UI5/UI6), to not introduce another experiment variable. We adopted EUBO so that the optimizer can infer a batch of outcomes based on ranking preferences.
To demonstrate the validity of our implemented Bayesian optimizer, we selected four standard synthetic functions\footnote{See \url{https://www.sfu.ca/~ssurjano/optimization.html} for exact definitions, \emph{last accessed \today}}: Weighted Vector Sum (5D), Hartmann (6D), Levy (10D), and Ackley (20D) for the benchmarks. The results in~\autoref{fig:listwise-bo} demonstrate that all of our implemented Bayesian optimizers outperform random exploration. Full implementation details can be found in the supplemental material.

\end{document}